THE GRAVITATIONAL WAVE INTERNATIONAL COMMITTEE ROADMAP

# The future of gravitational wave astronomy

GWIC

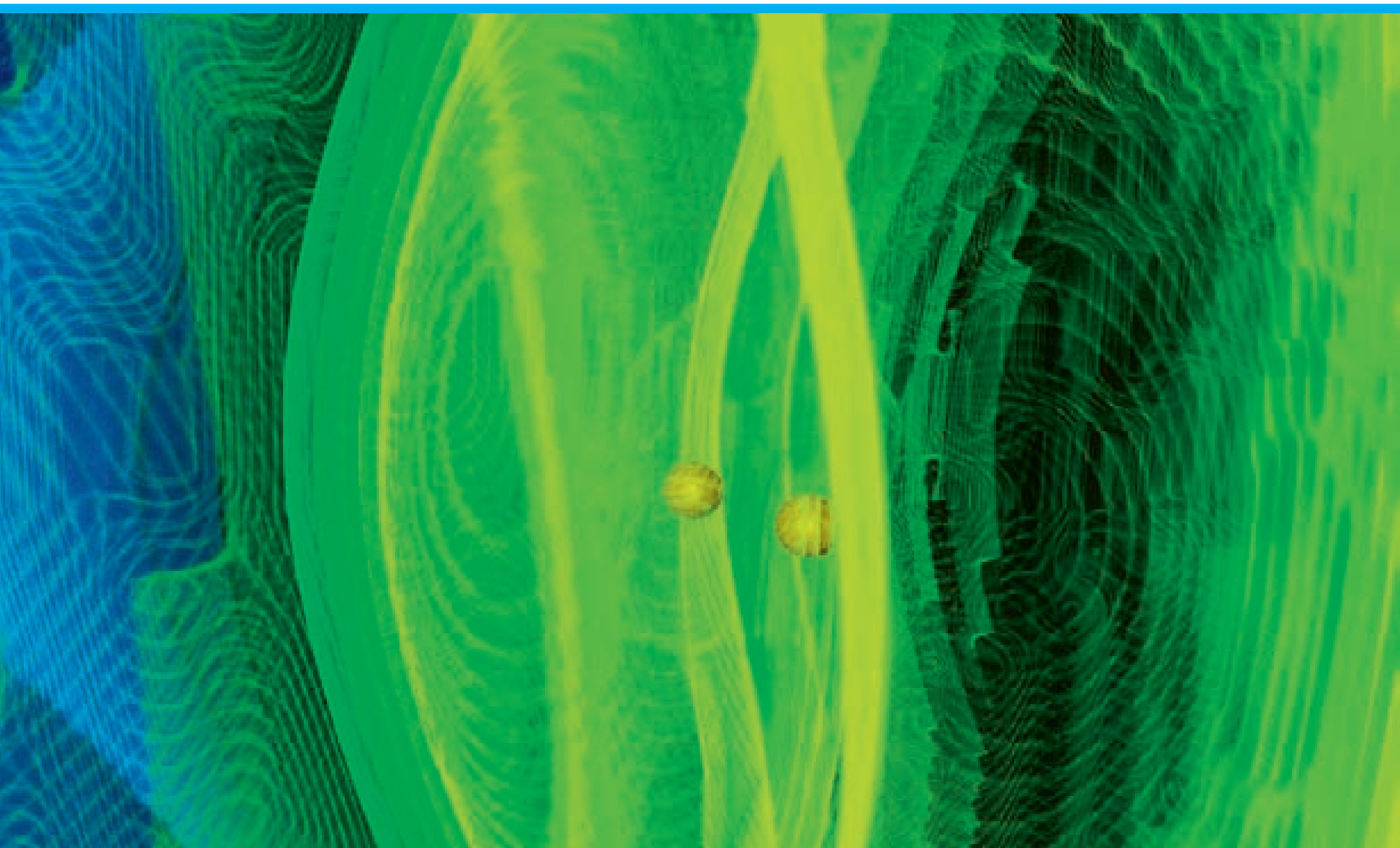

A global plan

**Title page: Numerical simulation of merging black holes**

The search for gravitational waves requires detailed knowledge of the expected signals. Scientists of the Albert Einstein Institute's numerical relativity group simulate collisions of black holes and neutron stars on supercomputers using sophisticated codes developed at the institute. These simulations provide insights into the possible structure of gravitational wave signals. Gravitational waves are so weak that these computations significantly increase the probability of identifying gravitational waves in the data acquired by the detectors.

Numerical simulation: C. Reisswig, L. Rezzolla (Max Planck Institute for Gravitational Physics (Albert Einstein Institute))
Scientific visualisation: M. Koppitz (Zuse Institute Berlin)

# GWIC

## THE GRAVITATIONAL WAVE INTERNATIONAL COMMITTEE ROADMAP

### The future of gravitational wave astronomy

June 2010

Updates will be announced on the GWIC webpage: https://gwic.ligo.org/

# CONTENT

# Introduction

Gravitational wave science is on the verge of direct observation of the waves predicted by Einstein's General Theory of Relativity and opening the exciting new field of gravitational wave astronomy. In the coming decades, ultra-sensitive arrays of ground-based instruments and complementary spaced-based instruments will observe the gravitational wave sky, inevitably discovering entirely unexpected phenomena while providing new insight into many of the most profound astrophysical phenomena known. This new window into the cosmos could revolutionize humanity's understanding of the Universe in which we live.

Recognizing that the field is approaching this historic moment, in July 2007 the Gravitational Wave International Committee (GWIC)[1] initiated the development of a strategic roadmap for the field of gravitational wave science with a 30-year horizon. The Roadmap Committee consists primarily of members of GWIC with representation from the ground and space-based experimental gravitational wave communities, the gravitational wave theory and numerical relativity communities, the astrophysics components of the gravitational wave community and major projects and regions participating in the field world-wide. The committee sought and received advice from many experienced practitioners in the field, as well as a number of highly regarded scientists outside the field. Input was also sought from the many funding agencies that support research and projects related to gravitational wave science around the world.

The goal of this roadmap is to serve the international gravitational wave community and its stakeholders as a tool for the development of capabilities and facilities needed to address the exciting scientific opportunities on the intermediate and long-term horizons.

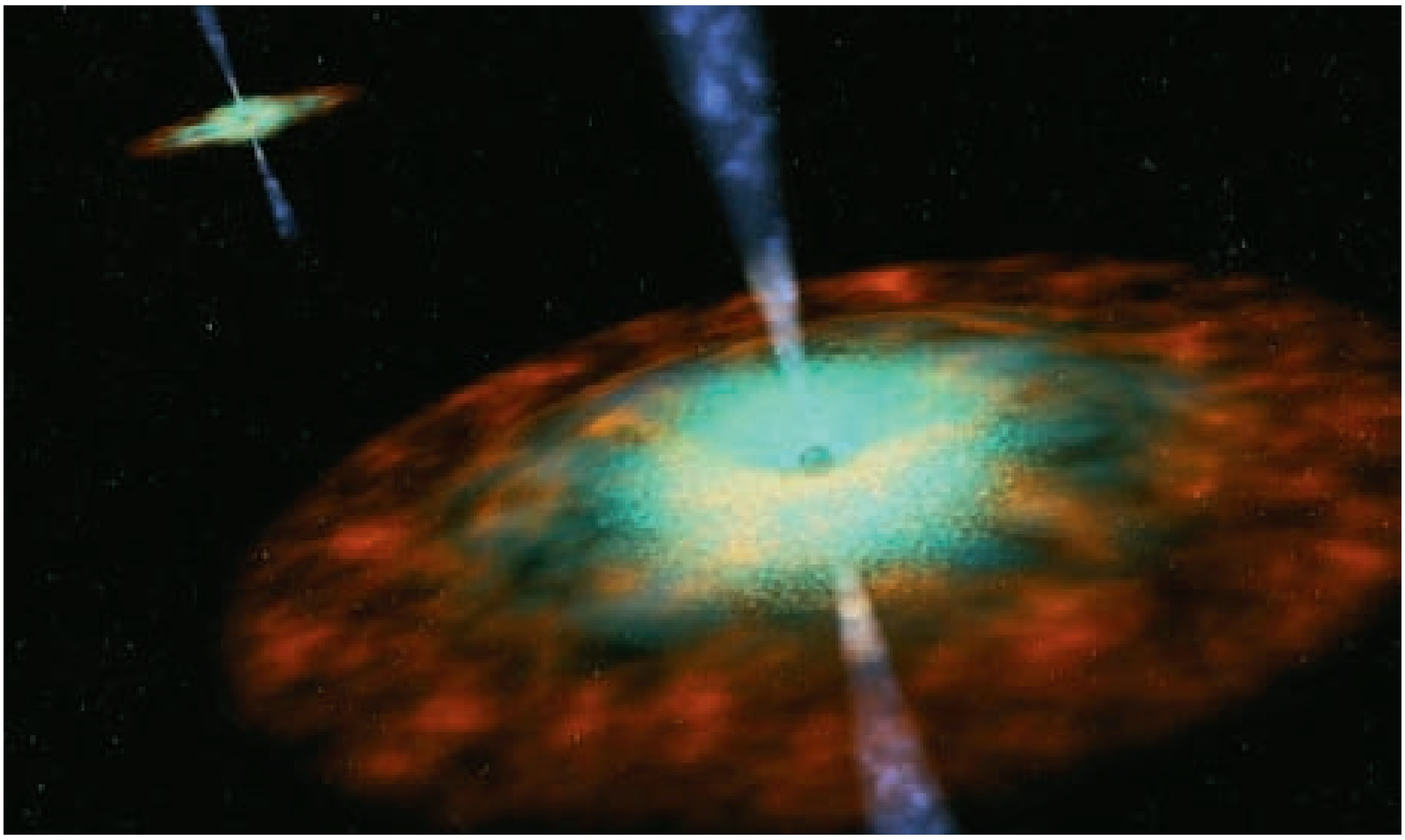

Illustration of a binary black hole system Courtesy NASA/JPL-Caltech

[1] GWIC, the Gravitational Wave International Committee, was formed in 1997 to facilitate international collaboration and cooperation in the construction, operation and use of the major gravitational wave detection facilities world-wide. It is affiliated with the International Union of Pure and Applied Physics as a sub-committee of IUPAP's Particle and Nuclear Astrophysics and Gravitation International Committee. The membership of GWIC comprises the directors or representatives of all major gravitational wave projects world-wide.

# Science Goals

The future development of the field should be driven by maximizing the discovery potential and subsequent exploitation of the field of gravitational wave observations. Gravitational wave detectors will uncover new aspects of the Universe by helping us to study sources in extreme physical conditions: strong non-linear gravity and relativistic motion, and extremely high density, temperature and magnetic fields. Because of the very weak nature of gravity it is the signals produced by astrophysical systems, where there are potentially huge masses accelerating very strongly, that are of the greatest interest. Gravitational wave signals propagate essentially unattenuated and are expected over a wide range of frequencies, from $10^{-17}$ Hz in the case of ripples in the cosmological background, through $10^3$ Hz when neutron stars are born in supernova explosions, with many sources of great astrophysical interest distributed within this range, including black hole interactions and coalescences, neutron star coalescences, ultra-compact binaries and rotating asymmetric neutron stars such as pulsars.

By observing the rich variety of signals from these sources our goal is to answer key scientific questions in:

### Fundamental physics and general relativity

- What are the properties of gravitational waves?
- Is general relativity the correct theory of gravity?
- Is general relativity still valid under strong-gravity conditions?
- Are Nature's black holes the black holes of general relativity?
- How does matter behave under extremes of density and pressure?

### Cosmology

- What is the history of the accelerating expansion of the Universe?
- Were there phase transitions in the early Universe?

### Astronomy and astrophysics

- How abundant are stellar-mass black holes?
- What is the central engine behind gamma-ray bursts?
- Do intermediate mass black holes exist?
- What are the conditions in the dense central cores of galactic nuclei dominated by massive black holes?
- Where and when do massive black holes form and how are they connected to the formation of galaxies?
- What happens when a massive star collapses?
- Do spinning neutron stars emit gravitational waves?
- What is the distribution of white dwarf and neutron star binaries in the galaxy?
- How massive can a neutron star be?
- What makes a pulsar glitch?
- What causes intense flashes of X- and gamma-ray radiation in magnetars?
- How do compact binary stars form and evolve and what can they tell us about the history of star formation rate in the Universe?

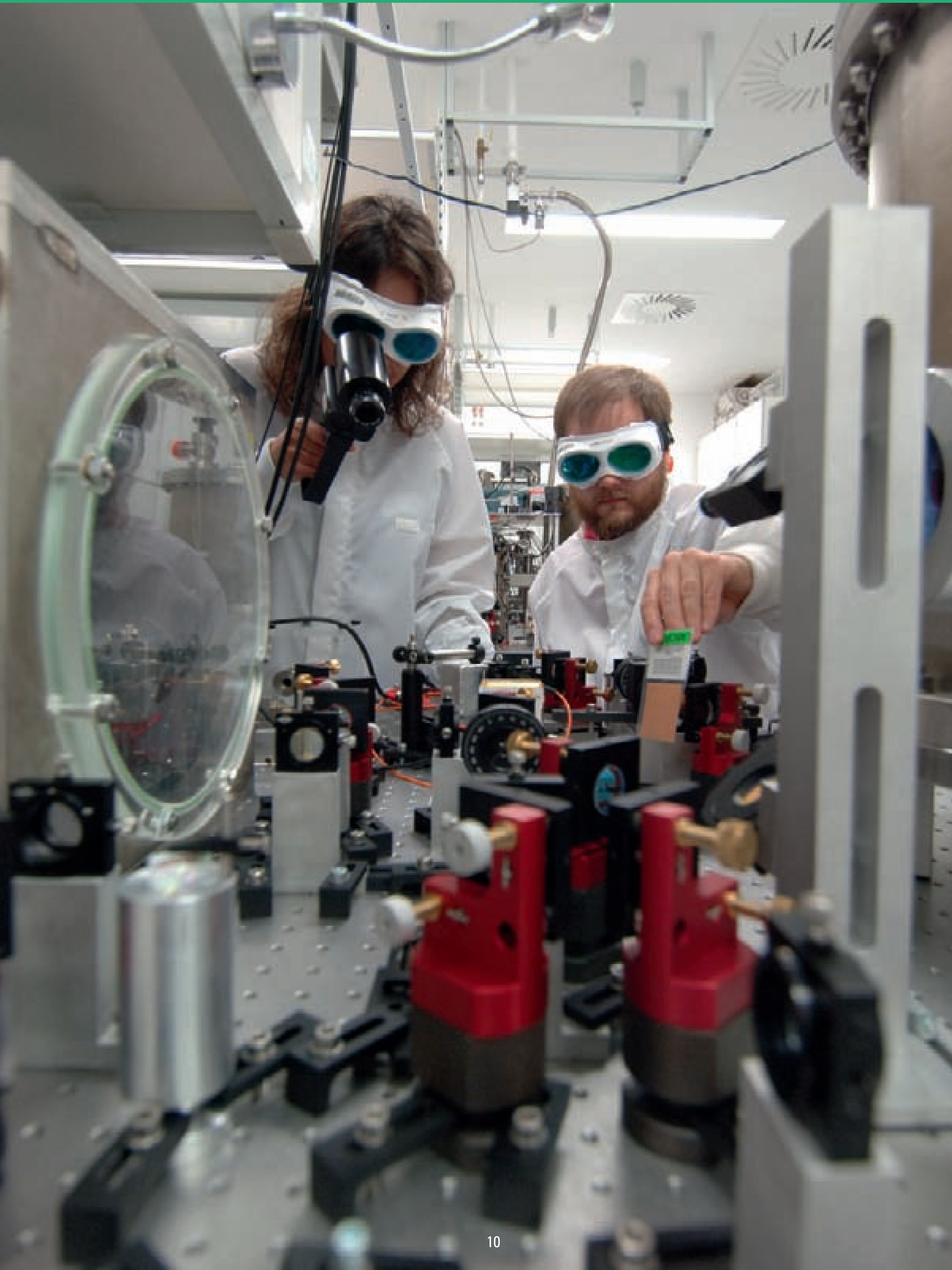

Credit: N. Michalke / AEI

# Ground-based, higher-frequency detectors

Recent years have seen a shift in the technologies used in gravitational wave searches as the first generation of large gravitational wave interferometers has begun operation at or near their design sensitivities, taking up the baton from the bar detectors that pioneered the search for the first direct detection of gravitational waves. These ground-based km-scale interferometers and their advanced upgrades will be critical in establishing the field of gravitational wave astronomy through the detection of high luminosity gravitational wave sources such as the merger of binary neutron stars and black holes.

In the US, the Laser Interferometer Gravitational-wave Observatory (LIGO) consists of three multi-kilometer scale interferometers, two in Hanford, Washington, and one in Livingston, Louisiana. In Europe, Virgo is a multi-kilometer scale interferometer located near Pisa, Italy, and GEO600, a kilometer-scale interferometer, is located near Hannover, Germany. The TAMA detector, located near Tokyo, Japan, is half the size of GEO600. It is hoped that the first direct observation of gravitational waves will be made in the next few years by this international network of detectors.

Major upgrades of LIGO (Advanced LIGO), Virgo (Advanced Virgo) and GEO600 (GEO HF) will be completed in the next half decade resulting in a network with the sensitivity needed to observe gravitational wave signals at a monthly or even weekly rate. To capitalize fully on this important scientific opportunity, a true global array of gravitational wave antennae separated by inter-continental distances is needed to pinpoint the sources on the sky and to extract all the information about each source's behavior encoded in the gravitational wave signal.  In the medium term this means the first priority for ground-based gravitational wave detector development is to expand the network, adding further detectors with appropriately chosen intercontinental baselines and orientations to maximize the ability to extract source information.

The most advanced plans along these lines are with the Japanese Large-scale Cryogenic Gravitational-wave Telescope (LCGT)[1] and the Australian International Gravitational Observatory (AIGO). Possibilities for a detector in India (INDIGO) are also being studied. An inter-continental scale network made up of advanced interferometers in the US, Europe, Asia and Australia would provide an all-sky array that could detect, decode and point to the sky-position of gravitational wave sources in the audio bandwidth where many of the most interesting sources are located.

There is a great opportunity to evolve the capabilities of this ground-based network in roughly the next 15 years by developing large underground observatories with greatly improved sensitivity, particularly at low frequency. Such underground facilities would operate together with Advanced LIGO, Advanced Virgo, LCGT and AIGO as the capabilities of these advanced instruments are further enhanced.

Successful deployment of third generation, underground gravitational wave observatories will require development of a number of new technologies by the gravitational wave community. Many of the necessary R&D programs are undertaken in a limited number of places, but with a growing level of coordination and communication. It is important that these developments are shared with the rest of the community and that the additional efforts required take place in all regions of the world so that a full range of robust technologies are ready when required for the third-generation facilities.

The most advanced concept for an underground low frequency detector is the Einstein Telescope (ET) project. A conceptual design study for ET was funded by the European Commission, within the Seventh Framework Programme (FP7), beginning in 2008 to assess the feasibility of a third-generation gravitational wave observatory, with a largely improved sensitivity, especially below 10 Hz due to an underground location.

[1] Approval of funding for the first phase of construction of LCGT was announced in June 2010.

The realization of the ET research infrastructure, allowing operations for many decades, will be triggered by the first gravitational wave detection with the start of the site preparation beginning as early as 2017 and with scientific data being available in the first half of the following decade. Similarly, there is the possibility for a high-sensitivity large-bandwidth observatory to be built in the US, potentially in the Deep Underground Science and Engineering Laboratory (DUSEL).

## Space-based and lower-frequency detectors

The low-frequency range, below about 1 Hz, includes a large and diverse population of strong gravitational wave sources that can only be observed at these frequencies. Detection technologies are diverse and range from polarization measurements of the cosmic microwave background and pulsar timing to spacecraft tracking and large baseline laser interferometry. All of these technologies will eventually be used to observe the complete gravitational wave spectrum covering more than 20 orders of magnitude in frequency.

LISA, a space-based interferometer funded by NASA and ESA will open the low-frequency gravitational wave window from 0.1 mHz to 0.1 Hz. The scientific objectives of space-based and ground-based instruments are complementary in the same way that optical and x-ray astronomy are complementary and have provided information about different types of astrophysical objects and phenomena. LISA is a project in NASA's Physics of the Cosmos Program, and the National Academy's Astro2010 decadal review of astronomy and astrophysics will prioritize LISA relative to other large astrophysics missions. LISA is the gravitational wave community's highest priority for a space-based mission. The goal of a launch of LISA in 2020 is technologically feasible and entirely timely, considering that the technology precursor mission LISA Pathfinder – which GWIC strongly endorses – will be launched in 2012.

The portion of the gravitational wave spectrum lying between the LISA band and that probed by ground-based interferometers also holds great potential, including observation of coalescences of intermediate black hole binaries and some of the deepest searches for stochastic backgrounds. The DECi-hertz Interferometer Gravitational wave Observatory (DECIGO) is a space gravitational wave antenna proposed in Japan, aiming for launch several years after LISA. The concept will be demonstrated with the DECIGO pathfinder mission that could be launched in around five years time. Also targeting this frequency range is the US-based ALIA mission concept.

In the longer time frame, the Big Bang Observer is conceived as a LISA follow-on mission targeted at detecting the gravitational waves produced in the big bang and other phenomena in the early Universe.

This committee also recognizes and supports the growing efforts to utilize radio astronomy for the detection of gravitational waves in the nanohertz frequency band, with the formation of the International Pulsar Timing Array (IPTA) collaboration. These efforts are complementary to those of the ground- and space-based laser interferometric projects and could well lead to observation of gravitational radiation in this band within the next decade.

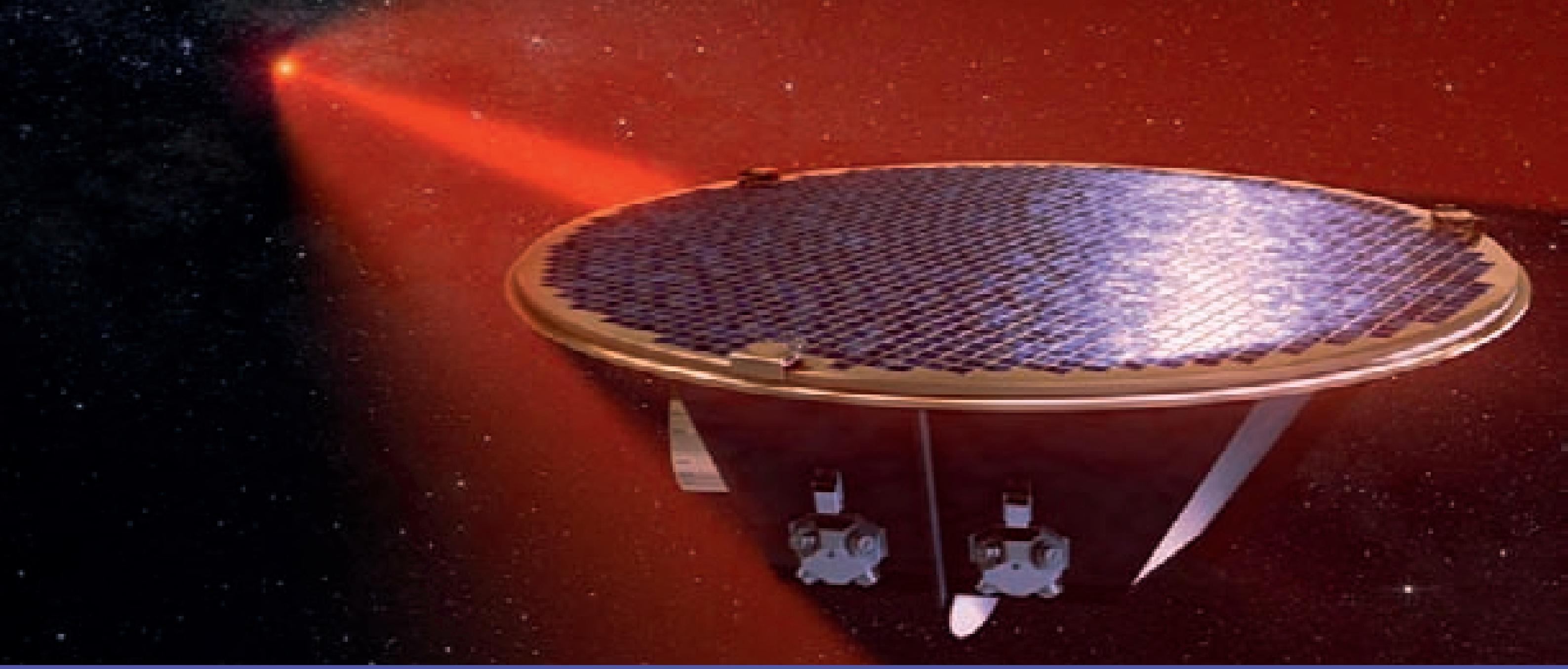

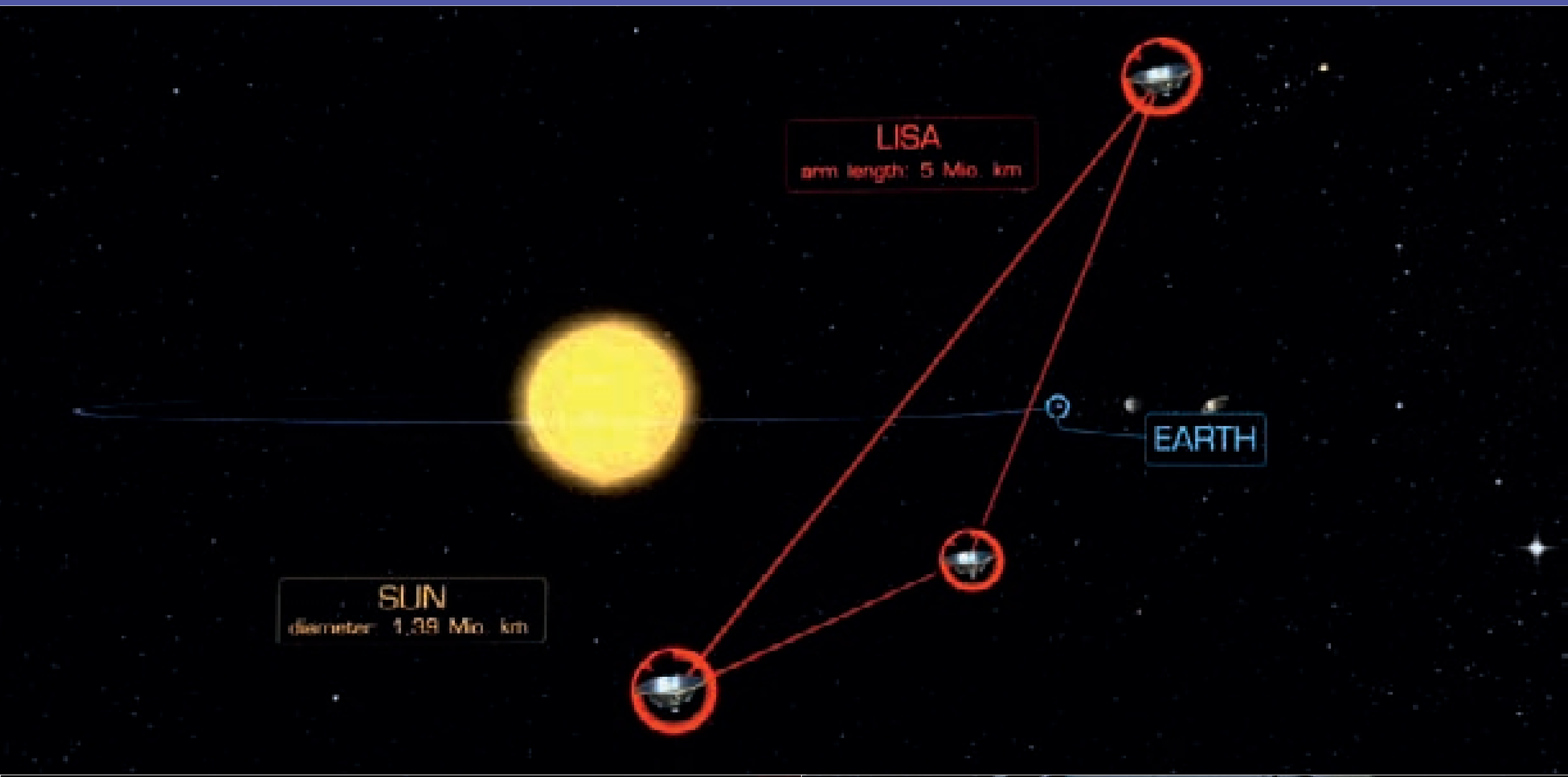
LISA
arm length: 5 Mio. km
EARTH
SUN
diameter: 1.39 Mio. km

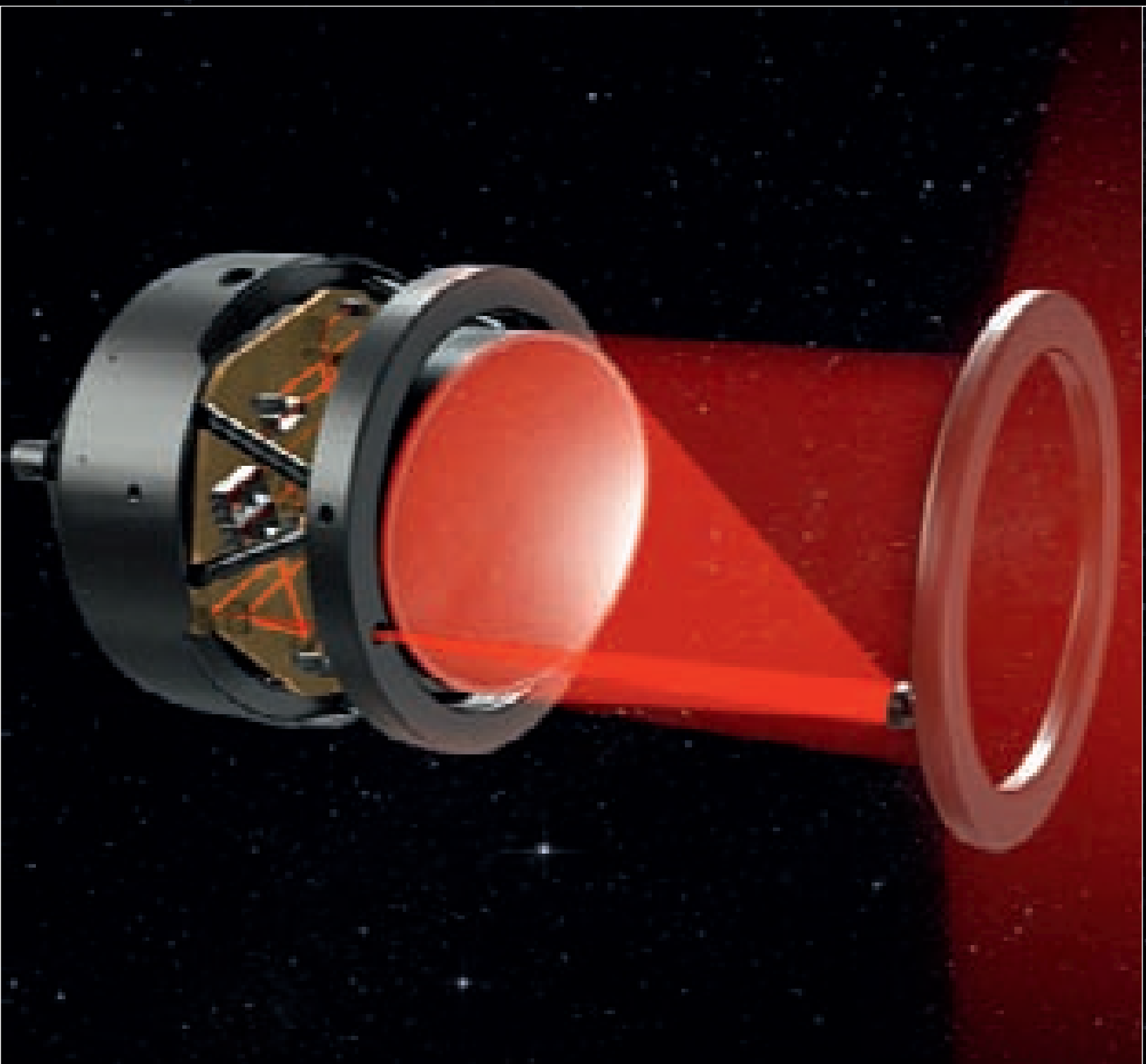

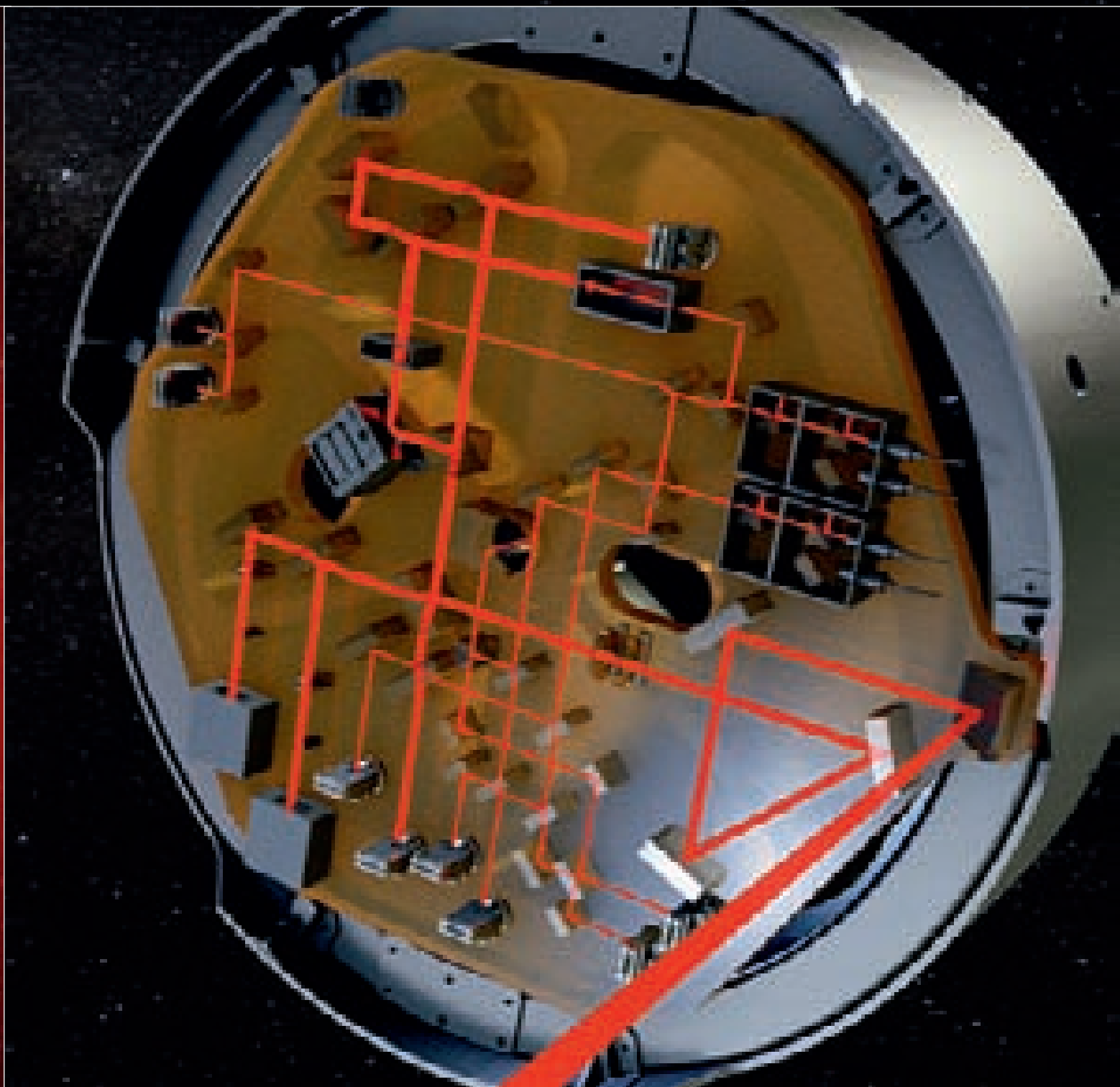

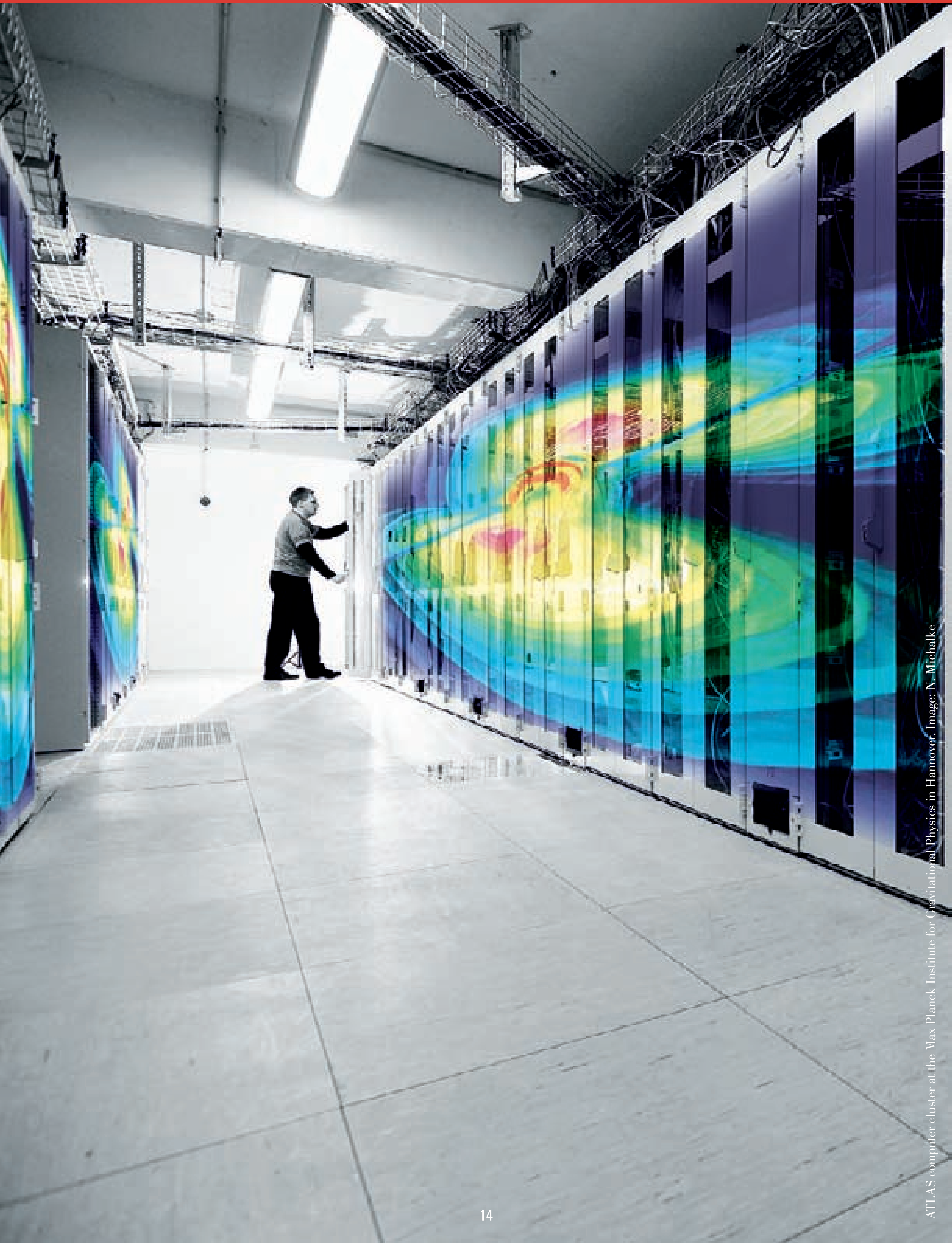

ATLAS computer cluster at the Max Planck Institute for Gravitational Physics in Hannover. Image: N. Michalke

# Theory, data analysis and astrophysical model building

Many of the most important sources of gravitational radiation involve strong-field dynamical general relativity and thus an essential component of this new astronomy will be continuing developments in theory and numerical relativity that will provide the framework for detailed understanding of the information encoded in the gravitational waves. Large scale, high-performance numerical computations will continue to play a critical role, and a continuing collaboration among data analysts, astrophysical theorists, and analytical and numerical general relativists will be essential.

The physical properties of gravitational wave sources can only be interpreted and understood using these theoretical tools. With them, light will be shed on such exciting objects as merging black holes and on some of the most fundamental questions in physics.

In the next decade the field of gravitational wave astronomy will become an important component of the worldwide astronomy effort; we anticipate a rich interplay between gravitational wave observations and more traditional electromagnetic observations and observations utilizing neutrino telescopes. As such, it will be important for data from gravitational wave instruments to be made available to the scientific community and the public as is now the case for publicly funded astronomical observatories in many countries. This will require a carefully implemented study on the part of the international gravitational wave community to determine the steps required to meet this goal, the appropriate deliverables, and the effort and cost involved, taking into account the requirements of funding agencies in different countries.

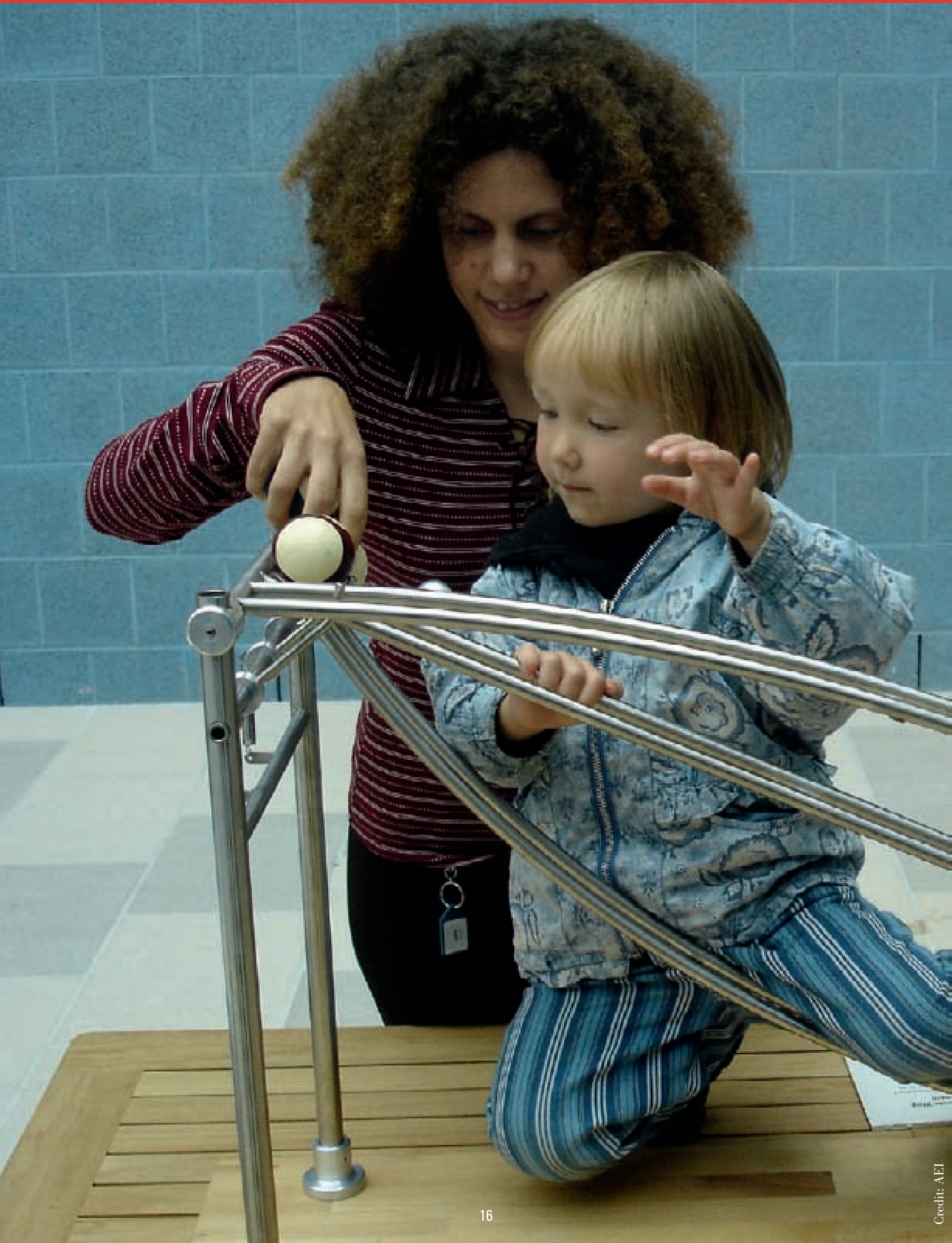

Credit: AEI

## Outreach and interaction with other fields

In the coming decade the scientific use of gravitational wave signals, both by themselves and in concert with other signals (electromagnetic, neutrino), will become part of mainstream astronomy and astrophysics. Given the timescale in which gravitational wave astronomy will produce results, there is now an excellent opportunity for our field to foster the interest of the current and coming generations of astronomers and astrophysicists in its scientific potential, as well as communicating the excitement of gravitational wave astronomy to the public.

## Technology Development

Finally, we note that gravitational wave research has always been a technology intensive field, driving technology in some areas while benefiting from developments in others. The need to stretch the limits of sensitivity has resulted in developments that have profited other fields of science and technology, notably precision measurement, metrology, optics, lasers, stable optical fabrication techniques, space technology, and computational techniques. At the same time, the field has been alert to technology developments in other realms that could enable new scientific directions in gravitational wave science. There is no doubt that this highly productive interchange between gravitational wave science and other spheres of science and technology will continue, providing tangible benefits to society, in addition to the intellectual satisfaction of an improved knowledge of our Universe.

# Priorities

**The Gravitational Wave community has identified its highest equally ranked priorities** so that the most compelling scientific opportunities utilizing gravitational waves can be realized:

- The construction, commissioning and operation of the second generation global ground-based network comprised of instruments under construction or planned in the US, Europe, Japan and Australia.

- The completion of LISA Pathfinder and a timely selection of LISA in 2012/2013 for a launch in 2020 to open the low frequency gravitational wave window from 0.1 mHz to 0.1 Hz.

- Coordinated R & D efforts in collaboration with existing design study teams to support the construction, beginning in the post 2018 time frame, of the Einstein Telescope, soon after the expected first gravitational-wave discoveries have been made.

- The continued development of an international pulsar timing array for the study of gravitational waves in the nano-Hertz band. This effort requires continued development of algorithms and data acquisition systems, and access to substantial amounts of time on the world's largest radio-telescopes.

**There are also high priorities that are essential to support continuing intellectual and technical development of forefront capabilities in the field.** They will also help to enable involvement of other scientific communities in the science:

- The timely start for technology development for LISA follow-on missions.

- The development of procedures that, beginning in the era of frequent ground-based detection of gravitation waves, will allow the broader scientific community to fully utilize information about detected gravitational waves.

- A strong and ongoing international program of theoretical research in General Relativity and astrophysics directed towards deepening our understanding of gravitational wave observations.

**This roadmap will serve the international gravitational wave community and its stakeholders as a strategic tool in planning for the development of capabilities and facilities needed to seize the tremendous scientific opportunities now on the horizon.** This material represents the consensus best judgment of GWIC following discussions with leaders in the gravitational-wave field and with other members of the gravitational physics and astrophysics communities, and with representatives of funding agencies.

Ideally the future development of gravitational wave astronomy will follow the overall direction presented herein so that the great discoveries possible with gravitational waves will become a reality.

# 1. Introduction

The field of gravitational wave science began in 1916, when the existence of gravitational waves was predicted by Einstein's General Theory of Relativity. In the early 1960's, Joseph Weber pioneered the use of resonant bars to attempt the direct observation of gravitational waves from cosmic sources. The field received a subsequent boost in the 1970's when the discovery and precise timing measurements of a binary pulsar system showed that the decay of the orbit of this system over 10 years could be accurately accounted for by the emission of gravitational waves, as predicted by Einstein's theory. A Nobel Prize was awarded in 1993 to Hulse and Taylor for this work.

The field has continued to mature to its current state today, when sophisticated instruments at sites around the world stand on the verge of direct observation and characterization of gravitational waves, and ushering in the birth of gravitational wave astronomy.

In recognition that the field is on the cusp of this historic development, the Gravitational Wave International Committee, GWIC, initiated the development of a strategic roadmap for the field with a 30-year horizon.

GWIC was formed in 1997 to facilitate international collaboration and cooperation in the construction, operation and use of the major gravitational wave detection facilities worldwide. It is affiliated with the International Union of Pure and Applied Physics as a sub-committee of IUPAP's Particle and Nuclear Astrophysics and Gravitation International Committee. The membership of GWIC comprises the directors or representatives of all major gravitational wave projects worldwide.

The goals of GWIC are to:

- Promote international cooperation in all phases of construction and exploitation of gravitational-wave detectors;
- Coordinate and support long-range planning for new instrument proposals, or proposals for instrument upgrades;
- Promote the development of gravitational-wave detection as an astronomical tool, exploiting especially the potential for coincident detection of gravitational-waves and other fields (photons, cosmic-rays, neutrinos);
- Organize regular, world-inclusive meetings and workshops for the study of problems related to the development and exploitation of new or enhanced gravitational-wave detectors, and foster research and development of new technology;
- Represent the gravitational-wave detection community internationally, acting as its advocate;
- Provide a forum for the laboratory directors to regularly meet, discuss, and plan jointly the operations and direction of their laboratories and experimental gravitational-wave physics generally.

In July 2007 GWIC chartered a committee to develop a strategic roadmap for the field of gravitational wave science. The committee was made up primarily of members of GWIC with representation from the ground and space-based experimental gravitational wave communities, the gravitational wave theory and numerical relativity communities, the astrophysics components of the gravitational wave community and major projects and regions participating in the field world-wide. The membership of the committee is presented in Appendix A.2.

The charge to the Roadmap committee from GWIC was to:

- Develop a global roadmap for the field with a 30-year horizon;
- Consider ground and space-based capabilities, theory and numerical relativity, and the possible impacts of new technologies and approaches;
- Take account of known national and regional planned projects;
- Identify relevant science opportunities and the facilities needed to address them.

The roadmap should address how and when major capabilities and infrastructure should be developed to address these opportunities.

During the year-long process in which the roadmap was developed, the committee sought and received advice from many experienced practitioners in the field, as well as a number of highly regarded scientists outside the sphere. Input was also sought from the many funding agencies that support research and projects related to gravitational wave science around the world. Representatives from four of those agencies met with the committee for fruitful discussions and individual members of the committee met with representatives of two other funding agencies.

This document is the result of the GWIC Roadmap Committee's deliberations. It defines a number of priorities and contains a number of recommendations from the Committee to GWIC on how to proceed. The Roadmap is intended for a wide audience including the gravitational wave community, the broader scientific community, funding agencies and government officials, and the general public. The document begins with an introductory chapter to provide background about the gravitational wave field, with subsequent chapters that directly address the committee's charge.

We hope this roadmap will serve the international gravitational wave community and its stakeholders as a strategic tool in planning for the development of capabilities and facilities needed to seize the tremendous scientific opportunities now on the horizon. This material represents the consensus best judgment of many of the leaders in the field and is endorsed by the gravitational wave community. Ideally the future development of gravitational wave astronomy will follow the overall direction presented herein so that the great discoveries possible with gravitational waves will become a reality.

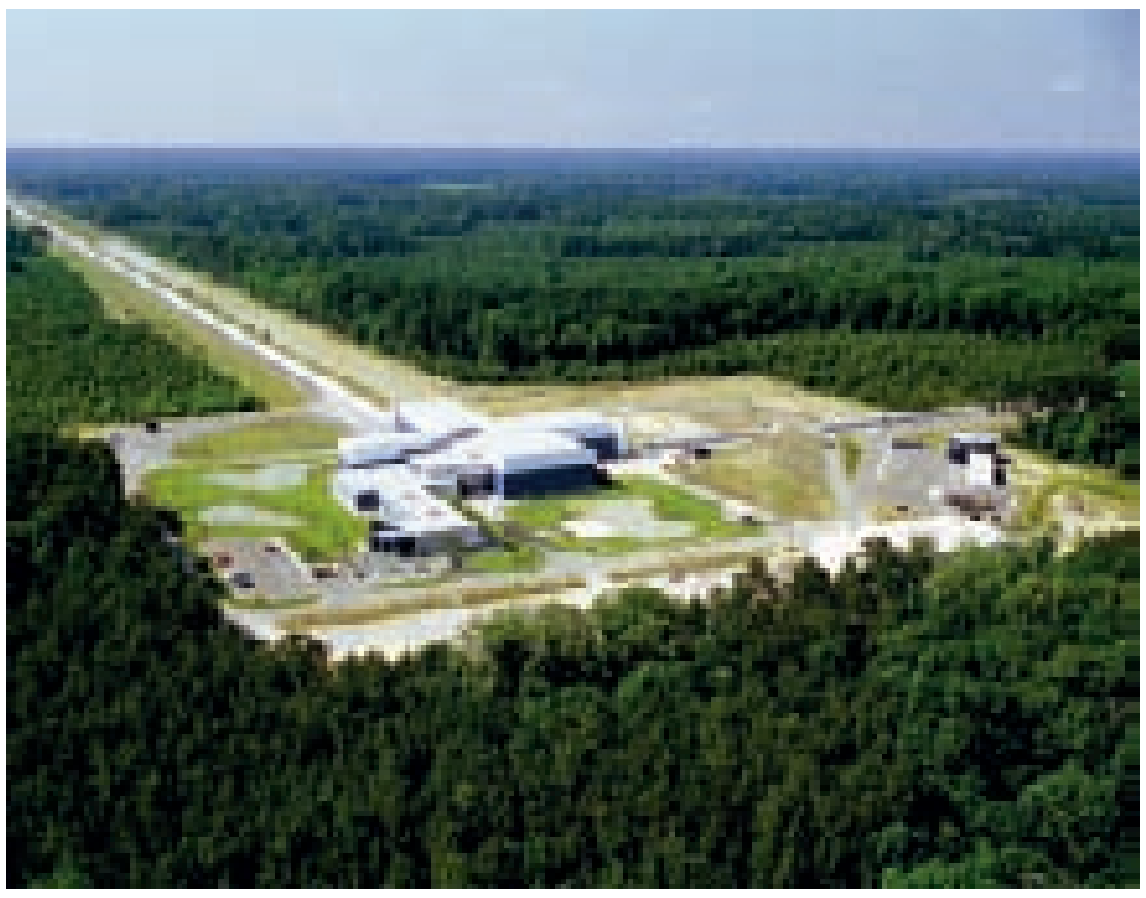

Fig. 1.1 – LIGO detector with 4 km arms at Livingston, Louisiana

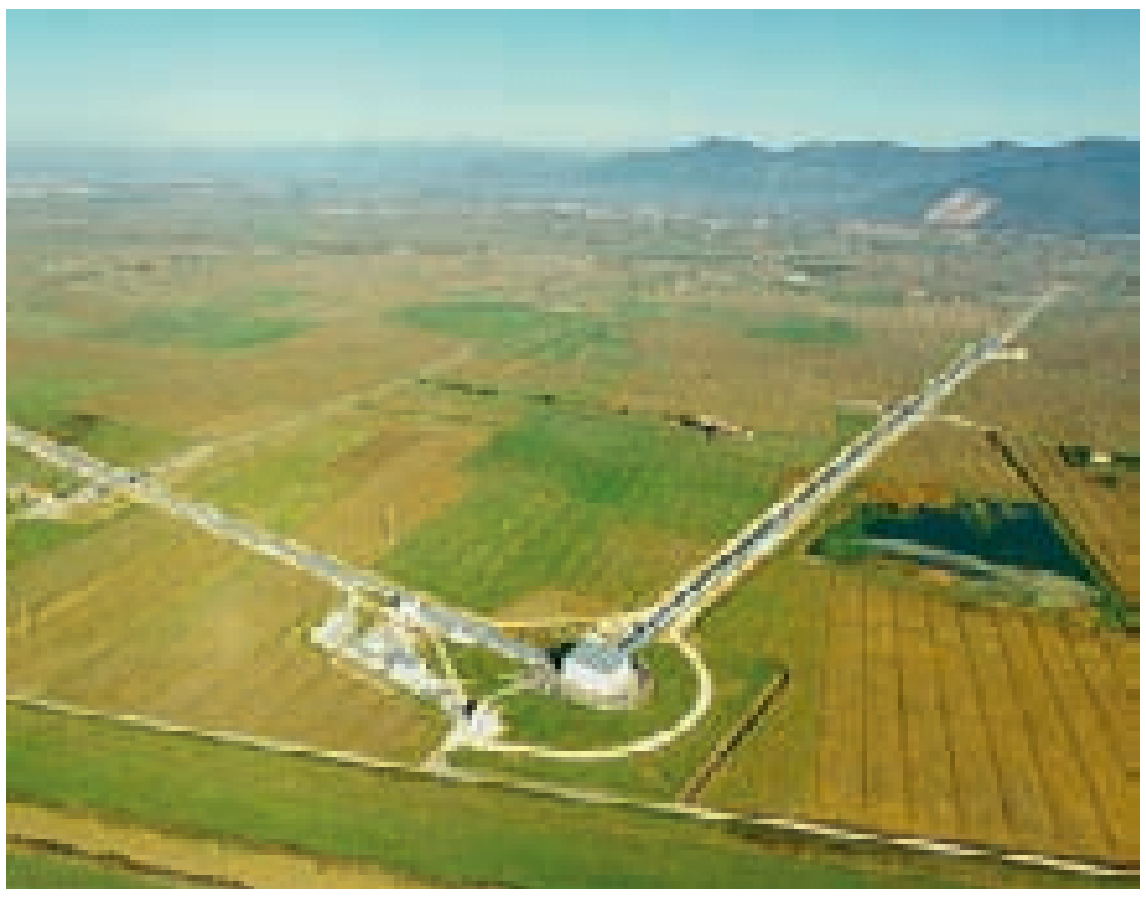

Fig. 1.2 – Virgo Detector, with 3 km arms, at Cascina, near Pisa

# 2. Introduction to gravitational wave science

## 2.1 Introduction

Galileo's application of the telescope to observe the heavens was the first stepping stone on a path which in the twentieth century led to a new era in observational astronomy. The construction of radio and microwave antennas, as well as gamma- and x-ray detectors, revolutionized astronomy and opened the entire electromagnetic window for observing and understanding the Universe. A remarkable revelation from these observations was the discovery that about 97 percent of our Universe is invisible and that the gravitational interaction powers the most luminous and spectacular objects and phenomena, such as quasars, gamma-ray bursts, ultra luminous X-ray sources, pulsars, even the big bang.

Einstein's General Theory of Relativity predicts that dynamical systems in strong gravitational fields will release vast amounts of energy in the form of gravitational radiation. These gravitational “waves” are among the most elusive signals from the deepest reaches in the Universe. They can be thought of as ripples in the curvature of spacetime.

Experiments aimed at searching for such waves have been in development for the last 40 years, but only now are sensitivities reaching levels where real detection is possible within a few years. The gravitational wave detectors currently collecting data, and the advanced detectors planned for the next decade, are the first steps in establishing the field of gravitational astronomy through their detection of the most luminous gravitational wave sources in the Universe, such as the merger of binary neutron stars and black holes.

These detectors are akin to the first telescopes, with more sensitive detectors to follow making it possible to observe a greater variety of phenomena and enabling a new tool for expanding our knowledge of fundamental physics, cosmology and relativistic astrophysics.

Was Einstein right? Is the nature of gravitational radiation as his theory predicts or is it something different? How did the black holes in galactic nuclei form? What were the physical conditions at the time of the big bang? What is the nature of quantum gravity and what is the origin of space and time? How many spatial dimensions are there? These are some of the key questions at the forefront of physics and astrophysics that gravitational wave observations may soon illuminate.

## 2.2 What are gravitational waves?

### 2.2.1 Gravitational radiation

To gain an impression of the nature of gravitational waves, we should remember that Einstein described the gravitational field in a geometrical way as the curvature of the spacetime “surface” caused by the presence of mass. Accelerations of any masses in the system thus can cause propagating gravitational disturbances, and these disturbances are gravitational waves.

The gravitational wave amplitude, $h$, can usually be interpreted as a physical strain in space or, more precisely,

$$h = 2\delta l / l$$

where $\delta l$ is the change in separation of two masses a distance $l$ apart.

### 2.2.2 Polarization

Gravitational waves are produced when mass undergoes acceleration, and thus are analogous to the electromagnetic waves produced when electric charge is accelerated. However, the existence of only one sign of mass, together with the law of conservation of linear momentum, implies that there is no monopole or dipole gravitational radiation. Quadrupole radiation is possible and the magnitude of *h* produced at a distance *r* from a source is proportional to the second time derivative of the quadrupole moment of the source and inversely proportional to *r*, while the luminosity of the source is proportional to the square of the third time derivative of the quadrupole moment.

For quadrupole radiation there are two orthogonal polarizations of the wave at 45 degrees to each other, of amplitude $h_+$ and $h_x$, and each of these is equal in magnitude to twice the strain in space in the relevant direction. The effect of the two polarizations on a ring of particles is shown in figure 2.1, and from this the principle of most gravitational wave detectors – looking for changes in the length of mechanical systems such as bars of aluminium or the arms of Michelson-type interferometers, as will be discussed below – can be clearly seen.

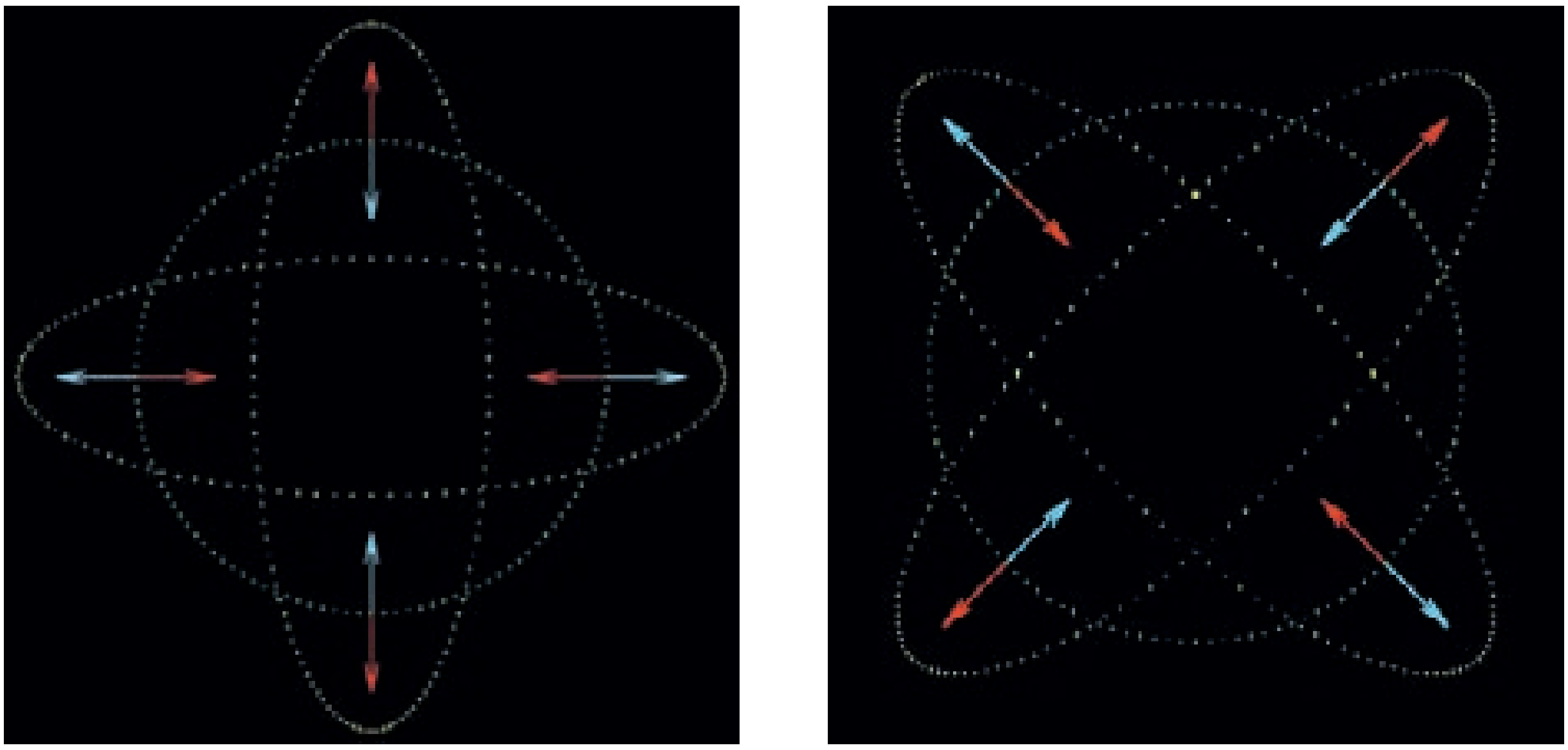

Fig. 2.1 – The effect of a gravitational wave on a ring of particles. The propagation direction is perpendicular to the ring. On the left the effect of a wave polarized in the '+' direction is shown. The figure on the right shows the effect of an 'X' polarized wave.

### 2.2.3 Expected amplitudes from typical sources

The challenge to be overcome in the successful operation of gravitational wave observatories is that predicted amplitudes or strains in space in the vicinity of the Earth caused by gravitational waves from astrophysical events are extremely small, of the order of $10^{-21}$ or lower in the case of ground-based observations. Indeed, current theoretical models of the event rate and strength of such events suggest that in order to detect, from the ground, a few events per year – from coalescing neutron star binary systems, for example – an amplitude sensitivity close to $10^{-22}$ over approximately 1000 Hz is required. Signal strengths at the Earth, integrated over appropriate time intervals, for a number of sources are shown in figure 2.2.

Gravitational wave signals are expected over a wide range of frequencies, from $10^{-17}$ Hz in the case of ripples in the cosmological background, through $10^3$ Hz when neutron stars are born in supernova explosions, up to possible signals at GHz – very difficult to detect but very rewarding if eventually detectable – from processes in the early Universe.

Because of the nature of gravity the only sources of gravitational waves that are likely to be detected are astrophysical, where there are potentially huge masses accelerating very strongly. There are many sources of great astrophysical interest, including black hole interactions and coalescences, neutron star coalescences, ultra-compact binaries, stellar collapses to neutron stars and black holes (supernova explosions), rotating asymmetric neutron stars such as pulsars, and processes in the early Universe.

This wide range of source types, and signal frequencies, means that a range of detectors is required to exploit fully the potential of gravitational wave astronomy. For example, at very low frequencies the effects of gravitational waves can be sought from studies of the polarization of the Cosmic Microwave Background, e.g. Planck, and future experiments such as CMBPol. At somewhat higher frequencies, phase fluctuations in pulsar timing measurements, e.g. the Parkes Pulsar Timing Array (PPTA), the European Pulsar Timing Array (EPTA) and the North American Nanohertz Observatory for Gravitational Waves (NANO-Grav), are a tool of interest and are of particular significance now that the International Pulsar Timing Array (IPTA), a collaboration of three groups, has been organized. Laser sensing of relative satellite positions, e.g. LISA and DECIGO, is planned for the frequency range $10^{-4}$ Hz to $10^{-1}$ Hz, a frequency range where Doppler tracking of spacecraft is currently employed, e.g. ULYSSES, Mars Observer, Galileo, Mars Global Surveyor and CASSINI. In the audio frequency range, ground-based observatories utilizing laser sensing for freely suspended test masses are operating – LIGO, Virgo, GEO600 and TAMA 300, with the Einstein Telescope (ET) planned for the future.

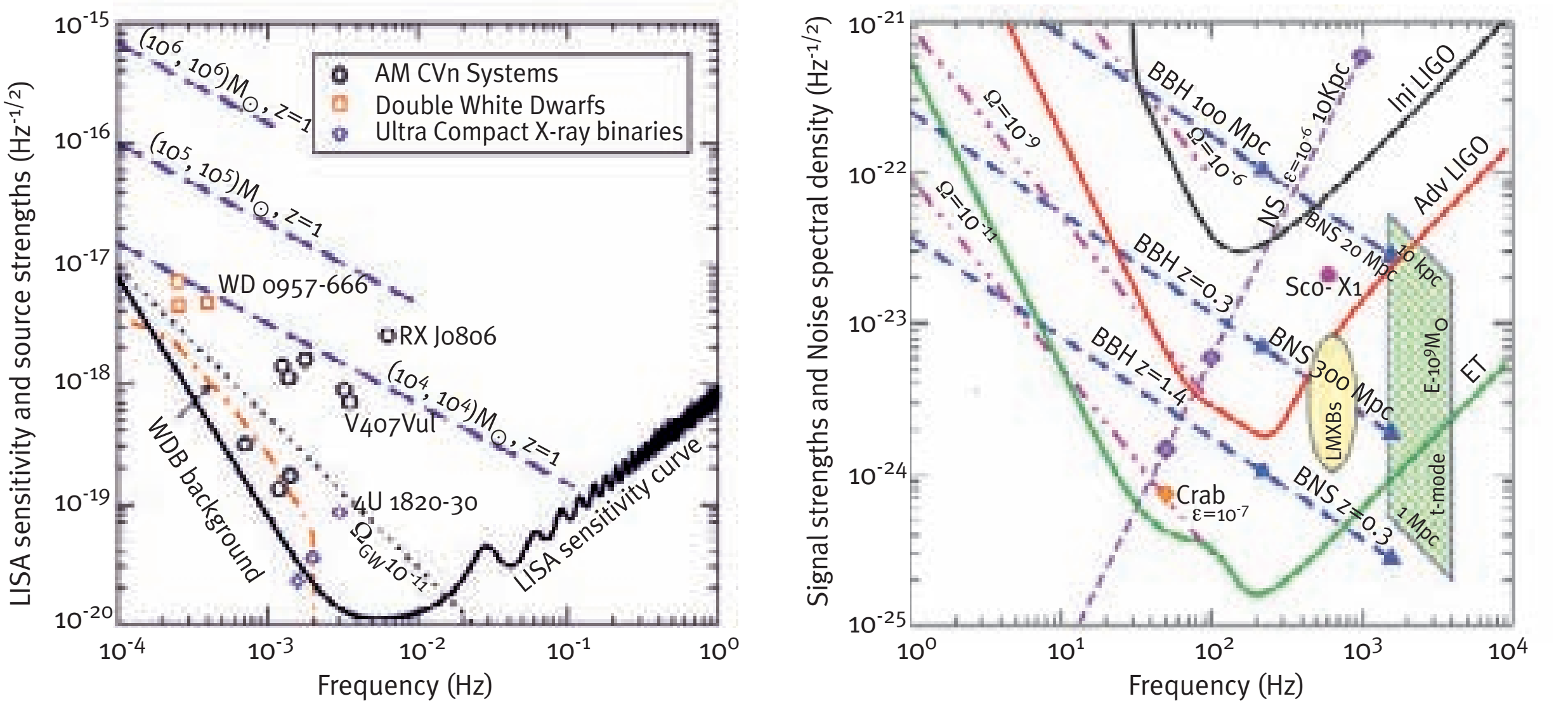

Figure 2.2 – The right panel plots the noise amplitude spectrum $\sqrt{S_h(f)}$, in three generations of ground-based interferometers. For the sake of clarity, we have only plotted initial and advanced LIGO and possible third generation detector sensitivities. Also shown are the expected amplitude spectra of various narrow- and broad-band astrophysical sources. The left panel is the same as the right except for the LISA detector. The supermassive BH sources are assumed to lie at a red-shift of $z = 1$ but LISA can detect these sources with a good SNR anywhere in the Universe. The masses labelling the curves correspond to the physical masses; the observed masses will be larger by a factor $(1+z)$. Also shown are examples of known AM CVn systems, ultra-compact X-ray binaries and double white dwarfs that stand above the confusion background from a population of Galactic double white dwarfs.

This range has also been the target for cryogenic resonant mass systems such as EXPLORER, NAUTILUS, AURIGA, and ALLEGRO, with the spherical detectors MiniGRAIL and the Mario Schenberg still under development. Resonant electromagnetic systems are being studied for the highest frequency signals.

## 2.3 The use of gravitational waves to test general relativity and other physics topics: relevance for fundamental physics

Astronomical sources of gravitational waves are often systems where gravity is extremely strong with relativistic bulk motion of massive objects. The emitted radiation carries the uncorrupted signature of the nature of the spacetime geometry and therefore is an invaluable tool to observe and understand the behavior of matter and geometry under extreme conditions of density, temperature, magnetic fields and relativistic motion. Here are some examples of how gravitational-wave observations will impact fundamental physics.

In Einstein's theory, the radiation travels at the speed of light and has two polarization states. In alternative theories of gravity neither could be true because gravitons (gravitational analogue of photons) might have a non-zero rest mass due to the presence of a scalar field (in addition to the tensor field) mediating gravity. Experimental tests of gravity, as well those afforded by the data from the Hulse-Taylor binary, are consistent with both Einstein's theory and one of its alternatives, the Brans-Dicke theory. Gravitational wave detectors will bring these theories face to face with observations that could decisively rule out one or the other.

According to Einstein's gravity, the spacetime in the vicinity of black holes is described by a unique geometry called the Kerr solution. Observation of the radiation from the in-fall of small black holes into super-massive black holes, phenomena that are believed to occur at the nuclei of most galaxies, will make it possible to test such uniqueness theorems. X-ray astronomy has provided firm indirect evidence that intense sources of x-rays may well host a black hole. An unambiguous signature of black holes, however, could eventually be provided by the detection of their quasi-normal modes – through gravitational radiation that has a characteristic frequency and decay time being emitted. Failure to detect such radiation from, for example, a newly formed black hole, would mean that gravity is more exotic than what we currently believe (e.g. gravitational collapse might lead to entities called naked singularities) and reveal new phases of matter at extremely high densities.

The most attractive versions of string theory require a ten dimensional spacetime, far larger than what we perceive. Certain phenomenological models at the interface of string theory and cosmology suggest that what we perceive as a four dimensional Universe (the brane) could indeed be a part of a higher dimensional Universe (bulk) with large spatial dimensions. The key feature of this theory is that gravitational interactions, and in particular gravitational waves, propagate in the bulk, while other interactions are restricted to the brane, which partly explains why gravity is so weak. Brane world models predict a specific signature in the spectrum of gravitational waves. Future ground- and space-based gravitational wave detectors offer the exciting possibility of observing radiation from the bulk, allowing us to explore whether the Universe has large extra dimensions.

## 2.4 Astrophysics/astronomy/cosmology with gravitational waves

### 2.4.1 Relativistic astrophysics

Electromagnetic astronomy has revealed a diverse and dynamic Universe that contains astrophysical sources for which strong gravity plays a critical role. These sources are of great current interest in astrophysical research. Examples include:

- Supernovae – end-states of stellar evolution resulting from sudden gravitational collapse followed by a huge explosion of in-falling matter;
- Gamma-ray bursts – intense sources of gamma radiation that last only a few seconds to minutes, yet emit more energy than would a star in its entire life time;
- End states of stellar evolution – white dwarfs, neutron stars, and black holes, both in binaries and as single objects, stellar "laboratories" that can be used to probe extremes of both fundamental physics and astrophysics;
- Super-massive black holes – believed to power most galactic nuclei, our own galaxy being a prime example with enigmatic massive young stars orbiting a central black hole four million times the mass of our Sun.

In each of these cases the source is believed to be couched in dense environs and strong gravitational fields and therefore a potential source of gravitational radiation. For example, sources of gamma-ray bursts could be colliding neutron stars which are electromagnetically invisible for most of their lives but which are very powerful emitters of gravitational waves. Transient radio sources could be the result of quakes in neutron stars with concomitant emission of gravitational radiation. Thus, the gravitational window on these phenomena will help us understand puzzles that have remained unsolved for decades.

The nucleus of nearly every galaxy is believed to host a compact object a million to a billion times as massive as our Sun and which is a powerful emitter of optical, radio and other radiation. What is the nature of this object? Is it a super-massive black hole or something even more exotic? How and when did it form? What is its relation to the size of the galaxy as a whole? These are but some of the questions which a model for the formation of structure in the Universe must answer. While electromagnetic observations have provided much valuable information to date, gravitational wave observations can help to address some of the key unanswered questions about the formation and nature of these objects – particularly in remote galaxies, where they will be extremely difficult to detect in the electromagnetic sector.

Future gravitational wave detectors will be sensitive to sources at very high red-shifts (great distances). Populations of different sources at such distances will help us understand, for example, the cosmological evolution of sources, the history of star and galaxy formation and their dependence on the matter content of the Universe, the development of large scale structure in the Universe and the formation of the network of super-clusters, the rich environs of galactic nuclei.

### 2.4.2 Cosmology

The latter half of the twentieth century witnessed the dawn of a golden age of cosmology. With the advent of radio and microwave astronomy it became possible to address key questions about the origin of the Universe and whether it truly originated in a big bang. The cosmic microwave background is a relic radiation from the big bang but because the early Universe was so dense, this radiation was in thermal equilibrium with matter for about $3.8 \times 10^5$ years after the birth of the Universe. It therefore cannot directly reveal the conditions in the very early phase of the Universe's history. The most direct way to observe the primeval Universe is via the gravitational window. Theoretical predictions based on fairly general assumptions indicate the production of gravitational waves in the early Universe which have been traveling to us unscathed as a consequence of their weak coupling to matter and other forms of radiation. A future space mission observing in the deci-Hertz band could detect this primordial stochastic background, and would surely represent one of the most exciting discoveries in cosmology.

The most surprising aspect of the Universe is that only about three percent of its content is in the form of visible matter; the rest is classified as dark matter and dark energy. One very effective method of probing the nature and distribution of these dark components is to make precise distance measurements of remote sources using "standard candles." These, by definition, are sources whose distance from the Earth can be inferred precisely from their apparent luminosity. Binary black holes may represent an astronomer's ideal standard candle. By measuring the signature of the gravitational radiation they emit it is possible to infer their intrinsic parameters (such as the masses and spins of the black holes) and thereby accurately deduce their luminosity distance. In fact, binary black hole sources eliminate the need to build a cosmic distance ladder (the process by which electromagnetic standard candles at different distances are combined in astronomy because there is no single source that may be suitably applied at all distances) and are therefore free from many of the systematic errors, such as obscuration by dust, that limit the use of standard candles in conventional astronomy. Observing the merger of binary super-massive black holes, and using them as precision cosmological distance indicators, will provide important clues about the nature of dark energy.

The early history of the Universe must have included several phase transitions during the process of evolution. Cosmic strings are one-dimensional topological defects that form at the boundaries of different phases of the Universe. Vibrations of these strings at the speed of light can sometimes form a kink which can then break, emitting a burst of gravitational radiation. The spectrum of the radiation from these strings has again a unique signature which can help us detect them and thus provide a glimpse of the Universe as it underwent phase transitions.

It is likely that the most exciting discovery made in the new gravitational window will be none of the above. If the history of conventional astronomy is an example, gravitational-wave astronomy should unravel phenomena and sources never imagined in the wildest possible theories – an eventuality of any new observational tool.

## 2.5 General background reading relevant to this chapter

Physics, Astrophysics and Cosmology with Gravitational Waves, B.S. Sathyaprakash and B.F. Schutz
Living Rev. Relativity 12, (2009), 2, http://www.livingreviews.org/lrr-2009-2 arXiv:0903.0338

# 3. Current state of the field

## 3.1 Introduction

The study of gravitational waves has come a long way since the first experimental searches began nearly half a century ago to the current worldwide network of kilometer-scale long-baseline interferometric detectors. This research has evolved to the situation today where detectors are starting to probe astrophysical predictions and detections are eagerly anticipated.

The first dedicated searches for gravitational waves looked for the effects of the waves on the resonant modes of massive objects, typically right circular cylinders of metal, a few meters long and of the order of 1000 kg, having a fundamental longitudinal resonant mode of ~1 kHz. A gravitational wave of amplitude $h \sim 10^{-21}$ will thus change the length of a m-long bar by $\sim 10^{-21}$ m. This is a very small displacement, and while bar detector sensitivity is today around a 1000 times better than the initial detectors in terms of gravitational wave amplitude – largely due to the use of cryogenic techniques – their current level of $h \sim 10^{-19}$ limits the likelihood of a bar detection.

Detector designs of increased sensitivity, spanning a broader frequency range than that accessible by resonant mass detectors, are now in operation. These are based on laser interferometric sensing to detect the changes, induced by gravitational waves, in the relative arm lengths of large-scale variants of Michelson-type interferometers, where the arms are formed between freely-hung mirrors.

After the study and construction in the 1970's and 80's of proof-of-principle interferometers of a few tens of meters length, creation of a new class of interferometer began in the 1990's for the purpose of gravitational wave searches. These instruments now form a worldwide network of detectors, and the TAMA interferometer was the first of this class. Operated at the Tokyo Astronomical Observatory, the TAMA detector has arm lengths of 300 m and was for a time the most sensitive interferometer. The German/British detector, GEO600, has 600 m long arms and operates on a site outside Hannover in Germany. It has the dual benefit of offering an interesting sensitivity for certain search classes and also offering a full scale test bed of innovative technologies such as low-thermal noise monolithic test-mass suspensions and advanced optical techniques. The American LIGO Laboratory comprises two detector systems with arms of 4 km length, one in Hanford, Washington State, and one in Livingston, Louisiana. Another half-length, 2 km inter-ferometer was also built inside the same evacuated enclosure at Hanford. Construction of LIGO began in the early 90's and subsequent progress has been outstanding, with the instruments now operating at design sensitivities. Finally Virgo, the French/Italian – and now also Dutch – detector of 3 km arm length at Cascina near Pisa, became the latest member of the worldwide network when it began collecting science data in 2007. Thanks to its advanced seismic isolation, Virgo offers the largest bandwidth for gravitational wave signals from 10 Hz to 10 kHz, a huge gain compared to the few Hz of the early resonant bar detectors. The current status of the interferometric detectors will be discussed in Section 3.4.

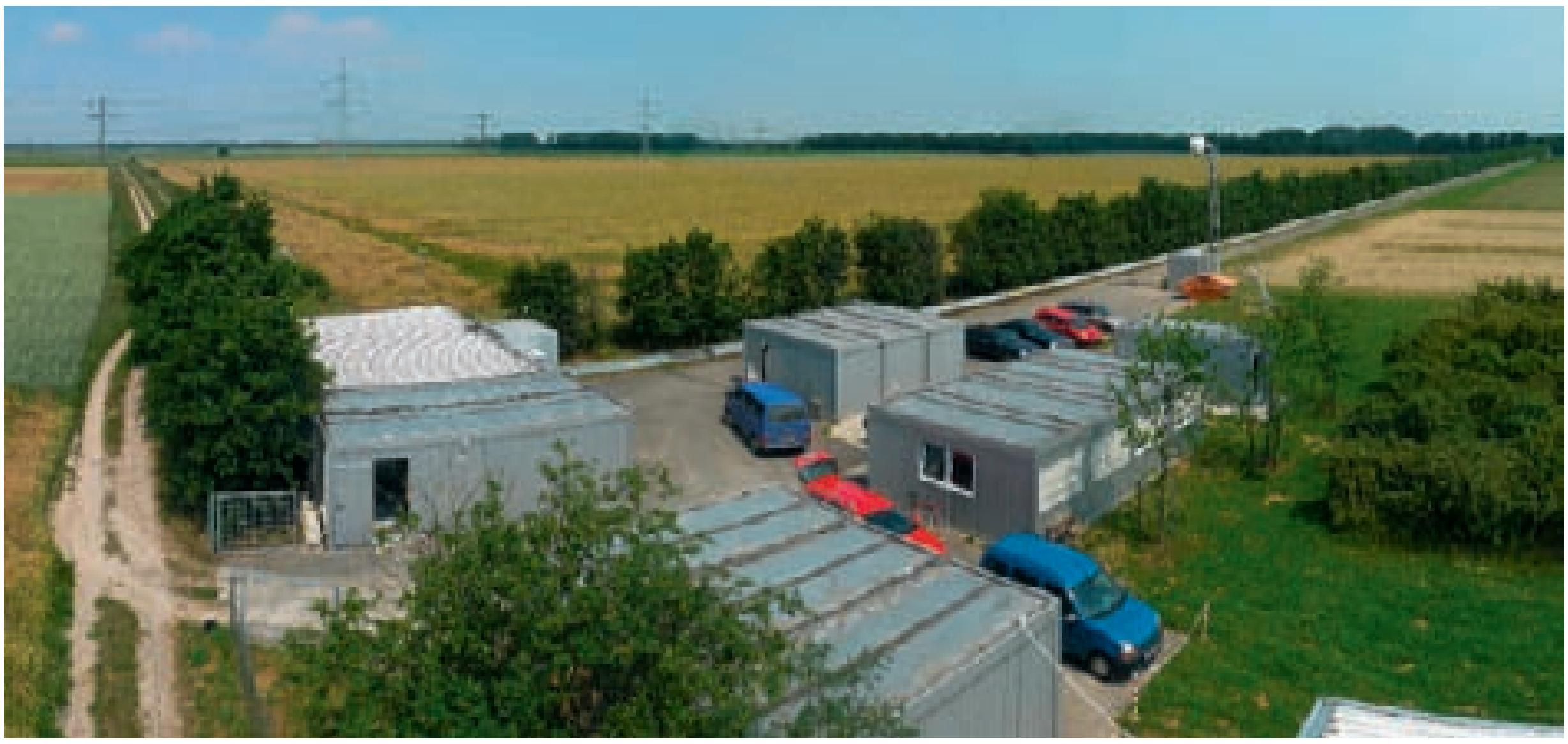

Fig. 3.1 – View of the German/British GEO600 detector site near Hannover, Germany.

The interferometric network has collected many months – and in some cases, years – of science data. The scope of the searches has been enlarged from the early goal of an opportunistic detection of a nearby supernova to the present systematic searches covering periodic sources, a stochastic gravitational wave background, and coalescing compact binary objects like neutron stars or black holes. The sensitivity of the instruments has improved by several orders of magnitude over the last five years, and is such that today the LIGO detectors are capable of observing the end of the life of a binary black hole system up to distances out to a few hundred Mpc (1 Mpc ~3 million light years).

The collected data sets are such that searches now challenge the upper bounds of the predicted event rates for such objects. Coincident searches with other sources of electromagnetic signals like gamma ray bursts, and the limit on the fraction of energy emitted through gravitational radiation which has been set on the crab pulsar, are other cases where the observations have produced interesting results for the astrophysical community. The status of these results is discussed in Appendix A.4.

Nevertheless, given our current understanding of the expected event rates, the detection of gravitational waves is not guaranteed with these initial interferometers. Thus plans are in place for upgrades to the existing detectors to create "enhanced" and "advanced" instruments, such that gravitational wave observation is expected within the first weeks or months of operating the advanced detectors at their design sensitivity. Advanced LIGO will incorporate significant major hardware improvements, with the installation of the upgrade expected to commence in 2011. Full installation and initial operation of the Advanced LIGO system is scheduled for 2014. On approximately the same timescale we can expect to see a similar upgrade to Virgo; the anticipated rebuilding of GEO as a detector aimed at high sensitivity in the kHz frequency region (GEO-HF); and the building of long-baseline detectors LCGT, in Japan, and AIGO in Australia in the coming decade. Further ahead lie the possibilities of "third generation" observatories in Europe (the Einstein Telescope, E.T.) and the USA.

With construction of the initial large interferometers completed, recent years have seen an evolution of the gravitational wave community towards operation of a true global array of gravitational wave antennae. This approach is driven by the nature of gravitational waves, which requires simultaneous observation from multiple antennae separated by continental distances to locate the sources on the sky, and to extract all information about the source characteristics encoded in the gravitational wave signal. As a key step in this direction, the LIGO, GEO and Virgo projects have joined forces to share their collected data and to carry out data analysis jointly.

Among the scientific benefits expected are: better confidence in signal detection, better duty cycle and sky coverage for searches, and better source position localization and waveform reconstruction. This collaboration is not limited to data analysis, and strong collaboration exists in the areas of hardware and technology development. This unified approach is of great benefit for the growth of the field. It is expected that as other instruments, such as LCGT, AIGO and next-generation instruments such as E.T., come online and reach design sensitivity, they will complement this global array of antennae. This alliance will result in a global gravitational wave observatory able to perform high quality gravitational wave astronomy and astrophysics, scanning the Universe from the 1 Hz to the several kHz band.

To access some of the most interesting gravitational wave signals (e.g. those resulting from the formation and coalescence of black holes in the range $10^3$ to $10^6$ solar masses) a detector is required whose strain sensitivity is approximately $10^{-23}$ in the region of $10^{-4}$ Hz to $10^{-1}$ Hz. The path to this goal is to fly a laser interferometer in space. The Laser Interferometer Space Antenna (LISA) is a proposed ESA/NASA mission comprising an array of three drag-free spacecraft at the vertices of an equilateral triangle of length of side $5 \times 10^6$ km placed at a distance of 1 AU from the Sun. Proof masses inside the spacecraft form the end points of three separate but not independent interferometers. LISA is a project in NASA's Physics of the Cosmos Program and is part of the ESA Cosmic Visions Program. The advisory committee reviewing the earlier Beyond Einstein Program recommended that LISA be the Flagship mission for the program. LISA is currently being reviewed by the Astro2010 decadal review committee of the US National Academy of Sciences. Thus LISA is expected to be launched as a joint NASA/ESA mission, possibly in 2020, and to produce data for five years or longer. A demonstrator mission, LISA Pathfinder, has a launch date in 2012. and, as discussed later, flight hardware is already being delivered.

The portion of the gravitational wave spectrum lying between 0.1 Hz and 10 Hz (between the LISA band and that probed by ground-based interferometers) also holds potential for observing a large number of interesting sources, including possible coalescences of intermediate black hole binaries. DECi-hertz Interferometer Gravitational wave Observatory (DECIGO) is a future planned space gravitational wave antenna first proposed in Japan, and aiming for launch after 2027. Also targeting this frequency range is the US-based ALIA mission concept.

Beyond LISA and DECIGO, the Big Bang Observer (BBO) is an interferometry-based mission in the NASA Beyond Einstein program, envisaged as a constellation of spacecraft aimed at seeing gravitational waves produced by the big bang inflation itself, searching for waves with frequencies of 0.1 - 10 Hz. The direct observation of inflating space-time will be a truly remarkable achievement.

This overview concludes by noting the searches for gravitational waves using data collected from pulsar observations. This technique makes use of the fact that pulsar rotation periods can be phenomenally stable, with the periods for millisecond pulsars stable to 1 part in $10^{14}$. In gravitational wave searches, a pulsar acts as a reference clock at one end of an arm sending out regular signals which are monitored by an observer on the Earth at the other end of the arm. Long term observations of pulsars have led to these searches setting upper limits on the energy density of a stochastic background of gravitational waves at very low frequencies, in the nano-Hz range, well below the ranges probed by other space- (or ground-) based detectors. Current results using pulsar timing already constrain the merger rate of supermassive binary black hole systems at high red shift, rule out some relationships between the black hole mass and the galactic halo mass, and constrain the rate of expansion in the inflationary era.

At present, pulsar timing studies are within an order of magnitude of detecting the stochastic gravitational wave background, given current estimates of the energy density from a population of supermassive black holes at $\Omega_{GW}$ of $\sim 10^{-9}$ at a frequency of ~10 nHz.

## 3.2 Status of numerical relativity and theory

The quest to detect gravitational waves is based on our understanding of general relativity (indeed of any theory of gravity that is compatible with special relativity), where the emission of gravitational waves is required by the existence of a fundamental limiting speed for propagation of information. However many of the most interesting sources involve extreme gravity and relativistic speeds, and considerable progress is being made on techniques for solving Einstein's equations to predict confidently the gravitational waves from these sources and to interpret the data taken. Furthermore, many important sources involve the behavior of matter under extreme conditions, and thus it is important that our knowledge of the laws governing that behavior be sufficiently robust.

During the past decade, a combination of analytical and numerical work has provided sufficient machinery to yield robust predictions from general relativity for the gravitational wave signals from the inspiral, merger and ringdown phases of compact binary systems of black holes or neutron stars. Indeed, recent breakthroughs in numerical relativity have been critical in providing solutions that link the inspiral signal, which is determined using analytical approximation techniques (commonly known as post-Newtonian theory), with the ringdown signal, which is determined from perturbation theory of black holes. These new methods are now being applied to the more complex and interesting case of mergers of rapidly spinning black holes, and substantial progress is likely during the next few years, in advance of the turning on of the advanced ground-based detectors or the launch of LISA. Meanwhile there has been considerable progress in simulating mergers involving neutron stars, including efforts to incorporate realistic equations of state for the nuclear matter, complex hydrodynamics, the effects of magnetic fields, and the microphysics of neutrinos.

A somewhat different problem is the extreme mass-ratio inspiral (EMRI) in which a small compact object inspirals into a massive black hole. Here the compact object can be viewed as a perturbation of the background spacetime of the large black hole, but one must take into account the "back-reaction" of the small body's gravitational field on itself, including the damping of the orbit due to the emission of gravitational waves. Considerable progress in this area has been made and is ongoing to develop waveform predictions for LISA that will cover the hundreds of thousands of expected orbits with sufficient accuracy.

There has been steady progress in recent years in computing the gravitational radiation from stellar collapse, in situations ranging from the accretion induced collapse of low-mass stars (< 8 solar masses) to the collapse of supermassive stars. The advances include the use of realistic models for the progenitor stars coming from stellar evolution codes, incorporation of realistic equations of state, use of advanced neutrino transport and interaction schemes, development of fully 3D Newtonian evolution codes and improvements in general relativistic codes. The historically wide range in estimates of gravitational-wave strength is beginning to converge, and efforts are being made to produce generic gravitational wave features as an aid to data analysis. Future progress will encompass the development of fully relativistic 3D codes, including the important effects of magnetic fields, and following the evolution of the post-collapse remnant.

The amplitude of gravitational waves from spinning neutron stars remains uncertain but is an area of considerable interest. The central issue is to what extent a neutron star can support a non-axisymmetric quadrupole or higher moment of its mass distribution large enough to generate detectable gravitational waves. A variety of mechanisms has been proposed, including unstable Rossby modes of oscillation in hot young neutron stars, anisotropic stresses due to magnetic fields, or free precessions. Underlying many of these phenomena are effects related to the microphysics of matter at supranuclear densities, such as superfluidity, color-superconductivity, or the appearance of strange particles or quark-gluon plasmas, which are as yet not well understood because they involve extrapolations beyond where experiments give reliable information on the behavior of nuclear matter.

For the more conventional sources, such as the galactic close binary systems, textbook general relativity is completely adequate. For LISA, for instance, such systems will serve as calibration sources.

There has been considerable theoretical work on estimating the spectrum of gravitational waves from processes in the very early Universe. Possible mechanisms include vacuum fluctuations in inflationary models, phase transitions from electroweak symmetry breaking, cosmic strings, and coherent excitations associated with branes residing in extra dimensions. Of course such models rely on extensions of the standard model of particle physics and thus are still somewhat speculative.

## 3.3 Status of resonant detectors

A resonant-mass antenna consists of a solid body that “rings like a bell” when a gravitational wave hits it; because of the quadrupole nature of the radiation only the odd-order vibrational modes are excited. The vibration is then converted into an electrical signal by a motion- or strain-transducer, and then amplified by an electrical amplifier. The thermo-mechanical properties of the body and its coupling with transducer/amplifier, together with external noises determine the sensitivity and bandwidth of the detector.

For the cylindrical resonant-mass detectors (ALLEGRO, Louisiana State University; AURIGA, Legnaro – *see picture, left side*; EXPLORER at CERN; NAUTILUS, Frascati – *see picture, right side*) the strain sensitivity is in the range of ($10^{-20}$ - $10^{-21}$) /√Hz at the resonant frequency (a little lower than 1 kHz) with a bandwidth increased in the last few years from some Hz to several tens of Hz. All these detectors have a similar central body, a 3 m long aluminium cylinder with a mass a little higher than 2.2 ton, and are cooled down to a few K or less.

With these characteristics, the target sources are galactic supernovae and millisecond pulsars but with a very low expected event rate, or probability of detection. A very good duty cycle allows long data-tak-ing runs, and long term joint observations were performed within the IGEC (International Gravitational Event Collaboration) including all resonant-mass detector groups. For the most recent observations the detectors were reoriented parallel to each other. The network analysis is based on a time coincidence search to reject spurious events and IGEC demonstrated a false alarm rate as low as 1/century. To date, no positive results have yet been reported.

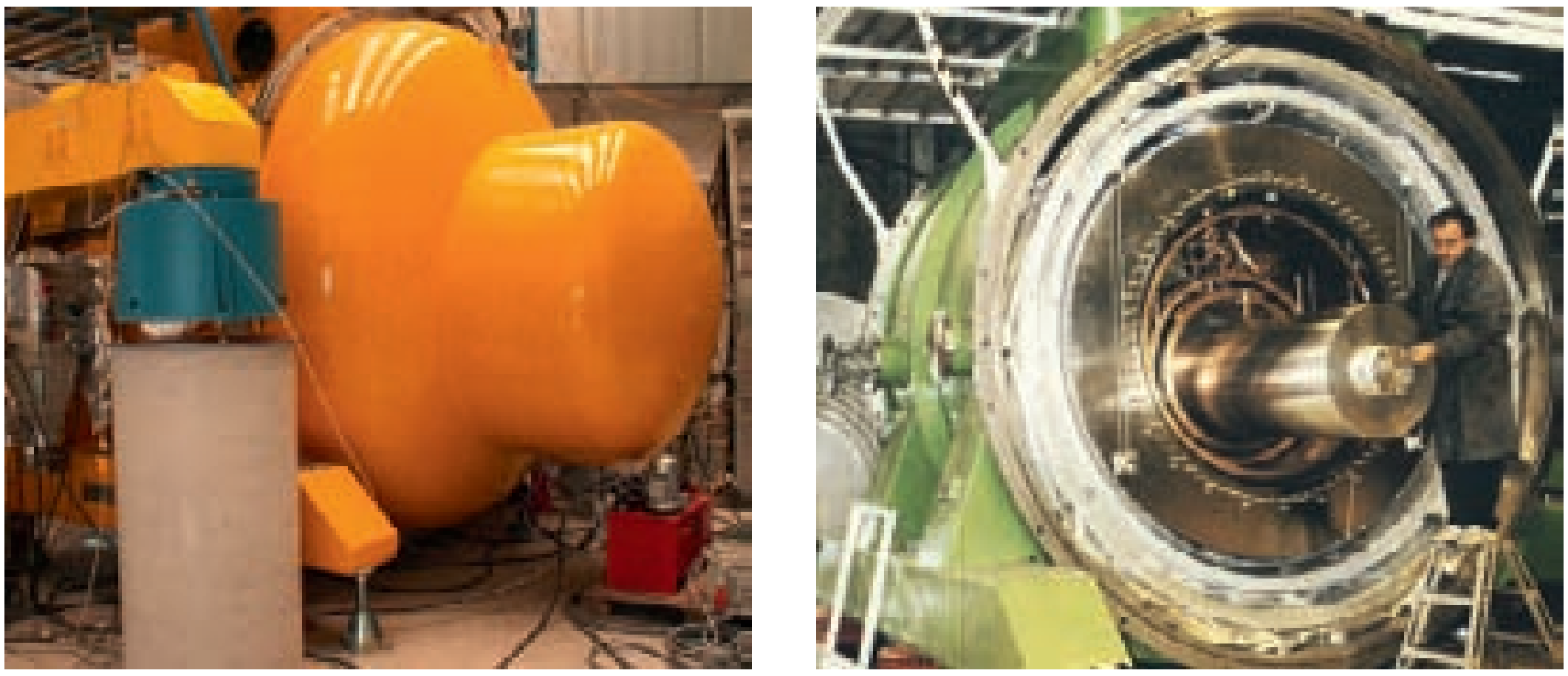

Fig. 3.2 – The AURIGA (left) and NAUTILUS (right) resonant detectors.

An interesting study has been performed with EXPLORER and NAUTILUS, collecting and analyzing the data of many events due to cosmic rays impinging on the bars. This study can be relevant for the technique of acoustic detection of particles with cryogenic detectors and for understanding the effect of cosmic rays on future interferometers.

A very innovative bar-type detector, still in the research and development phase, is represented by the DUAL concept. It consists of an elastic test mass (possibly of cylindrical shape) with low mechanical dissipation. The deformation induced by the gravitational wave is measured so as to select the superposition of the main gravitational wave sensitive resonant modes and reject the others. As a consequence some of the thermal and readout noise contributions are reduced. However there are problems with thermal noise in the optical readout system and these must be overcome to achieve an interesting sensitivity around 3 kHz.

## 3.4 Status of interferometric detectors

As mentioned, a network of interferometric gravitational wave detectors exists, where each instrument uses laser interferometric sensing to detect the changes, induced by gravitational waves, in the relative arm lengths of large-scale variants of Michelson-type interferometers, where the arms are formed between freely-hung mirrors. The optimum arm length is one quarter wavelength of the gravitational wave signal or $7 \times 10^5$ m for a signal centered on 100 Hz. However this is not achievable in practice on Earth. To compensate, in ground-based systems multiple reflections of light are used in the interferometer arms either by resonant Fabry-Perot cavities (Fabry-Perot Michelson arrangement) or by optical delay lines.

The main limitations to the sensitivities of current interferometric detectors are seismic noise, thermal noise in the mirrors and suspensions and photoelectron shot noise in the detecting photo-diodes. To reduce the last effect high laser powers (10s of Watts currently) are used, and further, to improve optical impedance matching of the laser to the interferometer a mirror of finite transmission is placed between the laser and the interferometer. This forms an overall resonant cavity with the interferometer system and increases the light intensity inside the interferometer. This impedance matching technique is known as power recycling and is used in all the interferometric detectors.

A further enhancement to sensitivity may be obtained by recycling a signal sideband by a mirror placed at the output of the detector before the detection photodiode. This has the effect of increasing sensitivity over a given bandwidth determined by the transmission of this mirror; the center frequency for enhancement is determined by the positioning of this mirror. This technique of “signal recycling” has been used to date only in the German/UK GEO 600 detector, but is planned for incorporation in future detector upgrades.

The main interferometer projects will be described in the following sections, and the status of the scientific results from the interferometer network discussed in Appendix A.4.

### 3.4.1 LIGO

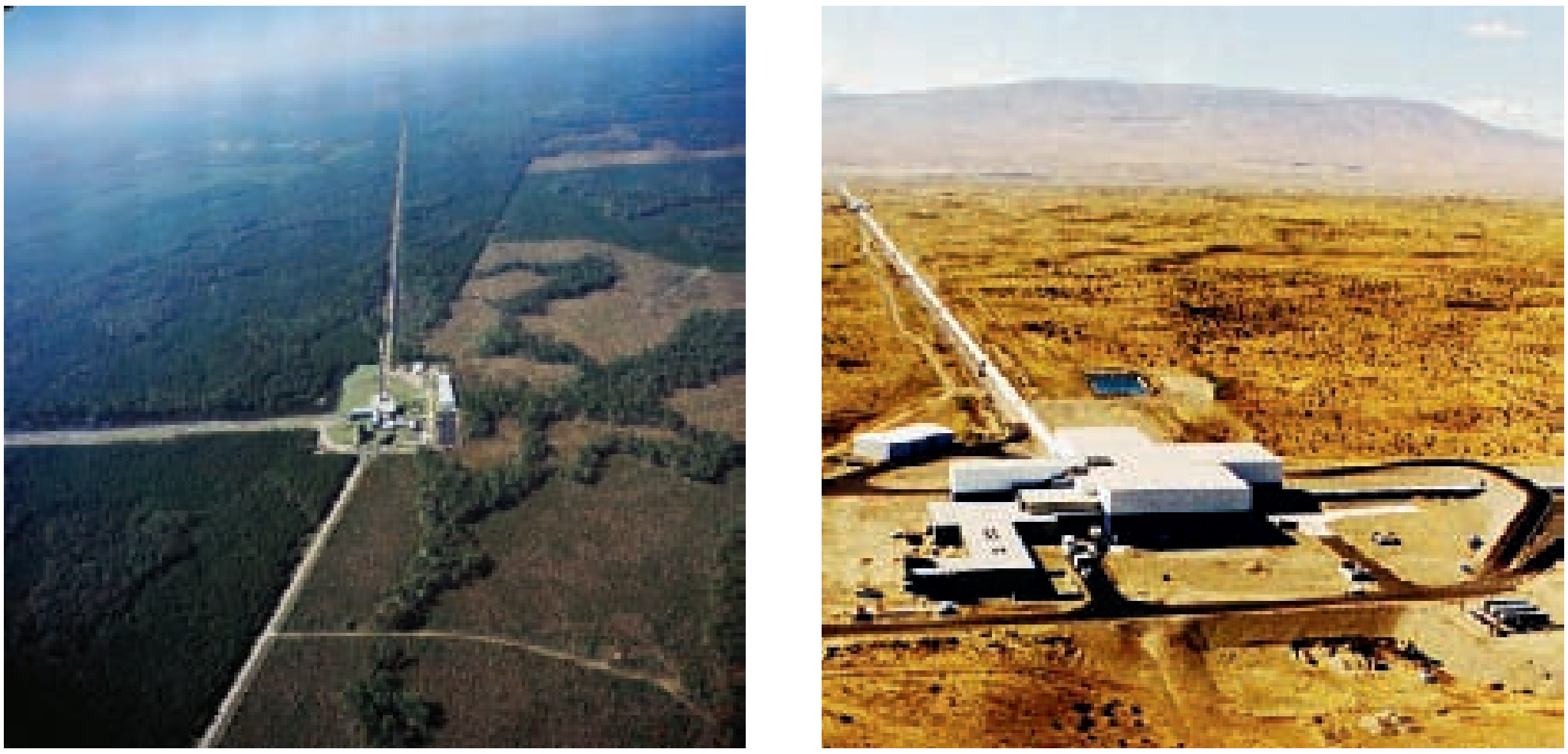

Fig. 3.3 – Bird's eye view of the LIGO detector sites at Livingston (left) and Hanford (right).

The Laser Interferometer Gravitational-wave Observatory (LIGO) is located in the US. LIGO comprises two 4 km long L-shaped facilities, one at Hanford, Washington and the other in Livingston, Louisiana (see figure 3.3). The site at Hanford also has a 2 km detector sharing the same vacuum system. LIGO is funded by the US National Science Foundation and managed by the California Institute of Technology and the Massachusetts Institute of Technology.

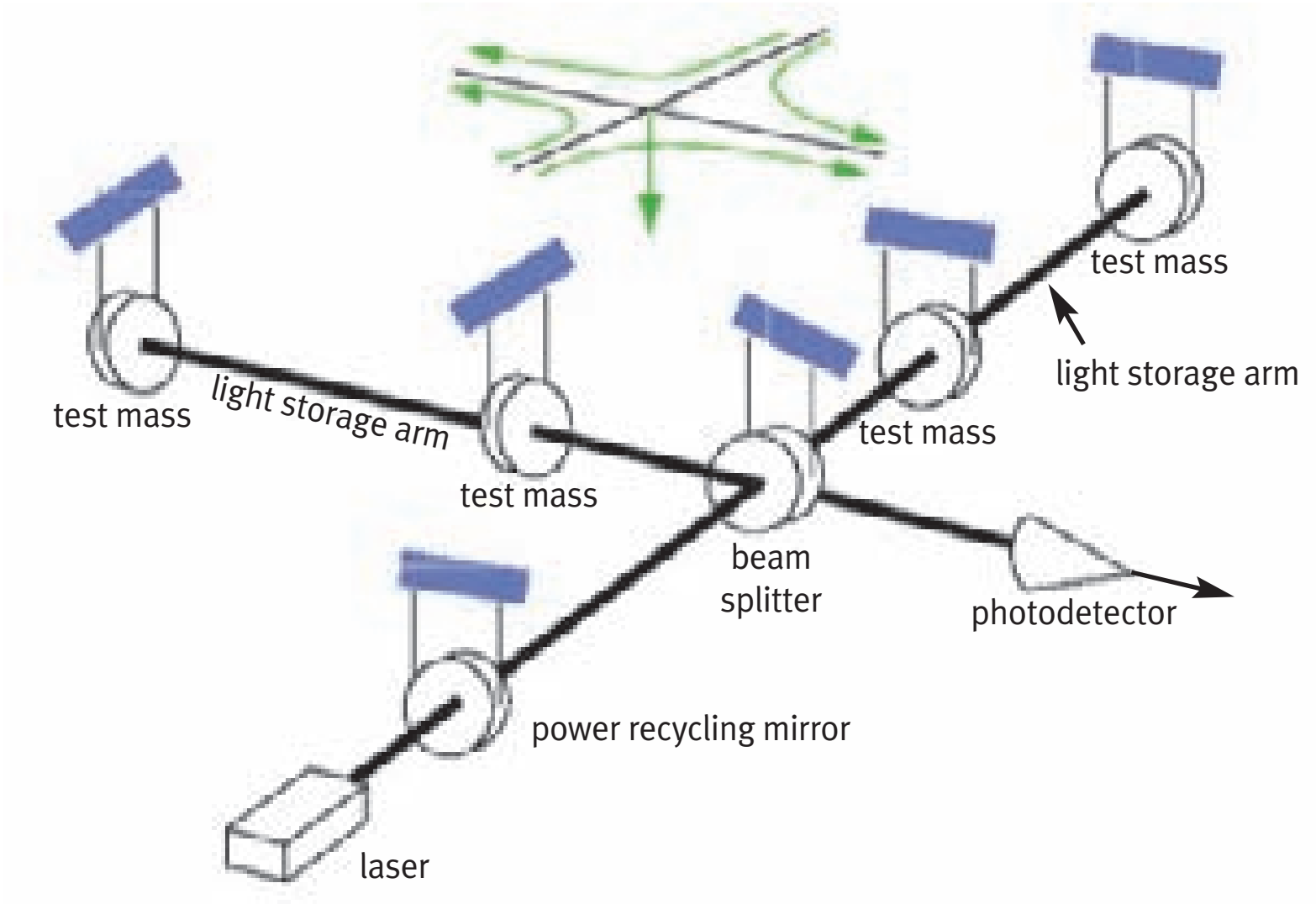

Fig. 3.4 – Schematic diagram of a LIGO interferometric gravitational wave detector.

The LIGO detectors have the Fabry-Perot Michelson configuration with power-recycling (figure 3.4). A highly stabilized 10 W Nd:YAG laser operating at 1.06 μm strikes the beamsplitter and is directed to the two arms. The interferometer is operated with constructive interference on the side of the beam-splitter which is illuminated by the laser, sending the light from the interferometer back toward the laser, and destructive interference on the other side where a photodetector reads out the difference in arm length. A partially transmitting mirror placed in the incident laser beam creates a second (compound) cavity, allowing the laser power incident on the beamsplitter to build up and increase the sensitivity of the interferometer. The interferometer mirrors are suspended in vacuum with simple pendulum suspensions using metal wires from a four-stage passive isolation system to protect them from ground motion which might interfere with the measurement. A complex control system maintains proper positioning and alignment of the optics. The LIGO detectors were designed so that the dominant noise would be seismic motion at low frequencies, thermal noise at intermediate frequencies and shot noise at the highest frequencies, to provide a sensitive band between approximately 40 Hz and 6 kHz. Figure 3.5 shows the strain sensitivity of one of the LIGO detectors compared with its design goal.

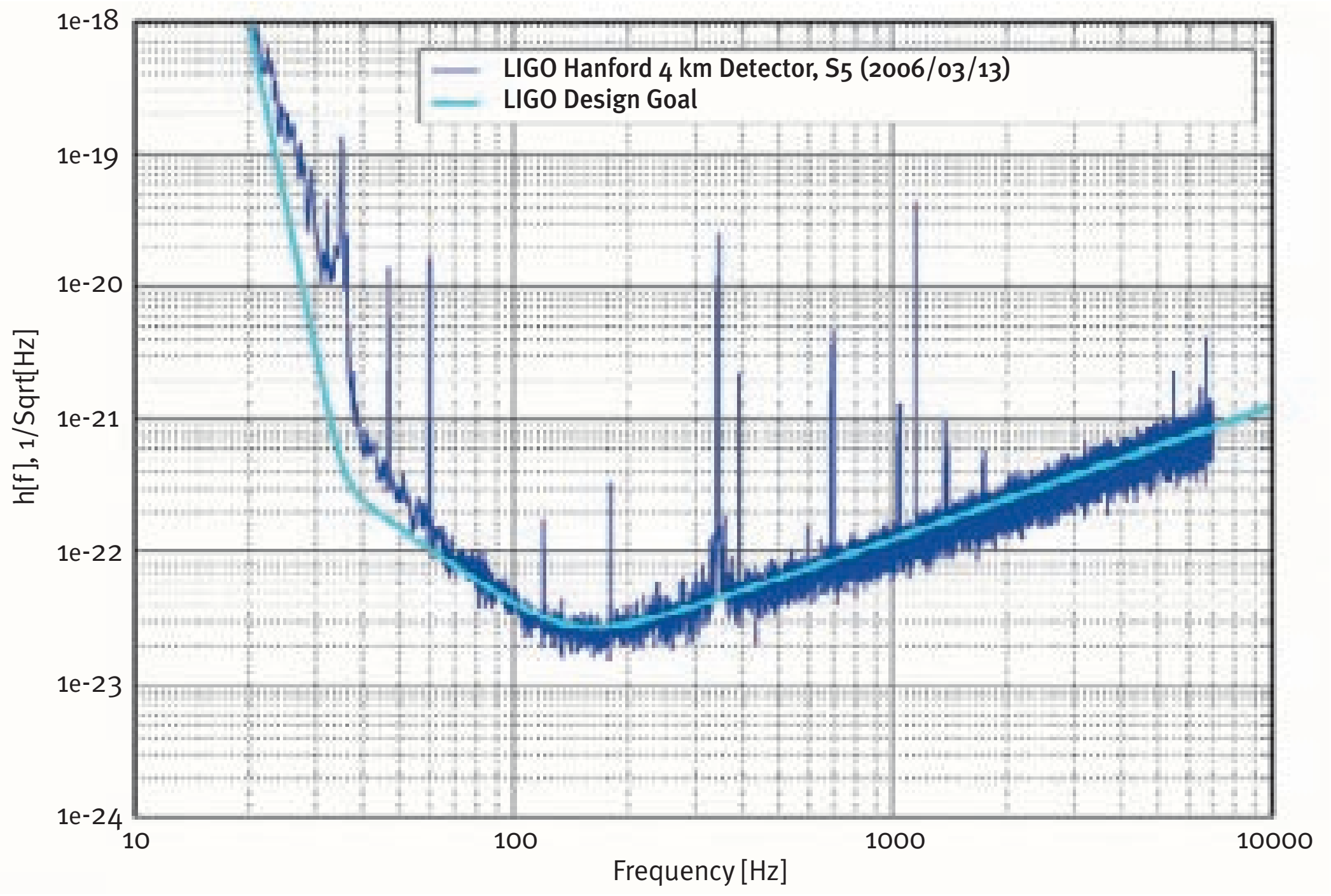

Fig. 3.5 – Strain sensitivity (noise level) versus frequency for one of the LIGO detectors, compared with the design goal. Except for a small range at low frequency, the detector meets or exceeds its goal.

In 2005, following five years during which the sensitivity of the instruments was improved by several orders of magnitude, the LIGO detectors reached their design sensitivity, making them, for example, capable of detecting the inspiral and merger of a neutron star binary system out to the distance of the Virgo cluster of galaxies (depending upon orientation). From November 2005 through September 2007, LIGO engaged in its fifth Science Run (S5) with the goal of collecting the equivalent of one full year of observation with all three of its detectors operating in coincidence. The LIGO research program is carried out by the LIGO Scientific Collaboration (LSC) with roughly 800 members at more than 70 institutions in eleven countries.

The successes of the initial LIGO detectors and a vigorous R&D program set the stage for an upgrade to the existing LIGO detectors, dubbed Advanced LIGO, which received its first funding in 2008. Advanced LIGO is designed to take advantage of new technologies and ongoing R&D leading to a substantial increase in sensitivity and bandwidth over the existing LIGO detectors. Virtually every part of the LIGO detectors, except the facilities and the vacuum system, will be replaced. New active anti-seismic systems, operating to lower frequencies, will provide the platforms for test mass suspensions. Lower thermal noise suspensions incorporating multiple pendulums and all fused silica final stages will be used. New low loss optics and a higher power laser will improve noise in the high frequency regions currently limited by shot noise. The Advanced LIGO optical configuration includes a signal recycling mirror to improve the sensitivity and to allow tuning of the sensitivity as a function of frequency to target particular types of sources.

The expected sensitivity of Advanced LIGO is shown in figure 3.6. In the most sensitive region Advanced LIGO will have a factor of ten times better sensitivity to gravitational wave strain, than initial LIGO. This improvement translates into a ten times larger range to detect sources, or a factor of one thousand larger volume in which gravitational wave sources can be observed. The extension of the sensitive frequency band to lower frequencies also greatly broadens the range of sources which may be detected. Construction of Advanced LIGO subsystems began in 2008, with their installation in the existing LIGO vacuum system expected to begin in 2011.

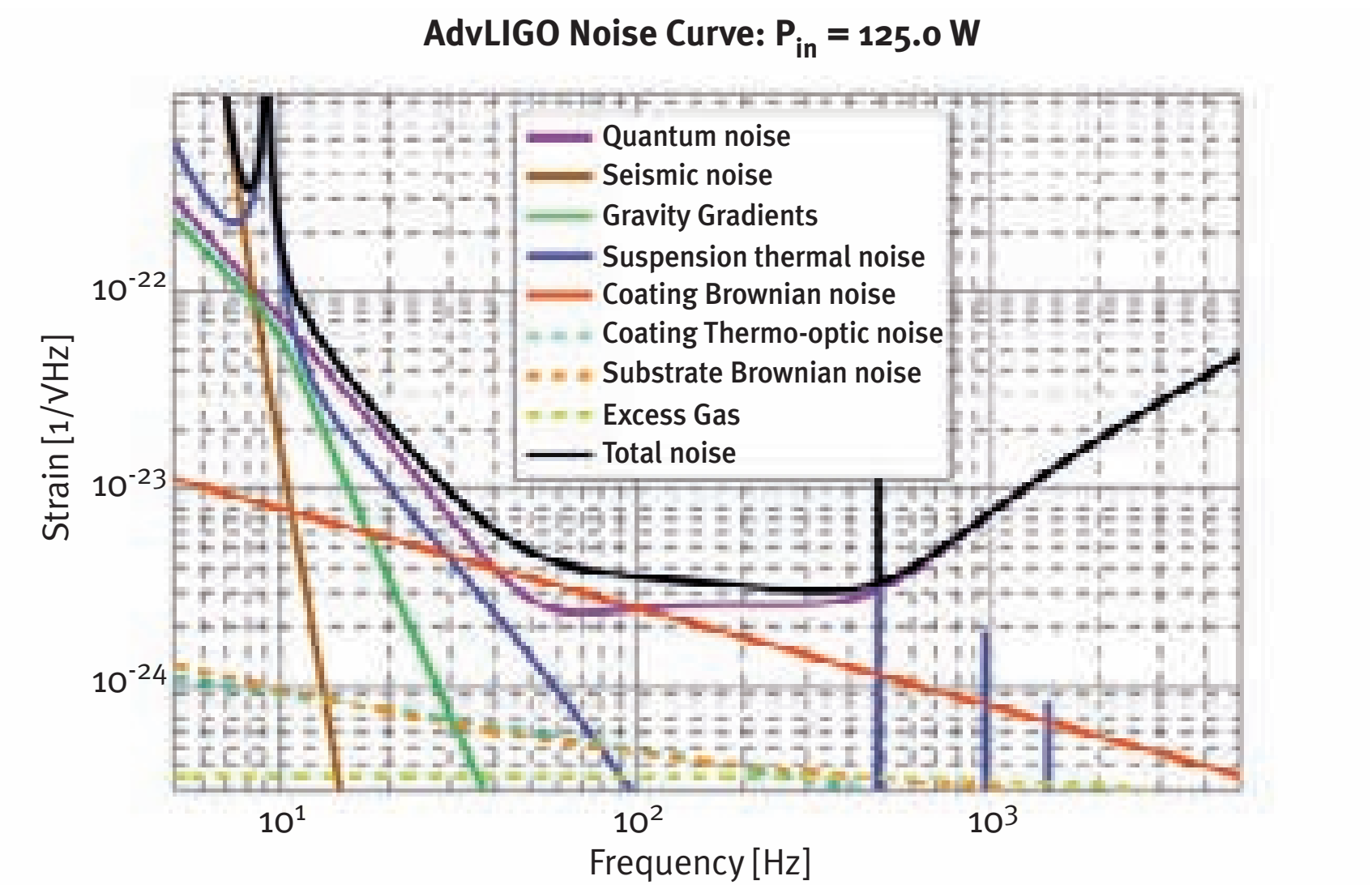

Fig. 3.6 – Design noise level for the Advanced LIGO detector. The Advanced LIGO noise curve shown is the sum of numerous contributions: seismic noise, suspension thermal noise, mirror thermal noise, quantum noise and gravitational gradients. (Figure courtesy of the LIGO Scientific Collaboration - please see https://dcc.ligo.org/cgi-bin/DocDB/ShowDocument?docid=2974 for further details.)

This schedule for Advanced LIGO leaves more than three years before the initial LIGO detectors must be removed, enough time for one significant set of intermediate enhancements. This project, called Enhanced LIGO, aims for a factor of up to two improvement in sensitivity over initial LIGO, corresponding to a factor of up to eight in the volume of the Universe observed. The main technical elements of Enhanced LIGO include a higher power laser (35 W versus the current 10 W), and a change in the readout scheme for the interferometer. The readout of the interferometer will be changed from a radio frequency scheme using phase modulated sidebands to a DC readout scheme using a tiny offset from the DC fringe minimum. An output modecleaner will be added to eliminate higher order transverse modes from the light incident on the photodetectors. Enhanced LIGO will give early tests of some Advanced LIGO hardware and techniques, thus reducing the risk and raising the probability of success of Advanced LIGO. Installation of Enhanced LIGO began in late 2007, after the completion of the S5 run with initial LIGO. LIGO's sixth science run (S6) using Enhanced LIGO technologies began in mid-2009.

### 3.4.2 Virgo

Virgo is a Michelson based interferometer with 3 km long arms located near Pisa, Italy. Like the LIGO interferometers, Virgo uses Fabry-Perot cavities in its arms and the technique of power recycling to enhance the sensitivity of the detector. However, Virgo has already the unique feature of using an advanced generation of seismic isolation, making the seismic noise contribution negligible above 10 Hz.

Construction of Virgo, approved by CNRS and INFN in 1993, started at the Cascina site in 1996. Virgo is managed by the European Gravitational Observatory (EGO), a private consortium with Dutch, French and Italian funding agencies as members. The construction was completed in June 2003 and was followed by a commissioning phase to optimize the control systems of the detector. In 2007 Virgo's sensitivity matched that of LIGO at high frequency and offered the best sensitivity at low frequency. The first Virgo science run started in May 2007 and lasted more than four months, in coincidence with part of the LIGO S5 data-taking run. This started the era of joint analysis between the LIGO and Virgo detectors.

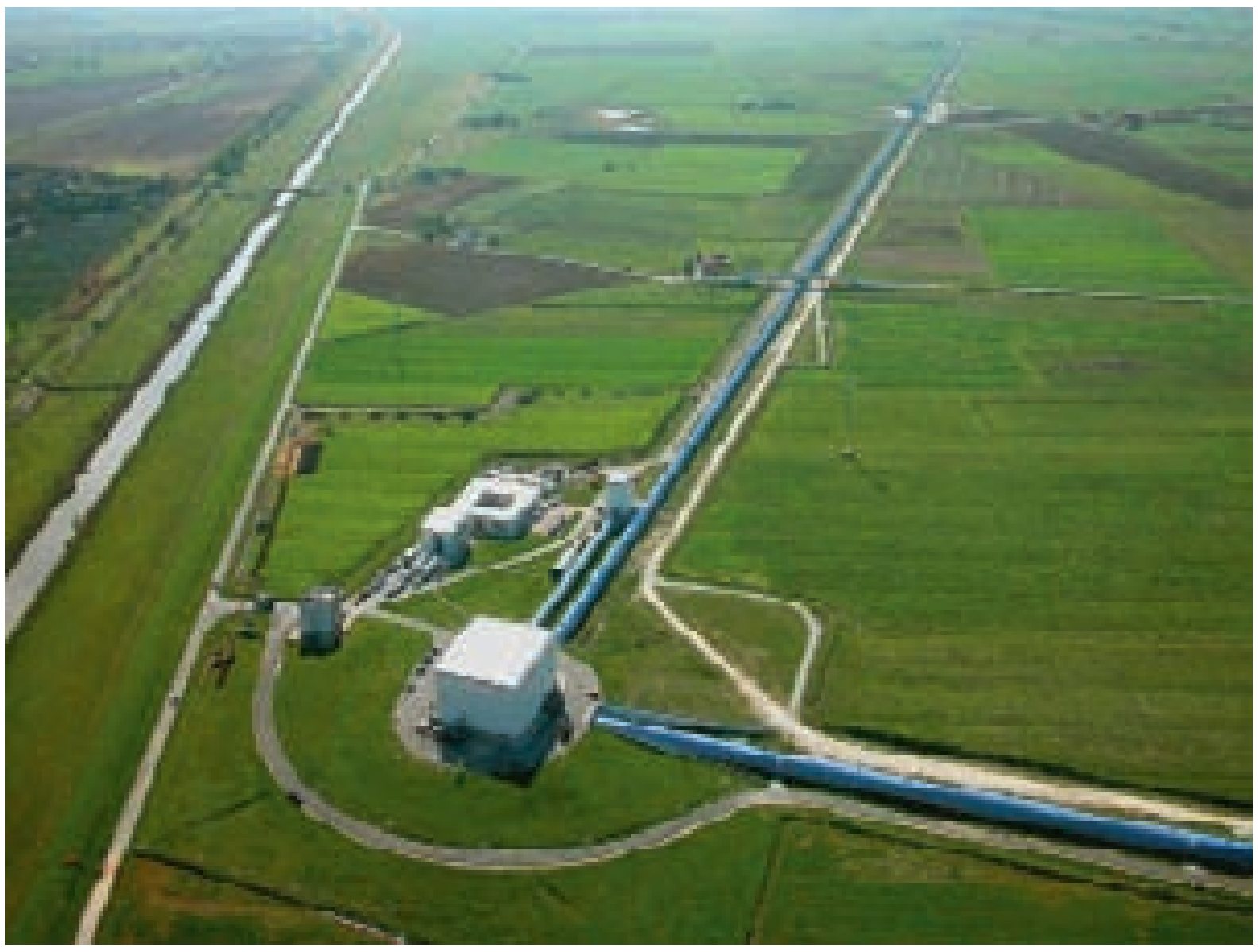

Fig. 3.7 – General view of the Virgo interferometer.

This run has been followed by another commissioning period and the beginning of an intermediate upgrade to Virgo, dubbed Virgo+. This upgrade initially involves increasing the laser power and modification of the control and readout electronics but does not modify the overall interferometer layout. Consequently, the upgrade period could be relatively short. The first Virgo+ upgrades were installed during spring/summer 2008, with the initial commissioning scheduled for completion by end 2008/early 2009 to allow the second Virgo science run to start in spring 2009, in coincidence with the LIGO S6 run. Replacement of the cavity mirrors and their suspension by a monolithic design as used by GEO 600 (section 3.4.4) and by an increase in cavity finesse is also scheduled as part of the Virgo+ upgrades but will be implemented later, in 2009/2010. Depending on the exact list of improvements installed during the Virgo+ phase, the sensitivity improvement will be between a factor of 1.5 to 4 times better than the nominal Virgo design level.

The next upgrade stage is the Advanced Virgo phase. Following Virgo+, the Advanced Virgo design foresees changes in the optical configu-ration. In addition to a change of beam topology, signal recycling will be introduced. The change of beam geometry will require an upgrade of the vacuum infrastructure and therefore requires a rather long shutdown. Heavier mirrors are also foreseen to reduce the thermal noise at low frequency, complemented by a new monolithic suspension. However, only a modest change of the seismic isolation system is planned for Advanced Virgo. The expected sensitivity is approximately a factor 10 better than the initial Virgo design, and therefore will match the Advanced LIGO sensitivity, maintaining a worldwide network of interferometers.

Procurement for Advanced Virgo construction is expected to start in early 2010 to enable hardware installation late 2011/2012. The following commissioning phase is expected to end at the same time as that of the Advanced LIGO detectors in order to start the first joint run of the Advanced LIGO/Virgo detectors around 2014.

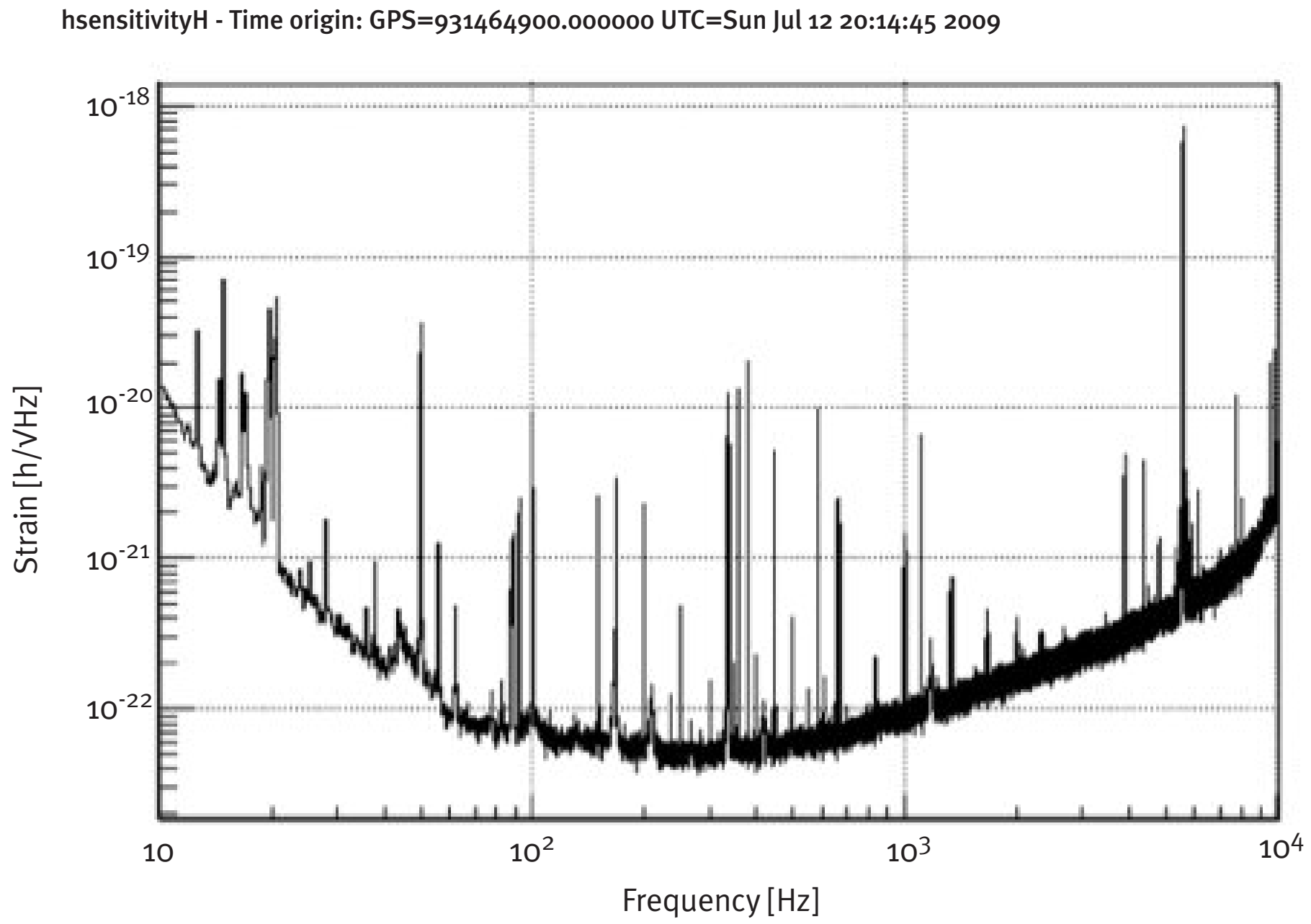

Fig. 3.8 – Sensitivity of the Virgo interferometer, July 2009

### 3.4.3 TAMA

The TAMA project was initiated in 1995 by combining the existing experimental gravitational wave groups in Japan. The TAMA interferometer was regarded as being a step towards reaching a large scale interferometer in terms of both technology and construction budget in Japan. The baseline length of TAMA is 300 m and it is placed underground at the Mitaka campus of the National Astronomical Observatory of Japan (NAOJ) in the west suburb of Tokyo. Like LIGO and Virgo, TAMA is a power recycled Michelson-Fabry-Perot system.

Nine data-taking runs were conducted by TAMA, several of these in coincidence with LIGO. TAMA's neutron star binary coalescence range achieved 70 kpc by the last run.

Still, a large gap was recognized between the sensitivity achieved and that possible with the detector infrastructure. After the last run, TAMA paused in its observational phase to allow a period of noise reduction efforts. The primary focus was to reduce the effect of stray light beams and install an improved system to further isolate the test mass mirrors in the detector from ground vibrations. This is in the final stage of commissioning to allow improved sensitivity for TAMA, and to study the physics of such interferometers. TAMA will be used as a test bench for technologies of the next generation detectors. One of these is a new long-baseline detector system proposed in Japan – the Large-scale Cryogenic Gravitational-wave Telescope (LCGT)[1].

The LCGT detector is proposed to be built underground to reduce seismic noise and the effects of gravitational gradients on test-mass displacement. It will also use cryogenically cooled test masses to reduce thermal noise. A series of instruments was used to study the feasibility of an underground cooled detector, which culminated in current studies using a 100 m cryogenic interferometer underground at the Kamioka mine, known as the CLIO project. Following successful cooling performance of test masses to 20 K, noise-hunting was started at room temperature, and research is ongoing with an upper limit on gravitational wave amplitude emitted from the Vela pulsar of $5.3 \times 10^{-20}$ being set at 22 Hz at a 99.4% confidence level.

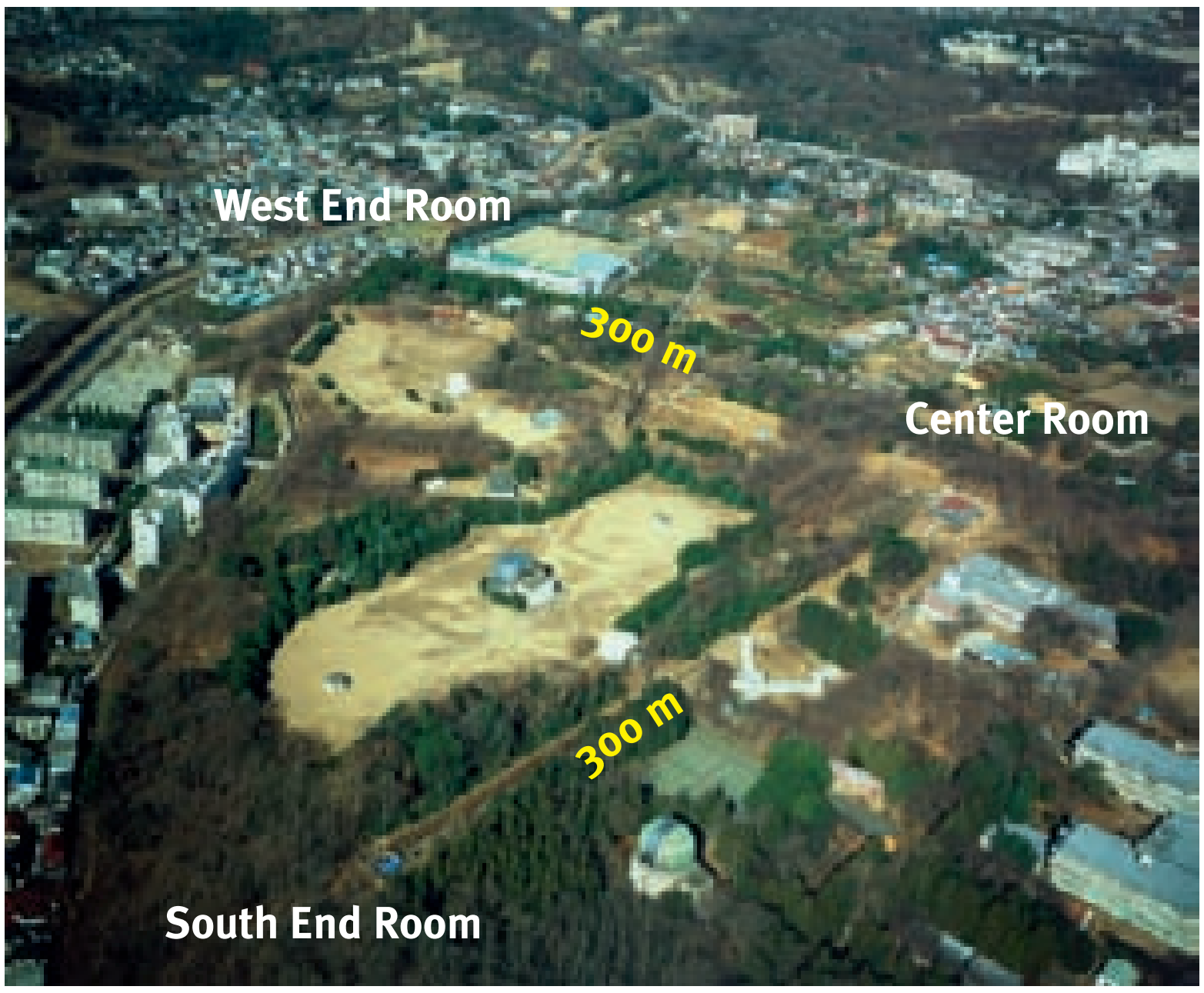

Fig. 3.9 – Aerial view of the TAMA site.

[1] subsequent to the preliminary version of this roadmap being made available internationally, approval of funding for the first phase of construction of LCGT was announced in June 2010.

### 3.4.4 GEO

GEO600 is a German/UK interferometer of 600 m arm length situated near Hannover – see figure 3.10 below. The project is funded by STFC, MPG and the State of Niedersachsen.

Construction and installation of GEO600 concluded in 2002. GEO uses a three-bounce, four-beam, delay line in each arm and both power recycling and signal recycling are incorporated as shown in figure 3.11.

Between 2002 and 2007 GEO600 participated in four science runs, extended periods of data-taking, in coincidence with the US LIGO detectors.

The GEO collaboration is a member of the LIGO Scientific Collaboration, and the data collected by the network are analyzed by members from LIGO and GEO within the LSC.

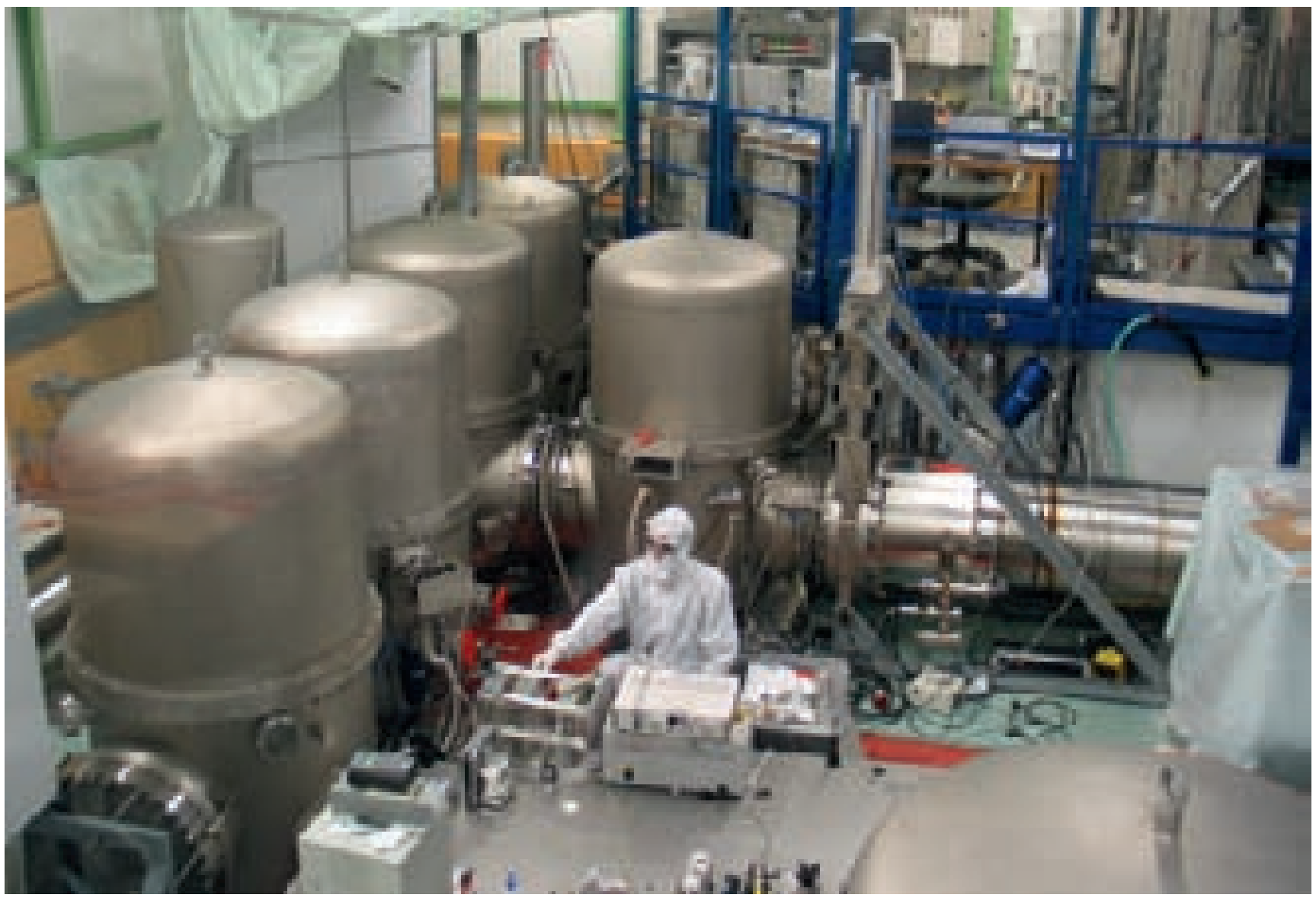

Fig. 3.10 – GEO600 vacuum system.

In 2006 and 2007, work at GEO600 focused on the fifth LIGO Scientific Collaboration data taking run (S5) in coincidence with the three LIGO detectors.

#### Astro-watch engagement and further commissioning

After S5, the LIGO and Virgo full-length detectors began an upgrading procedure and from November 2007 until July 2009 were not capable of collecting data. Along with the 2km Hanford instrument, GEO600 covered this duration as much as possible, operating in 24/7 mode with short commissioning periods. The goal of these commissioning periods was to maintain the sensitivity and high duty cycle as well as to perform short investigations to prepare for the GEO-HF upgrades to be installed immediately after the *astro-watch* period. In the high frequency range the aim was to gain improved sensitivity by about a factor of two by changing the optical readout system and injecting squeezed light through the output port. Once GEO600 runs in a stable way with squeezed light injection, the light power in the interferometer will be increased by replacing the laser with a higher power version. With its improved sensitivity GEO-HF can either then join the LSC S6 data taking run, or perform a second astrowatch program during the Advanced LIGO and Advanced Virgo upgrade period.

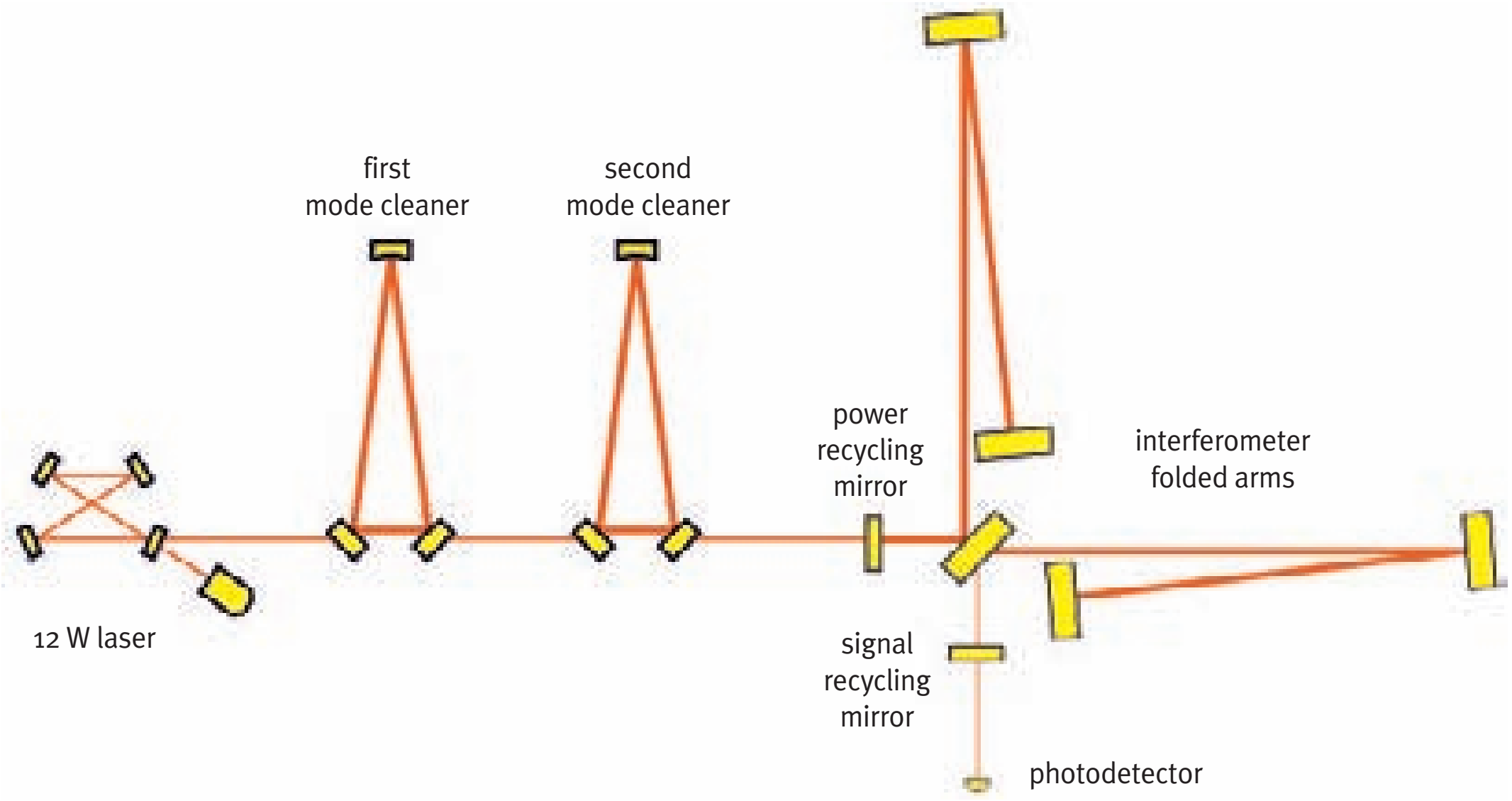

Fig. 3.11 – Optical layout of the GEO interferometer.

### 3.4.5 LISA / LISA Pathfinder

The Laser Interferometer Space Antenna (LISA) is an ESA-NASA project to develop a space-based gravitational wave detector that will operate in the band between 3 x $10^{-5}$ and 0.1 Hz. LISA detects the gravitational-wave induced strains in space-time by measuring changes of the separation between free flying test masses in three spacecraft 5 million kilometers apart.

LISA will produce a catalog of extraordinary astrophysical sources: tens to hundreds of inspiraling and merging binaries composed of intermediate-mass and massive black holes out to z ~ 20; tens of stellar-mass compact objects spiraling into central massive black holes out to z ~ 1; tens of thousands of close, compact binaries in the Galaxy; a sky map of the background made by millions more; and possibly backgrounds of cosmological origins. For many of those sources, the catalog will include precise measurements of the astrophysical parameters. For the more energetic mergers, the mission will release sky position and luminosity distance weeks to months in advance to trigger electromagnetic searches for counterparts.

This catalog and the underlying data will enable studies of: the formation and growth of massive black holes and their co-evolving host galaxies; structure formation; stellar populations and dynamics in galactic nuclei, compact stars, the structure of our Galaxy, severe tests of General Relativity in the most extreme conditions, cosmology and searches for new physics. When used in conjunction with electromagnetic observations, these observations promise spectacular breakthroughs in astrophysics.

The LISA mission architecture has been stable since 1997 and under continuous development and analysis for more than two decades. The architecture is well defined and extensively analyzed. The requirements have been flowed down, and the error budget is well understood and labo-ratory tested. The LISA mission uses three identical spacecraft whose positions mark the vertices of a equilateral triangle five million km on a side, in orbit around the Sun. LISA can be thought of as a giant Michelson interferometer in space, with a third arm that provides independent information on the two gravitational wave polarizations, as well as redundancy. The spacecraft separation – the interferometer armlength – sets the range of gravitational wave frequencies LISA can observe (from about 0.1 mHz to above 0.1 Hz). LISA will be sensitive enough to detect gravitational wave induced strains of amplitude $h = \Delta l / l < 10^{-23}$ in one year of observation, with a signal-to-noise ratio of 5. The center of the LISA triangle traces an orbit in the ecliptic plane, 1 AU from the Sun and 20° behind Earth, and the plane of the triangle is inclined at 60° to the ecliptic. The natural free-fall orbits of the three spacecraft around the Sun maintain this triangular formation throughout the year, with the triangle appearing to rotate about its center once per year.

The LISA concept relies on unusual flight technologies like “drag-free” flight and precision measurement techniques that are well established in the laboratory, like laser metrology. LISA Pathfinder is a technology demonstration mission for LISA, led by ESA, scheduled for launch in 2012. LISA Pathfinder will test technology designed for LISA, i.e. drag-free control, microthrusters and interferometry technology. If successful, no new developments regarding these technologies are required for LISA, so Pathfinder can be considered as part of the LISA project itself. The Pathfinder ground development program is complete; engineering models of the science instruments have been built and flight qualified. Flight hardware is now under construction.

The LISA partnership has an effective international team based on agreements that have been in place for more than five years. The Project is in the Mission Formulation Phase at both ESA and NASA, and ESA has had a formulation study contractor, Astrium GmbH, for nearly four years. The Project team and independent organizations have estimated the mission cost and schedule multiple times since 2001, and the current estimates are robust and reliable for the current state of development. In Europe, LISA and LISA Pathfinder are integral components of ESA’s Cosmic Vision Scientific Programme announced in 2005 (ESA 2005). LISA Pathfinder is expected to launch in 2012 and LISA is an L1 candidate in Cosmic Vision for a launch in 2020. In the US, LISA has been endorsed as a high priority mission in several influential reports. In particular, most recently in 2007, the Beyond Einstein Program Assessment Committee, commissioned by the National Research Council of the Na-tional Academies gave a glowing recommendation for LISA. Currently, LISA is under review in the Astro2010 Decadal Survey of Astronomy and Astrophysics.

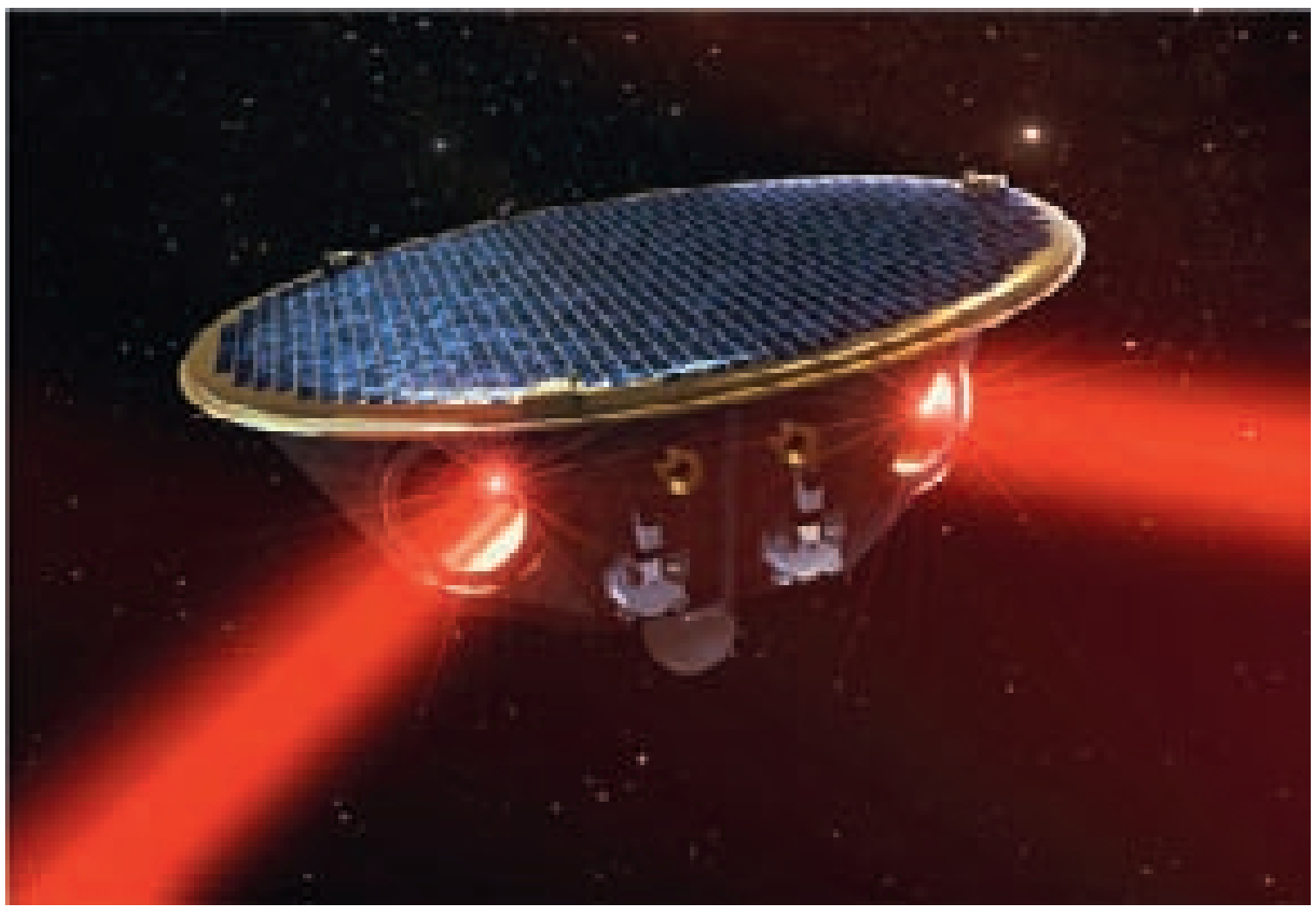

Fig. 3.12 – Artist’s impression of one LISA spacecraft

As mentioned above, to pave the way to LISA, ESA and NASA are implementing a precursor mission, LISA Pathfinder (LPF), with the aim of demonstrating a large fraction of LISA metrology in interplanetary orbit.

LPF consists of one entire LISA arm, except for the 5 million kilometer laser transponder, which is taken out in order to accommodate the full instrument within one spacecraft (SC).

The scope is to quantitatively validate all possible aspects of the model for the physical disturbances that affect the LISA arm described so far, with particular emphasis on those that cannot be tested in 1 g on ground. In addition this is achieved by using as much as possible the exact hardware to be used on LISA.

Thus LPF flies, in interplanetary orbit, the LISA technology Package (LTP): two test masses (TM), and two local interferometers measuring both the TM relative displacement and their displacement relative to the SC. TM, electrode-housing, TM launch-lock device, UV charge management system, and micro-Newton thrusters, are nominally equal to those to be used on LISA. All in all then the requirement for LPF is to demonstrate a differential acceleration noise of the arm of better than

$$S^{1/2}_{\delta a_{tot}}(f) \leq 3\times 10^{-14} \left(ms^{2}/\sqrt{Hz}\right)\left[1+\left(f/3\,mHz\right)^{2}\right]$$

at frequencies between 1 and 30 mHz. It must be stressed that even if LISA could only achieve the performance represented by the LPF requirement, a large fraction of the galactic binary signals, and those from super-massive black-holes, would indeed still be detectable with significant signal to noise ratio.

LISA Pathfinder is currently in an advanced implementation state for a launch in 2012.

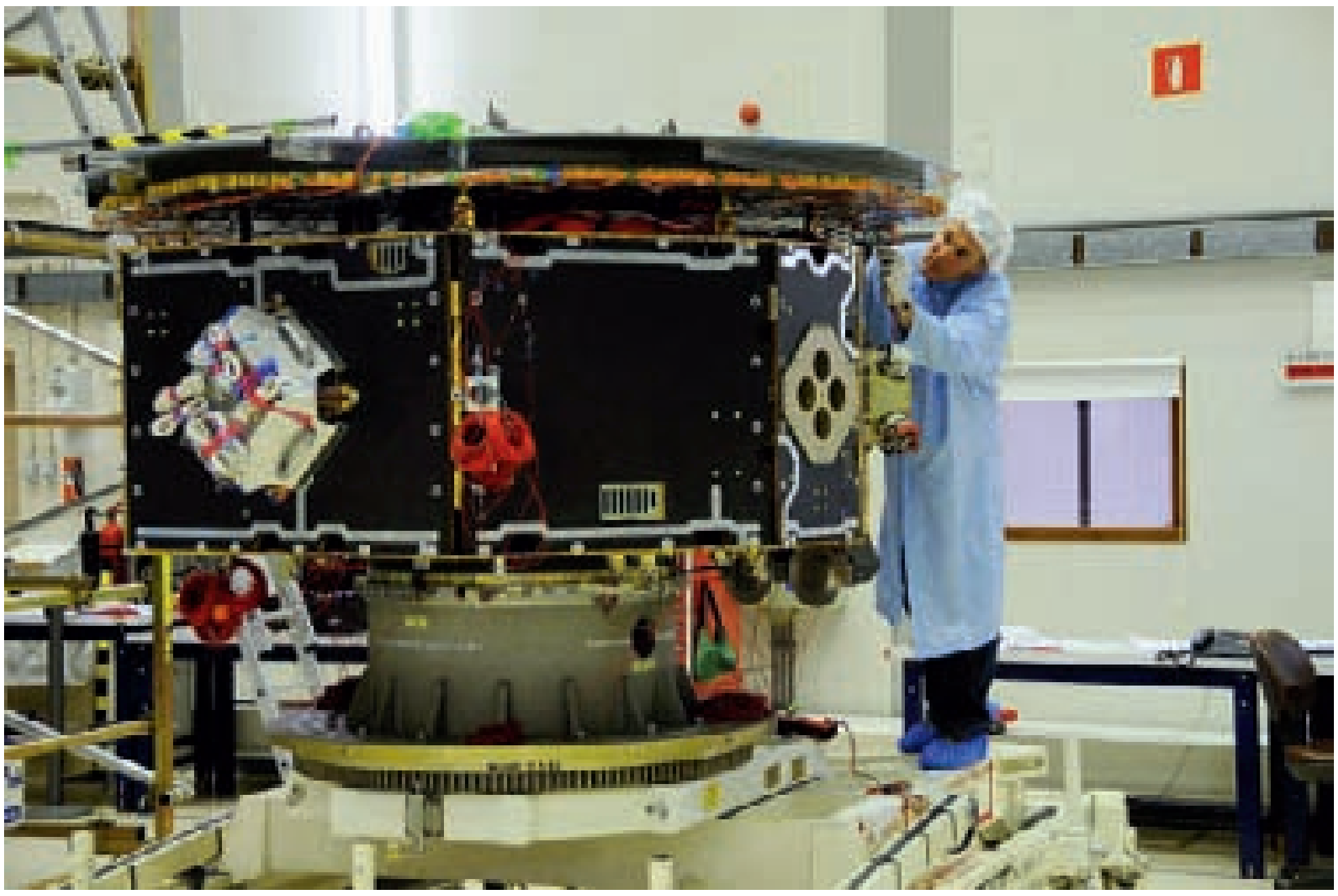

Fig. 3.13 – LISA Pathfinder spacecraft under integration

## 3.5 Doppler tracking measurements and pulsar timing

The tidal force experienced by separated masses is the basic effect of a gravitational wave. This means that Doppler Tracking of Spacecraft involved in a range of missions can be used for gravitational wave searches. In these projects the Earth and an interplanetary spacecraft are test masses, whose relative position is determined through a nearly monochromatic microwave signal sent from a ground station to the spacecraft and (coherently) sent back to the Earth. The ground station compares the frequency of the transmitted signal with the frequency of the signal it is receiving. The output is a Doppler frequency time series $\Delta\nu / \nu_o$, where $\nu_o$ is the central frequency of the signal from the station. Anomalous perturbations (caused by gravitational waves for example) can be revealed as peculiar signatures in the Doppler time series, with the frequency band studied here ranging from approximately $10^{-5}$ to 0.1 Hz, with the low end of this band set by the reciprocal of the round-trip light-time for spacecraft in the Solar System. In the past decades many experiments (Ulysses, Mars Observer, Galileo, Mars Global Surveyor) have collected data; and more recently Cassini performed two 40-day data-taking campaigns (2001-2003), and one 20-day reduced campaign, with a sensitivity (for nearly monochromatic signals) around $10^{-16}$ over a broad frequency band. No dedicated experiments are planned at present, but adding Doppler Tracking as one of the experiments in a planetary mission would permit coverage of the low frequency range, while waiting for laser space interferometers such as LISA to come online.

The measured ratio $\Delta\nu / \nu_o$ is used also in Pulsar Timing Experiments, but in a very different context. Millisecond Pulsars (MSPs) are incredibly precise clocks: a gravitational wave passing over the pulsars and over the Earth modulates the received pulsar period; this can be seen as a slight variation in the measured frequency of the pulsar.

Pulsar timing experiments measure variations of pulse phase relative to a model prediction. However measurements of just a single pulsar or even a small number of pulsars can only set a limit on the strength of a gravitational wave signal; many sources of noise exist and to detect gravitational waves measurements using an array of pulsars is needed. The fundamental point is that signals from different sources have different spatial correlation signatures; clock errors will produce the same residual for all pulsars (monopole signature); planetary ephemeris errors are equivalent to an error in earth velocity, leading to 180° anti-correlated residuals (dipole signature); the quadrupolar symmetry of gravitational waves leads to different modulations for different pulsars, with the antenna pattern having a four-lobed shape, reminiscent of a quadrupole. It is worth noting that MSPs are not perfectly stable, with intrinsic timing noise. However timing noise in different pulsars is uncorrelated. So, an array of pulsars widely distributed in the sky and observed at regular time intervals allows separation and detection of the different sources of noise. The weakness of gravitational wave signals results in very small contributions to the residuals, and so long-term observations are needed.

Pulsars are typically observed every few weeks for many years; the peak sensitivity being reached at a frequency f = 1/T or in the $10^{-9}$ Hz band. Even for these long data spans, the expected time residuals are very small (order of nano-seconds for binary BHs of $10^9$ solar masses in a galaxy at red shift 0.5 with a four-year orbital period, while typical rms timing residuals are 2 - 3 orders of magnitude larger). But the expected level of the stochastic background from binary super-massive BHs in galaxies is considerably higher, with millions of galaxies in the Universe contributing. Predicted backgrounds from other sources, such as cosmic strings or relic gravitational waves from the Big Bang, are at a comparable level. The present upper limit on the gravitational wave energy density relative to the closure density of the Universe obtained with pulsar timing is about $2 \times 10^{-8}$. This is the most stringent limit to date on the existence of a gravitational wave background in the nano-Hertz frequency band. The direct detection of gravitational waves is the main goal of the European Pulsar Timing Array (EPTA), the North American Nanohertz Observatory of Gravitational Waves (NANOGrav), and the Parkes Pulsar Timing Array (PPTA) projects. These three collaborations aim to observe a sample of 20 MSPs with a timing precision of 100 ns over more than five years and have joined forces in the International Pulsar Timing Array (IPTA).

The European Pulsar Timing Array (EPTA) collects pulsar data on a large sample of pulsars using a combination of 100-m class radio telescopes at Jodrell Bank (UK), Effelsberg (D), Nancay (F), Westerbork (NL) and (later) Cagliari (I). The North American project, NANOGrav, has more than 20 years of data on the 305-m Arecibo and 100-m Green Bank Radio telescopes. The PPTA project is using the 64-m radio telescope in Parkes, Australia (350 km from Sydney), operated by the Australian Telescope National Facility (ATNF).

Looking ahead, the EVLA will provide another highly sensitive telescope in the Northern Hemisphere on a timescale of two years, and the proposed Allen Telescope Array 350-dish build out (USA) would add a similarly sensitive telescope within five years. The proposed Square Kilometer Array (SKA) will have very high sensitivity and is expected to observe 100 millisecond pulsars with rms residuals of the order of 50 ns for observing periods of 10 years, pushing the detection limit for the stochastic background at 3 nHz to $\Omega_{GW} \sim 10^{-13}$. The site of the SKA is not yet chosen but it is planned that it will be operational by the year 2020. Given the importance of the IPTA's science, the community may seriously consider building a dedicated pulsar timing facility.

## 3.6 Cosmic microwave background polarization measurements

Microwave background radiation is described by its temperature and polarization as a function of sky direction. Measurements of the polarization of the cosmic microwave background (CMB) radiation provide a means for detecting primordial gravitational waves generated by inflation. The ratio of tensor (gravitational-waves induced) perturbations to scalar (density induced) perturbations in the polarization is a measure of the energy scale of inflation and thus a diagnostic of the nature of the field giving rise to inflation.

The results of the 2008 WMAP set a limit on a tensor-scalar ratio R of around 0.2. A number of future space (e.g. Planck, CMBPOL), balloon (e.g. EBEX, SPIDER) and ground based (e.g. QUIET, POLARBEAR) experiments are in development and should greatly increase achievable sensitivities for gravitational wave effects.

## 3.7 Other research and development

There is an 80m interferometer research facility in Western Australia near Gingin – currently used for high power optics developments relevant to Advanced LIGO. This site could be the basis for a large southern hemisphere dedicated interferometric gravitational wave detector, expanding the existing network of long-baseline instruments.

Several processes active in the very early Universe, for example, parametric amplification of quantum fluctuations during inflation, are thought capable of generating gravitational waves at very high frequencies, with the predicted gravitational wave spectrum having maximum amplitudes in the MHz to GHz region, necessitating the development of specialized instrumentation for their detection. Projects at the University of Birmingham (UK) and INFN Genoa (I), are targeted at fabricating high-frequency gravitational wave detectors using resonant electromagnetic techniques, however significant enhancements of current performance would be required to obtain useful sensitivities.

## 3.8 General background reading relevant to this chapter

*http://www.ligo.caltech.edu/*

*http://www.virgo.infn.it/*

*http://geo600.aei.mpg.de/*

*http://gw.icrr.u-tokyo.ac.jp:8888/lcgt/*

*http://www.gravity.uwa.edu.au/*

*http://www.universe.nasa.gov/gravity/research/numrel.html*

*http://lisa.jpl.nasa.gov/*

*http://sci.esa.int/science-e/www/area/index.cfm?fareaid=27*

*http://cgwp.gravity.psu.edu/*

*http://cgwa.phys.utb.edu/*

*http://www.skatelescope.org/*

*www.nanograv.org*

*http://www.atnf.csiro.au/research/pulsar/ppta/*

*http://www.astron.nl/~stappers/epta*

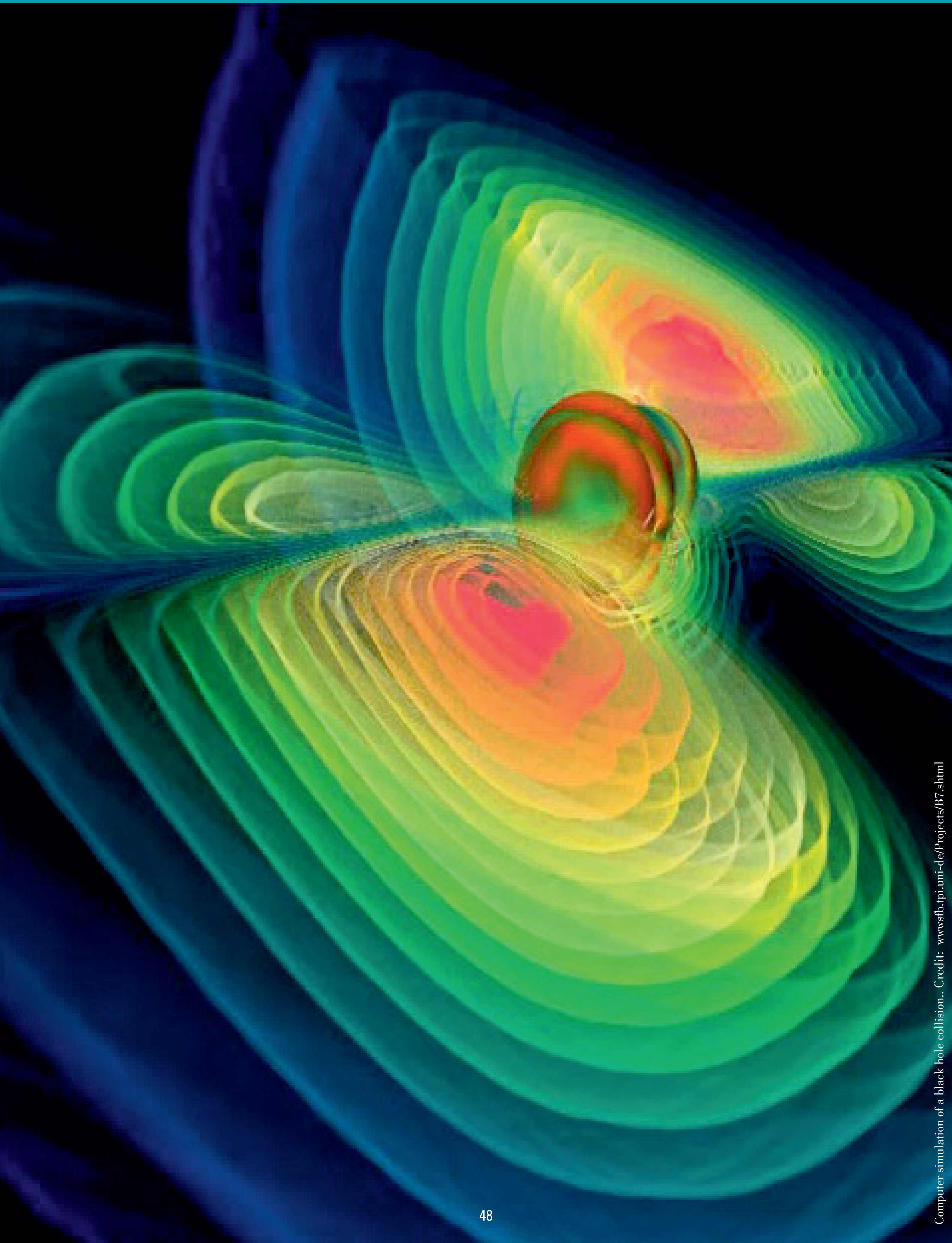

Computer simulation of a black hole collision.. Credit: wwwsfb.tpi.uni-de/Projects/B7.shtml

# 4. Scientific opportunities in gravitational wave science now and in the next several decades

## 4.1 Introduction

The next three decades of research in gravitational waves promise to be rich in scientific discoveries, both in the realm of fundamental physics and general relativity, as well as in astronomy, astrophysics and cosmology. Gravitational wave astronomy will open a unique window on the Universe, with the capability of detecting phenomena unobservable by electromagnetic or other means. Because gravitational waves are not absorbed by matter, they can be detected in principle from very great distances, and from processes occurring in the very early Universe.

If the initial period covered by this roadmap sees the first direct detection of gravitational waves by ground-based interferometers, then a stunning confirmation of one of the central predictions of Einstein's general relativity will have been accomplished. But beyond the detection of waves are opportunities for major scientific discoveries, both in fundamental physics and in astrophysics and cosmology as are fully discussed below.

## 4.2 Questions for fundamental physics and general relativity

### What are the properties of gravitational waves?

General relativity makes two firm predictions for gravitational waves, that their speed is identical to the speed of light, and that they come in only two polarizations (see Chapter 2 for a discussion). To date there is no direct observational evidence that addresses either of these fundamental properties.

If a detected burst or inspiral chirp of gravitational waves should be accompanied by an electromagnetic signal such as a gamma-ray burst from the same source, then a comparison of the arrival times of the two bursts will give an immediate, high-precision measurement of the speed of gravitational waves relative to that of light. For example, if ground-based interferometers detect a signal from a source at 200 Mpc whose arrival is within minutes of the arrival of an electromagnetic pulse from the same source, then a bound of parts in $10^{15}$ on any difference in speed between the two waves could be achieved. Space-based interferometers such as LISA could also bound the speed difference. Such a test depends crucially on having sufficient angular resolution for the gravitational wave signal to identify reliably that the electromagnetic pulse came from the same source.

From the perspective of elementary particle physics, the prediction that the speed of gravitational waves is the same as that of light goes hand-in-hand with the concept that the "graviton," the hypothetical carrier of the gravitational interaction, has zero mass. If, by contrast, the graviton were massive, then the speed of gravitational waves would depend on wavelength, and the detected chirp-like signal from inspiraling binaries would be distorted compared to what would be expected if the graviton mass were zero. By searching for such distortions in the signal using matched filtering techniques one can place stringent bounds on the graviton mass. The advanced ground-based interferometers could place bounds comparable to existing bounds derived from examinations of the orbits of outer planets in our solar system, and LISA, observing the waves from inspirals of massive black hole binaries, could do four orders of magnitude better. Such bounds could constrain a variety of theories of gravity designed to go beyond general relativity, including theories where the non-gravitational interactions reside on a four-dimensional "brane" in a higher dimensional spacetime, while gravity extends to all the dimensions.

If a gravitational wave is detected simultaneously in three independent detectors of the ground-based array, and the source direction is determined by other means, then it will be possible for the first time to test whether the waves are composed of only two polarization states. General relativity predicts that only two modes are present irrespective of the source, while scalar-tensor theories predict a third polarization mode. Similarly LISA, whose spacecraft orbit the Sun, will be sensitive to varying mixtures of the polarization modes in the waves from a sufficiently long-lasting source, and thus will also be able to test the polarization content of the waves as well as determine the source direction. Any confirmed evidence of a third mode of polarization would cast doubt on the validity of general relativity.

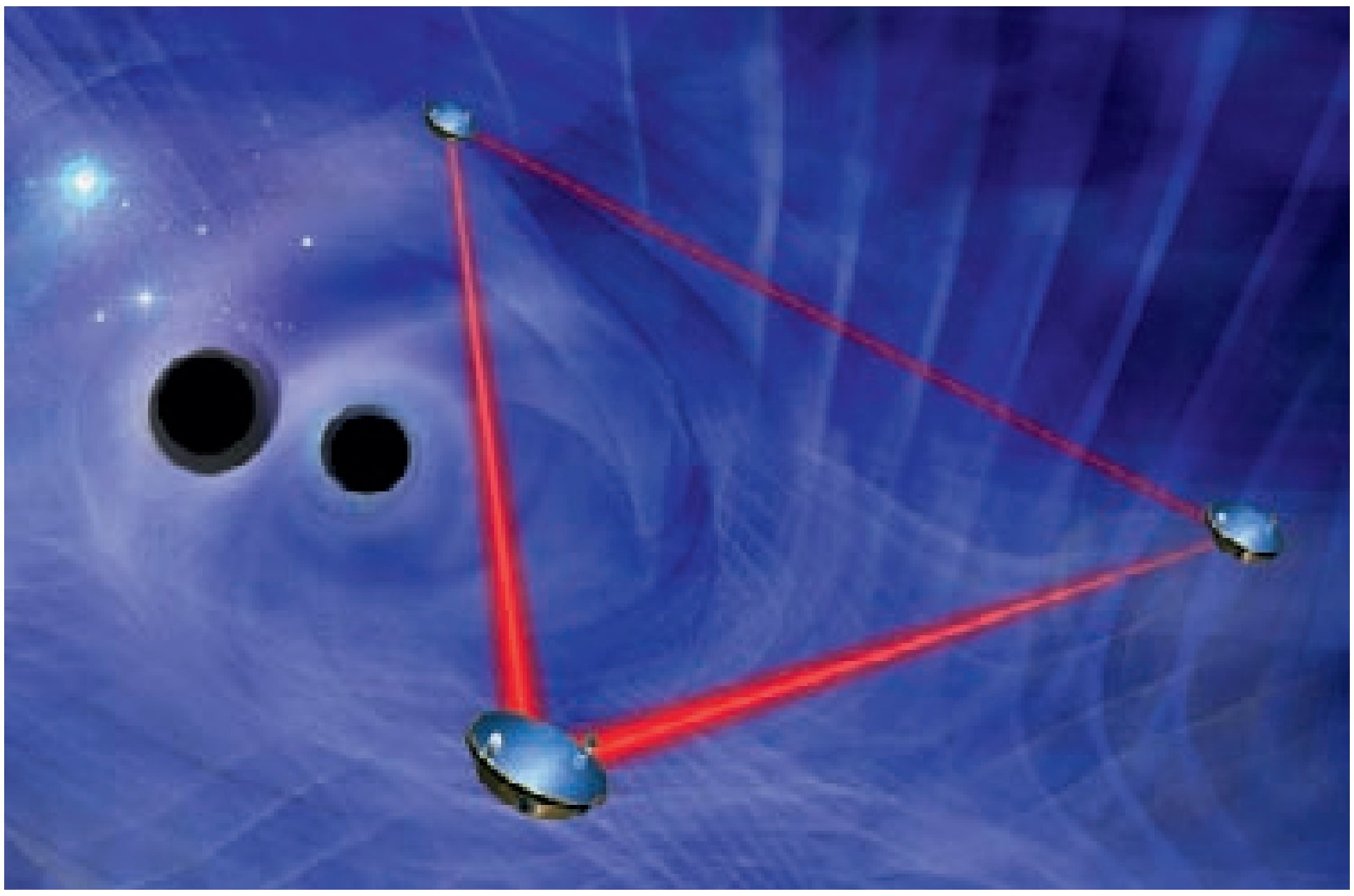

Fig. 4.1 – Artist's impression of the three spacecraft of the LISA mission and gravitational waves from a binary black hole system

## Is general relativity the correct theory of gravity?

Despite decades of experimental tests supporting general relativity, numerous alternative theories or extensions of general relativity exist. Some are motivated by attempts to unify gravitation with the other interactions of physics, such as string theory, while others are motivated by a desire to avoid dark matter or dark energy by modifying general relativity at large distance scales.

Gravitational waves provide entirely new ways to test general relativity. We have already mentioned the possibility of bounding the mass of the graviton and of measuring the polarizations of the waves, as tests of general relativity.

The inspiral of a binary system of compact objects is governed by a relatively small number of param-eters, such as the mass and spin of the two bodies, the orbital eccentricity and the orbital frequency. Much of the inspiral phase, characterized by slow motions and weak gravitational fields, can be described with exquisite precision by analytic solutions to general relativity that make use of the so-called “post-Newtonian” approximation. Specifically, the evolution of the phase of the gravitational-wave signal can be calculated to extremely high precision. A precise comparison between the observed gravitational-wave phasing and the predicted phasing using such techniques as matched filtering can search for any failure of general relativity. Such tests can be made using both ground and space-based observations.

With the regular observation of BH-BH inspirals, it will also be possible to detect the effects of the dragging of inertial frames, an important general relativistic effect hinted at in observations of accretion onto neutron stars and black holes. This has only recently been observed directly in the solar system by tracking LAGEOS satellites and by the Gravity Probe B mission. This effect likely plays a central role in producing relativistic jets observed in quasars, for example. In inspiraling binaries, when the two bodies have significant spin, there can be precessions of the body’s spins as well as of the orbital plane, and modifications of the rate of inspiral. These phenomena result in modulations of the gravitational waveform and alterations in the phase evolution that can be measured precisely using interferometric detectors, both on the ground and in space. For example, there is now evidence from numerical solutions of Einstein’s equations that, depending on the magnitude and alignment of the spins, the mergers could be very rapid or could experience a momentary “hang-up,” with significant consequences for the observed waveform. Observing the effects of frame dragging in such an extreme environment would be a stunning test of general relativity.

In some alternative theories of gravity, there is the additional possibility that systems could emit “dipole” gravitational radiation, in addition to the uniquely “quadrupolar” radiation that general relativity predicts. For inspiraling binary systems, this additional form of radiation can modify the decay of the orbit and thus distort the gravitational waveform compared to what general relativity would predict. Observations by ground-based detectors and also by LISA could strongly constrain such alternative theories.

### Is general relativity still valid under strong-gravity conditions?

Most tests of general relativity have taken place under the relatively weak-gravity conditions of the solar system or of binary pulsars. Although there is some evidence of strong-field general relativistic phenomena, notably from observations of the inner regions of accretion disks near neutron stars or black holes, the evidence is not conclusive. The final merger of two black holes is a quintessentially strong-field phenomenon. Recent breakthroughs in numerical relativity have led to increasingly robust theoretical predictions for the shape of the merger waveform. The comparison of such predictions with observed merger waveforms would provide unique tests of general relativity in the highly dynamical, strong-gravity regime. The ground based interferometers will provide initial evidence of such strong-gravity behavior, while LISA, with large SNR expected for massive binary black hole inspirals, will yield quantitative tests of strong-field general relativity.

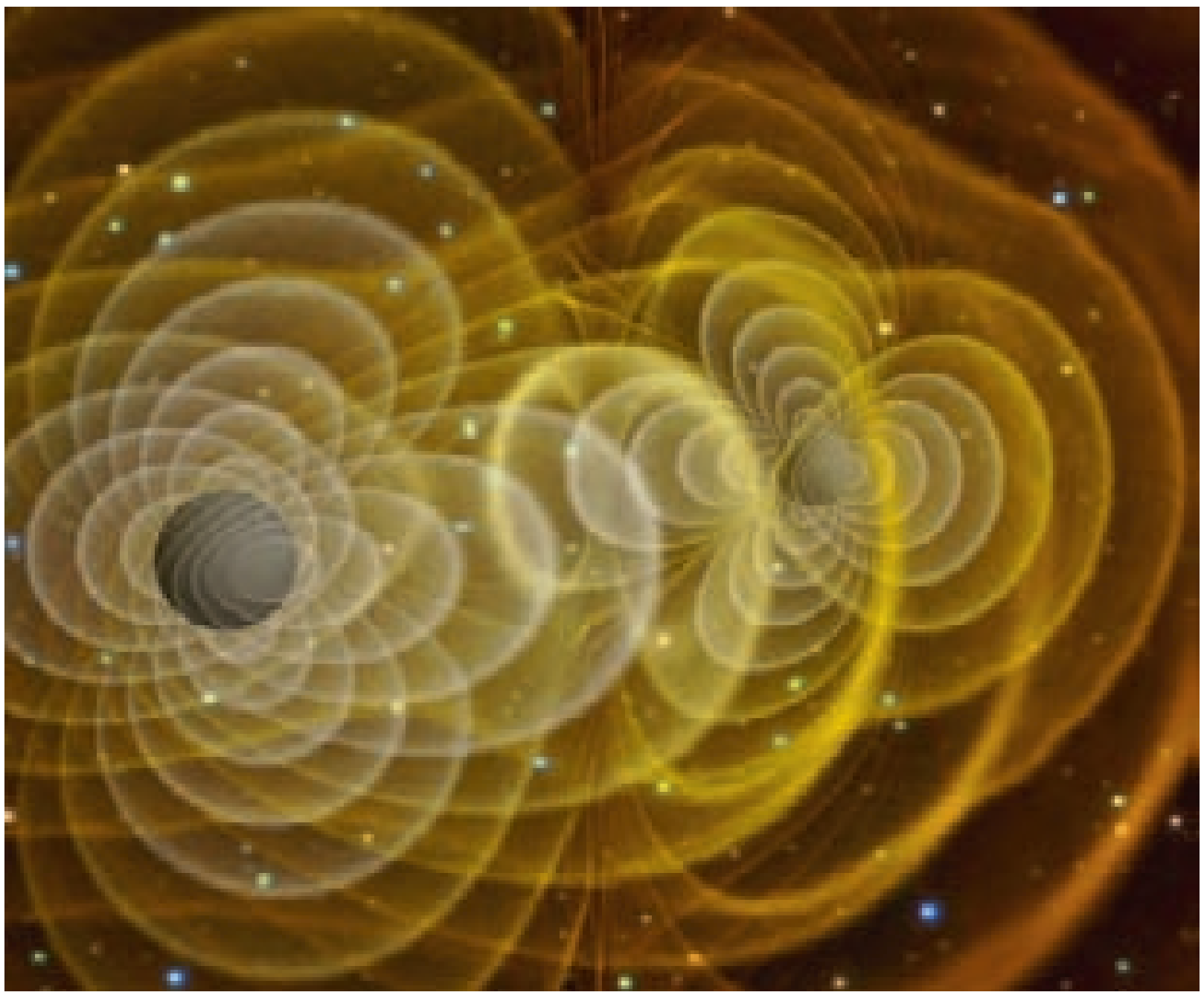

Fig. 4.2 – Snapshot from a 3-D simulation of merging black holes (Credit: Henze, NASA)

## Are Nature's black holes the black holes of general relativity?

To date, the observational evidence for black holes consists entirely of evidence that a large amount of mass is confined to a small region of space, too small to be compatible with any other stable object governed by general relativity and the normal laws of matter. However, general relativity predicts that black holes have a unique property, sometimes called the "no-hair" property. The external space-time of a black hole is completely characterized by its mass and angular momentum; for example the multipole moments of its external field are all determined by these two parameters. Specifically, according to general relativity, all undisturbed black holes are described by the exact solution of Einstein's equations known as the Kerr metric. This is in complete contrast to normal gravitating bodies, whose gravity field may depend on their internal structure, on whether they are partly solid and partly liquid, and on surface deformations such as mountains. Black holes have no such "hair." Gravitational waves provide a number of ways to test the no-hairness of black holes.

One class of gravitational-wave sources, called extreme mass-ratio inspirals (EMRI), will provide no-hair tests of black holes. These involve a stellar-mass compact object (black hole, neutron star or white dwarf) spiraling into a massive or intermediate mass black hole. Over the $10^4$–$10^5$ eccentric, precessing orbits traced out by the smaller mass, the emitted waves encode details about the space-time structure of the larger hole with a variety of distinct signatures. In addition to providing determinations of the black hole's mass and angular momentum to fractions of a percent, the observations can also be used to test whether the space-time that determines the orbits is the unique Kerr geometry. These sources are detectable by LISA and will be studied in detail. However, third generation ground-based detectors might also detect them if intermediate black holes of one to ten thousand solar masses exist at cores of globular clusters and smaller galaxies.

There is also the possibility, if a binary merger is observed, of detecting the so-called "ringdown" gravitational waves emitted by a distorted final black hole. Since the frequencies and damping times of each of the individual normal modes of oscillation of the black hole depend uniquely on the mass and spin of the hole, it would be possible to measure the mass and spin, and possibly even to test the "no-hair" theorems of black holes, which require these unique dependences on mass and spin. Such no-hair tests would be carried out by ground-based interferometers observing the merger of two intermediate mass black holes, and LISA could perform even more precise tests by observing massive black hole mergers.

In the case of inspirals of binary black holes of comparable masses, LISA and third generation detectors will be able to measure accurately the masses and spins of both the inspiraling holes and the final hole. This could provide a test of the famous "area increase" theorems of general relativistic black holes.

### How does matter behave under extremes of density and pressure?

The detection by ground-based interferometers of continuous wave signals from spinning neutron stars (pulsars), from stars spun up by accretion in low-mass X-ray binaries, or from newly born neutron stars, would have major implications for fundamental physics. A variety of mechanisms have been proposed by which a spinning neutron star could support a non-axisymmetric deformation, based either on extrapolations of our knowledge of the behavior of matter at nuclear densities, on ideas from the standard model of particle physics, or on the inclusion of such effects as strong magnetic fields on the internal structure. If large enough, such deformations would lead to the emission of detectable gravitational waves. Detection of such waves, or even improved upper bounds, will help to constrain the possible viable models.

Tidal distortion or tidal disruption is likely to play a role in the late inspiral and merger phase of NS-NS systems, and the resulting imprints on the gravitational waveform as measured by ground-based interferometers could provide constraints on the equation of state of the high-density nuclear matter making up the stars.

If waves are detected from a range of NS-BH inspirals, additional constraints on equations of state for neutron star matter could be obtained, for example by determining the onset of tidal disruption of the neutron star as a function of the mass of the black-hole companion, and combining that information with improved theoretical or numerical models for tidal disruption in NS-BH inspirals. It may also be possible, through precise matching of theoretical templates with the observed signals, to measure the small effects on the gravitational-wave phase of tidal deformations of the neutron star, whose magnitude is sensitive to the star's radius and equation of state.

## 4.3 Questions for astronomy and astrophysics

### How abundant are stellar-mass black holes?

Gravitational radiation from stellar-mass black holes is expected mainly from coalescing binary systems, when one or both of the components are black holes. However, there is considerable uncertainty as to whether the processes of gravitational collapse of the massive stars that make up the progenitor binary system will lead to complete disruption of the system, or to a pair of compact objects formed so close to each other that they merge on too short a timescale. Globular clusters may be efficient factories for black-hole binaries through various capture processes, although, being only loosely bound to the cluster, the binary may be ejected and complete the merger process in free space.

The larger mass of stellar black hole systems means that black hole events may be detected much more frequently than those involving pairs of neutron stars. It is very possible that the first observations of binaries by interferometers will be of black holes. Detection of the gravitational waves from the inspiral and merger could provide information on the population of such systems in galaxies and on the distribution of masses and spins.

### What is the central engine behind gamma-ray bursts?

One candidate for the underlying power source behind many of the observed short, hard gamma-ray bursts is the inspiral and merger of a neutron star and a black hole. Such a system is also a strong source of gravitational waves. Simultaneous gamma-ray and ground-based gravitational wave detections would settle the issue and open the way to more detailed modeling of these systems. Further, an advanced space gravitational-wave interferometer with sensitivity in the 0.1 to 10 Hz range and good angular resolution (like the proposed DECIGO instrument discussed in Chapter 7) could detect the inspiral part of a NS-BH system, and give a year's advance notice of the occurrence of a gamma-ray burst. Third generation ground-based detectors have the potential to detect such a large number of these systems that it will help in identifying different populations, shed light on the physical mechanism behind the intense gamma-ray flashes and enable a very detailed study of the progenitors.

### Do intermediate mass black holes exist?

Intermediate-mass black holes, with masses between 100 and $10^4$ solar masses, are expected to exist on general evolutionary grounds, but have proved hard to identify because of their weaker effect on surrounding gas or stars. If such black holes are reasonably abundant they could be detectable by ground-based interferometers with improved low-frequency sensitivity. This would tell us how important black holes were in the early stellar population, and whether these holes have anything to do with the massive central black holes in galaxies. They will also be LISA sources when they capture a stellar-mass black hole or a neutron star from the surrounding cluster or when they are captured by a massive black hole in the center of a galaxy.

## Where and when do massive black holes form and how are they connected to the formation of galaxies?

There is strong and growing observational evidence for the existence of massive astrophysical black holes. Formation and interaction of these provide motivating sources for space-based gravitational wave detectors. The most convincing case comes from our own Galaxy, where a population of stars is seen orbiting a compact object of 3.7 million solar masses, but evidence supports the conclusion that black holes with masses between $10^5$ and $10^9$ solar masses reside in the centers of nearly all nearby massive galaxies. There is also a robust correlation between the mass of the central black hole and both the luminosity and velocity dispersion of the host galaxy's central bulge. How such massive holes formed and what is the origin of this correlation is still not certain. The leading scenario involves the repeated mergers of, and gas accretion by, galactic-center black holes following the merger of their respective host galaxies. However, it is not known whether the original "seed" black holes were 30 to 300 solar mass holes which formed from the collapse of heavy-element-free Population III stars in the early Universe (redshift ~20), or $10^5$ solar mass holes that formed much later from the collapse of material in protogalactic disks.

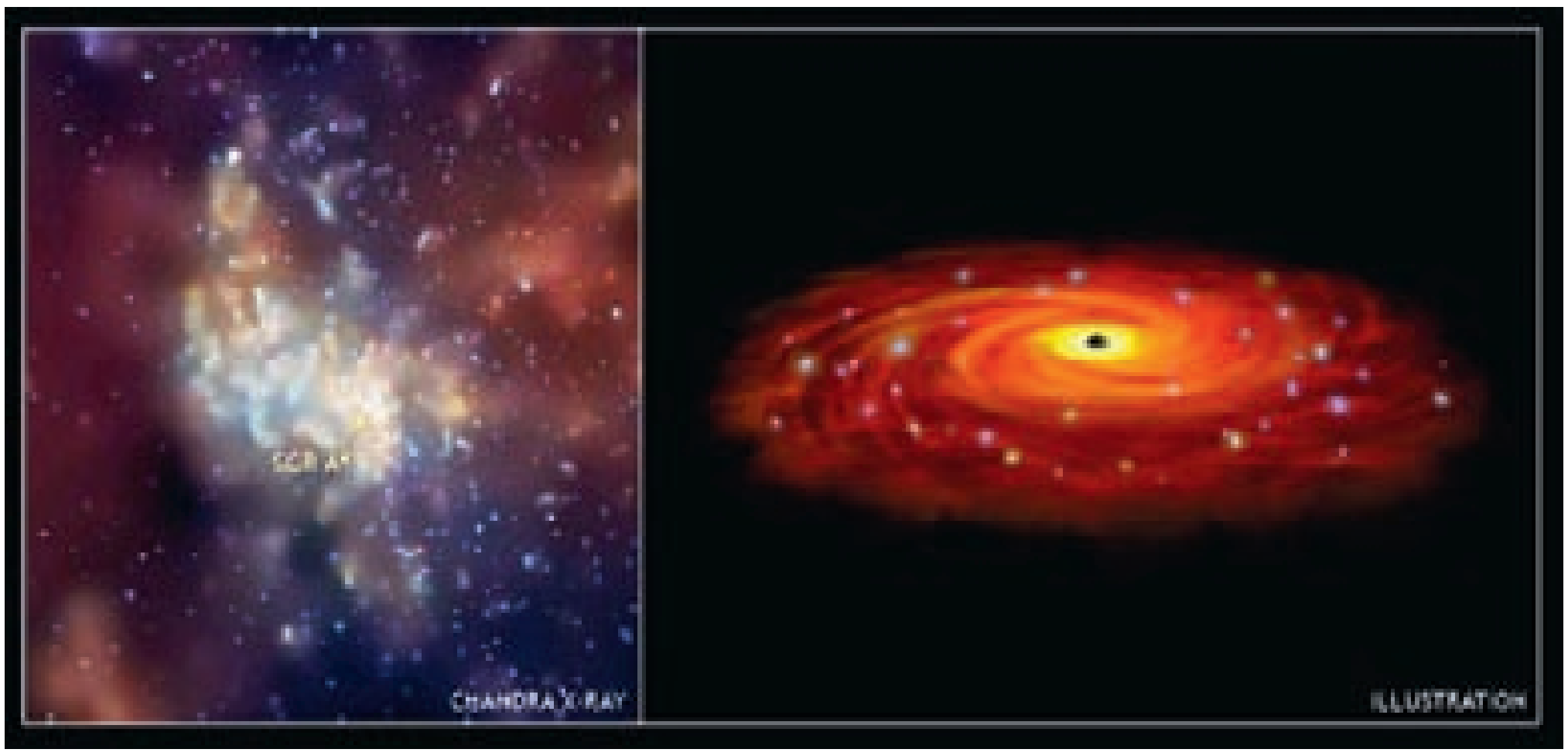

Fig. 4.3 – Chandra image of the Galactic Center (left). Illustration of massive stars formed from a large disk of gas around Sagittarius A*, the Milky Way's central black hole (illustration on right). Credit: X-ray: NASA/CXC/MIT/F.K.Baganoff et al.; Illustration: NASA/CXC/M.Weiss

By studying massive black hole mergers beyond redshift 10 for holes between $10^5$ and $10^7$ solar masses and to redshift 10 to 20 for holes between 100 and $10^5$ solar masses, LISA will be able to search for the earliest seed black holes.

In addition, LISA will be able to make very precise measurements of black hole masses and distances. Furthermore, in the hierarchical merger scenarios, the rate of detectable mergers may be as high as two per week. Thus, LISA will be able to trace the history of the growth of black hole masses and thereby shed direct light on how their formation and growth may be linked to the evolution of galaxies.

The capture of a small compact object by a massive black hole in a galactic center can also result in observable radiation. The tidal disruption of main sequence or giant stars that stray too close to the hole is thought to provide the gas that powers the quasar phenomenon. These disruptions are not expected to produce observable gravitational radiation. But the central regions of the galaxies should also contain neutron stars and stellar-mass black holes, which will spiral undisrupted into the hole. By observing the gravitational waves from these EMRIs, LISA will provide information about the stellar population near central black holes. When combined with modeling and spectroscopic observations, this will provide a deep view of the centers of galaxies and their evolution.

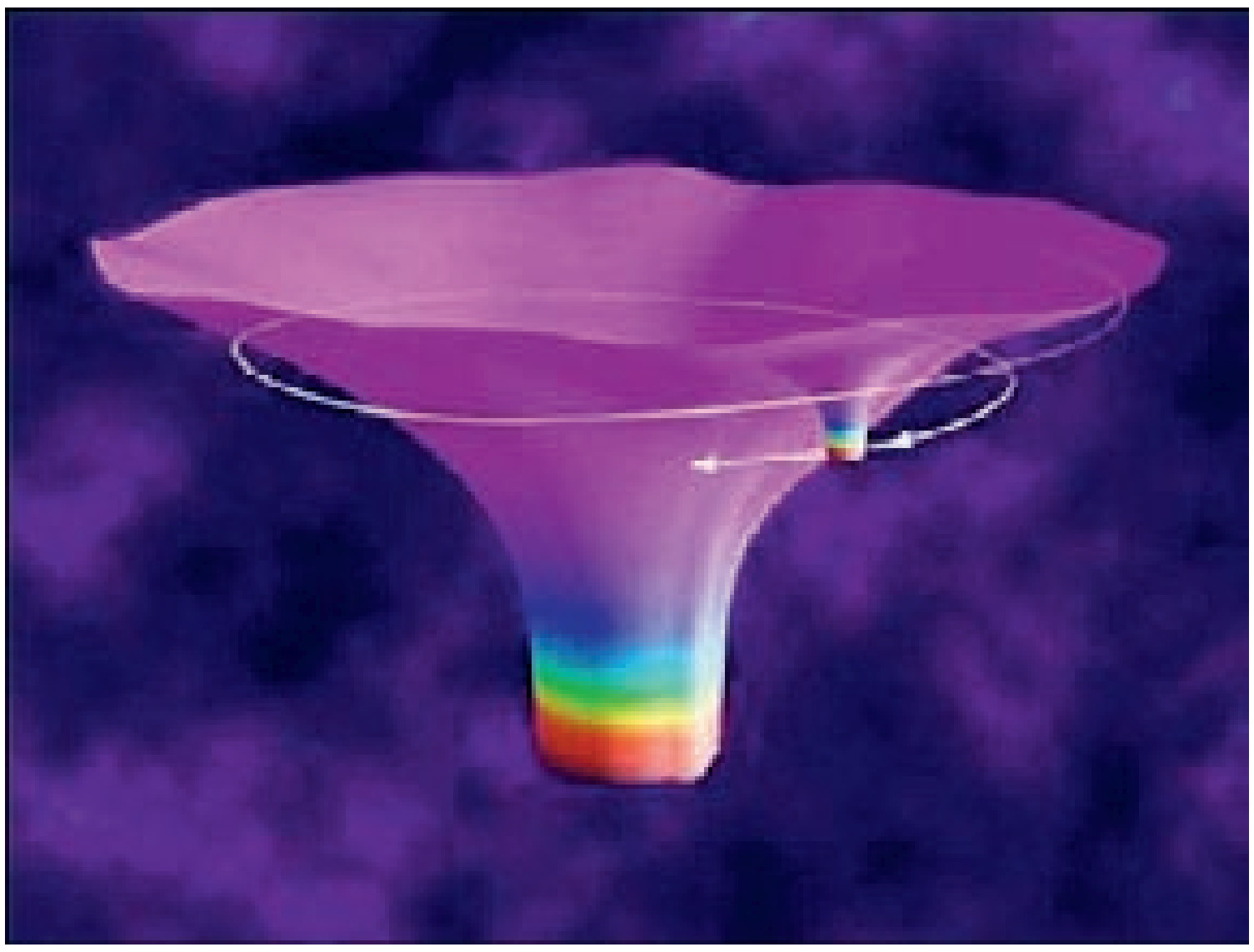

Fig. 4.4 – Illustration of the distortion of spacetime associated with the inspiral of a compact object around a much larger black hole

## What happens when a massive star collapses?

The event that forms most neutron stars is the gravitational collapse that results in a supernova. Despite many decades of work, it is still difficult to predict with certainty the waveform or amplitude expected from this event. The burst might be at any frequency between 100 Hz and 1 kHz, and it might be a regular chirp (from a rotating deformed or bifurcated core) or a more chaotic signal (from convective motions in the core). Based on current models, such signals might be detectable by second-generation ground-based detectors from a supernova in the Galaxy, but not from much greater distances. When they are finally detected, the gravitational waves will be extremely interesting, providing our only information about the dynamics inside the collapse, and helping to determine the equation of state of hot nuclear matter.

## Do spinning neutron stars emit gravitational waves?

Over 2000 spinning neutron stars have been detected in our galaxy, through their emission of radio waves as pulsars. But there may be as many as $10^8$ spinning neutron stars in the galaxy, most too old to be active as pulsars. They could be emitting gravitational waves, if they develop some kind of non-axisymmetric deformation. Currently there is lively debate as to the magnitude of deformations possible. But if gravitational waves were detected from spinning neutron stars, not only would they shed light on the underlying physics of such distortions, they would also permit mapping the distribution of neutron stars in the galaxy. These form an interesting target for advanced and third generation ground-based instruments.

If gravitational collapse forms a hot neutron star spinning very rapidly, then it may be followed by a relatively long period (perhaps a year) of emission of nearly monochromatic gravitational radiation, as the r-mode instability[1] forces the star to spin down to speeds of about 100–200 Hz. The accretion of matter onto a neutron star in a low-mass X-ray binary (LMXB) system could also induce a deformation on the neutron star that would lead to gravitational radiation in an amount that correlates with the X-ray flux.

[1] A mode of a rotating neutron star, driven unstable by the gravitational radiation reaction

## What is the distribution of white dwarf and neutron star binaries in the galaxy?

Within our own galaxy, LISA will measure the orbits and determine the locations of ~ 10,000 close binary systems consisting mainly of white dwarfs and neutron stars. Because such systems are the precursors of Type Ia supernovae and millisecond pulsars, such a census will aid in understanding the evolution of such systems.

## How massive can a neutron star be?

An unsolved long-standing question in astronomy relates to the maximum mass a neutron star could have. Because of the complexity of the physics at such high densities, temperatures and magnetic fields as can be expected in a neutron star core, the answer to this puzzle remains elusive even in theory, although there are some crude estimates. Because neutron stars are tidally disrupted close to the merger, the last few cycles of gravitational waves from a binary in which one or more members is a neutron star is fundamentally different from a system whose components are both black holes. This will make it possible to measure the mass-spectrum of neutron stars and ask how massive a neutron star could be and in what range of masses neutron stars are stable.

## Probing neutron star interiors with gravitational waves

Many years after their discovery properties of pulsars have remained as unexplained puzzles. Even very quiet pulsars seem to occasionally show a sudden increase in their pulse rate. Although the precise mechanism still remains a puzzle, it is expected that the transfer of angular momentum from the core to the crust, believed to be differentially rotating with respect to the core, might be responsible for pulsar glitches. In the process, the core could undergo oscillations leading to gravitational wave emission at normal mode frequencies of the core. Expected energies in gravitational waves of $10^{-12}$ solar masses in a typical glitch makes it possible to observe these phenomena in advanced instruments like GEO-HF (in the case of Vela pulsar) and third-generation detectors (up to 10,000 light years).

A class of neutron stars with extraordinarily large magnetic fields of ~ $10^{14}$ – $10^{15}$ G, called magnetars, could be behind what are called soft gamma repeaters. These sources sporadically emit short bursts of hard X-rays and soft gamma rays, with luminosities of L ~ $10^{41}$ erg $s^{-1}$. Occasionally, they emit giant flares with luminosities a million times larger. Such events are relatively rare and their origin is of great interest. The flaring activity is probably due to sudden and violent reconfigurations of complex magnetic field topologies. Magnetar flares contain clear signatures of several quasi-periodic oscillations in the ~ 20 - 2000 Hz band. Some of the oscillations appear to match well the seismic mode frequencies of the crust, an interpretation also consistent with the large scale crust fracturing expected to occur during a giant flare.

Magnetar flares could become a prime target for third generation gravitational wave detectors like the Einstein Telescope. It is likely that any vibration in the crust will quickly develop to a global magneto-elastic crust-core oscillation, since these two regions of the star are efficiently coupled by the strong magnetic field. The gravitational wave signal emitted in the process would carry precious information about the (largely unknown) interior properties of magnetars such as the strength and topology of the magnetic field and the elastic properties of the crust. As in the case of pulsar glitches, observing such signals could tell us about the structure of the core and its equation of state.

## What is the history of star formation rate in the Universe?

The birth rate of compact binaries is intimately related to the birth rate of massive stars. In turn the birth rate of massive stars can tell us about the formation of galaxies and proto-galaxies. Together with the knowledge of when the central black holes at galactic nuclei formed we can begin to understand the formation and evolution of large scale structure in the Universe. Third generation detectors will map out the density of compact binaries up to a red-shift of z=2 and can therefore address the question of what the formation rate of massive stars is. This would also address the question of what the stellar mass function is as a function of red shift.

## 4.4 Questions for cosmology

### What is the history of the accelerating expansion of the Universe?

Because binary black hole inspirals are controlled by a relatively small number of parameters, such as mass, spin, and orbital eccentricity, they are good candidates for standard candles. This is because the frequency and frequency evolution of the waves are determined only by the system's parameters, while the wave amplitude depends on those same parameters and on the luminosity distance to the source. No complex calibrations are needed.

A third generation detector like the Einstein Telescope will measure the luminosity distance to a gamma-ray burst source at a red-shift z =1 to an accuracy of 2 - 5%, with the host galaxy of the burst providing the red-shift. Therefore, third generation ground-based detectors will enable precision cosmology, avoiding all the lower rungs of the cosmic distance ladder. Over the course of a year, a third generation ground-based detector will accumulate enough statistics to measure the Hubble parameter, the dark energy and dark matter content of the Universe and the dark energy equation of state to a good accuracy.

Further, LISA can measure luminosity distances to a few percent at redshift 2 and to tens of percent at redshift 10. At the same time, because of the changing orientation of the LISA array with respect to the source, it can also determine the orientation, with precision of 1 to 10 arcminutes for massive inspirals at z = 1. If this angular and distance resolution were enough to link a LISA event with a corresponding electromagnetic event in a host galaxy or quasar and thereby to yield a redshift, LISA would contribute a direct, absolute calibration of the Hubble constant that relies only on fundamental physics.

Gravitational wave measurements have the potential to measure the parameter w, which characterizes the "dark energy" supposedly responsible for the acceleration of the Universe, to an accuracy better than 10%, for third generation ground-based detectors, and around 4%, for LISA.

Proposed LISA follow-on missions that are highly sensitive in the 0.1 - 1 Hz band (such as DECIGO or BBO), as well as the Einstein Telescope, will be able to measure cosmological parameters far better than LISA, and indeed far more accurately than other proposed cosmology missions. These projects will be able to detect ~100,000 coalescing compact binaries per year, and measure the luminosity distance to each to several percent (an error dominated by weak lensing). Moreover, for the proposed next-generation space missions, the typical error box on the sky will be only ~10 arcsec$^2$: small enough to uniquely identify the host galaxy in most cases, and then obtain the host's redshift by optical follow-up. Consequently, a second-generation space-mission would obtain hundreds of thousands of independent measurements of the luminosity distance-redshift relation, which together (and combined with expected priors from the Planck mission) would determine the Hubble constant to ~0.1%, $\Omega_M$ and $w_o$ (the current value of the w parameter) to ~1%, and wo (which describes the time-evolution of w) to within ~0.1. Moreover, since compact binaries are very simple, clean systems, these measurements would not suffer from the unknown systematics that potentially limit the accuracy of some other methods (e.g., the unknown time-evolution of the light curves of type Ia supernovae).

The Einstein Telescope could observe double neutron star binaries in coincidence with short hard gamma-ray bursts and measure the luminosity distance (to within a few percent) and red shift and infer the dark matter energy density and w to within 10%. It could measure w alone to within 1% and its variation with redshift to within 10%.

#### Were there phase transitions in the early Universe?

First-order phase transitions of new forces or extra dimensions in the early Universe could produce a detectable background of gravitational waves. Such events would occur between an attosecond ($10^{-18}$ s) and a nanosecond after the big bang, a period not directly accessible by any other technique. Other potential exotic sources include intersecting cosmic string loops or vibrations and collapses of "boson stars," stars made of hypothetical scalar-type matter. There could also be a background at cosmological wavelengths associated with inflation.

Such a background could be detectable by a single detector if it is stronger than instrumental noise, but a weaker background could still be detected by using a pair of detectors and looking for a correlated component of their "noise" output, on the assumption that their instrumental noise is not correlated. The first-generation LIGO detectors could reach a sensitivity in the effective energy density in background waves as a fraction of the cosmic critical density of $10^{-6}$ at 100 Hz, and second-generation LIGO could reach $10^{-10}$ at around 40 Hz and a third generation detector could reach $10^{-12}$ at 100 Hz. LISA could also approach $10^{-10}$ at about 1 mHz.

Exploiting the fact that millisecond pulsars are such stable clocks, the International Pulsar Timing Array (IPTA) can in principle detect gravitational waves with periods longer than a year by looking for induced fluctuations in the arrival times of pulses. A single pulsar can set limits on a stochastic background by removing the slow spindown and looking for random timing residuals. Arrays of pulsars offer the possibility of cross-correlating their fluctuations, making it possible to distinguish between intrinsic and gravitational-wave induced variabilities. It will soon be possible to monitor many pulsars simultaneously with multi-beam instruments, such as LOFAR and the Square Kilometer Array. Pulsar timing could push the limits on a background at 3 nHz to $\Omega_{GW} \sim 10^{-13}$.

Observations of the cosmic microwave background (CMB) may in fact make the first detections of stochastic gravitational waves. As mentioned in the last chapter, this will require detecting fluctuations in the polarization of the CMB. The Wilkinson Microwave Anisotropy Probe (WMAP) made the first measurements of polarization, but was not sufficiently sensitive to the gravitational-wave part. Future ground-based and space-based experiments (notably Planck and CMBPOL) will continue to improve the sensitivity to gravitational waves.

**Priority — A strong and ongoing international program of theoretical research in General Relativity and astrophysics directed towards deepening our understanding of gravitational wave observations.**

## 4.5 General background reading relevant to this chapter

Physics, Astrophysics and Cosmology with Gravitational Waves B.S. Sathyaprakash and B.F. Schutz
Living Rev. Relativity 12, (2009), 2, http://www.livingreviews.org/lrr-2009-2 arXiv:0903.0338

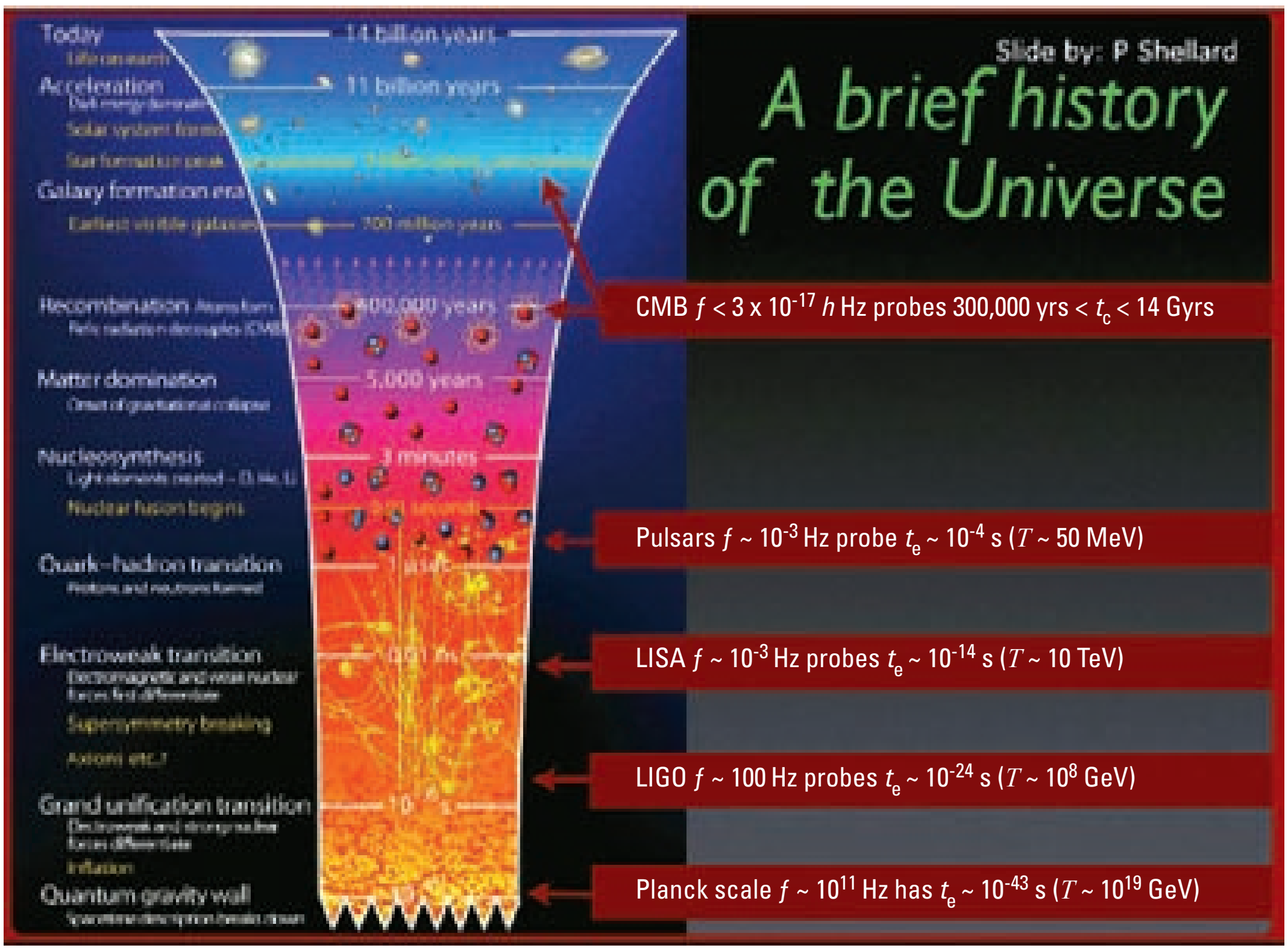

Fig. 4.5 – A brief history of the Universe

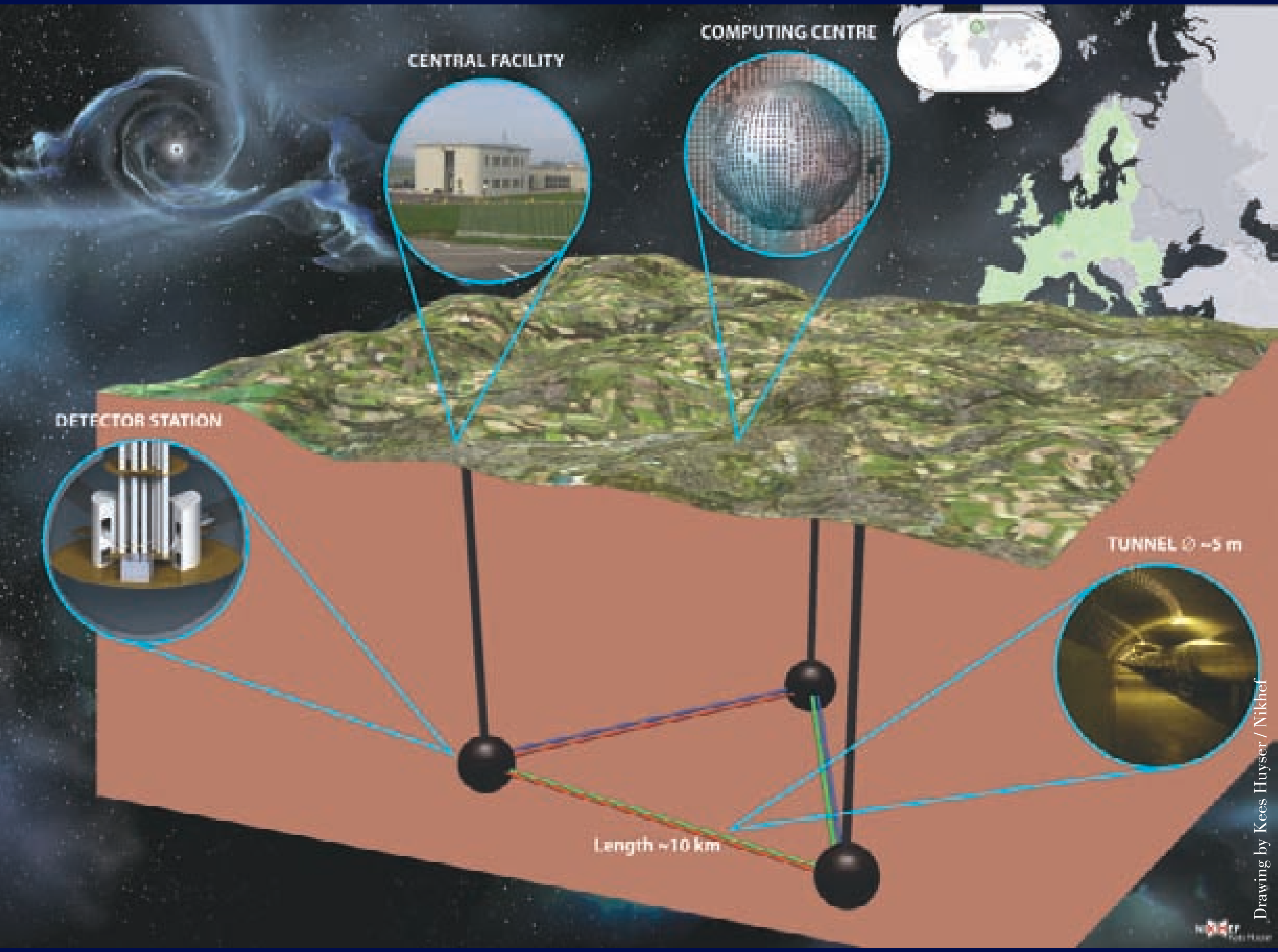

The proposed Einstein Telescope will be a third generation gravitational wave detector.

# 5. The future of the field in response to anticipated scientific opportunities—on the ground/higher frequencies

## 5.1 Introduction

The preceding chapters sketched the current status of ground-based detectors, describing their performance and plans for future improvements. Now we will outline a vision for the future development of the ground-based portion of gravitational wave detection. The timescale we have chosen is roughly 30 years. While this may seem long-term, the life cycle of gravitational wave detectors has typically been 15 years or more, and even longer if one considers the enabling R&D.

For the high frequency region of the spectrum (from approximately 1 Hz to 10 kHz), we foresee the evolution and expansion of the current LIGO-Virgo-GEO network into a worldwide alliance of interferometric detectors, operating in coordination to extract the maximal scientific return. The second generation detectors (Advanced LIGO, Advanced Virgo and GEO-HF) set the standard for the types of detectors which will form the core of this second generation network, and supplant the current bar detector network. Even now, as these second generation detectors undergo their final design steps, R&D and planning for the third generation is underway. That network will incorporate the upgraded second generation detectors, as well as new low frequency detectors (the third-generation detectors) whose conceptual designs are just underway.

The second area of ground-based gravitational wave science involves continued development of the pulsar timing array for studying gravitational waves in the nanoHertz regime. The Square Kilometer Array offers a solid advance in capability for detecting gravitational waves via pulsar timing, and represents the core of the future programs in this area.

These two developments will unleash the use of gravitational waves as an invaluable new tool for astronomy, and it is important to plan now to provide appropriate access to the gravitational wave data for the broader astronomical community. This topic is addressed at the end of this chapter.

The field of gravitational wave astrophysics is still in infancy, and early detections in each of these frequency bands will likely bring some expected confirmations and some complete surprises. We will no doubt discover pointers toward new observational directions. Some technology limitations may prove more intractable than expected, while other breakthroughs may open promising new avenues of development previously thought impenetrable. Planning for the time scale envisaged in this roadmap must remain flexible allowing the field to capitalize on the span of possibilities.

## 5.2 Interferometric Detector Networks

The existing terrestrial network of interferometric gravitational wave detectors, with its planned upgrades (Advanced LIGO, Advanced Virgo, and GEO HF), forms a solid basis for the first detections and surveys of gravitational wave sources in the 10 Hz - 10 kHz regime. However, as the focus shifts from these first encounters to more in-depth astrophysical studies, this network will need to be bolstered. Source localization and identification, along with full extraction of all data from the gravitational wave waveforms, will become paramount goals, particularly as joint gravitational wave - electromagnetic studies come to the fore.

For many sources (particularly short duration bursts and inspirals), the network derives its directional sensitivity by comparing arrival times at widely spaced detectors, Thus the angular resolution is inversely proportional to the separation of the detectors in the network. Similarly, extraction of full information concerning the waveforms in both polarizations of the incident wave requires a network with detectors of different orientation in all three dimensions. The two LIGO sites are located relatively close together (measured on a global scale) and with similar orientations; thus they provide limited directional and polarization information. Consequently, the network formed by just LIGO, Virgo and GEO has good directional sensitivity along one axis but poor resolution in the other. Limited information about the wave strength in the two polarizations will leave many questions about the sources unanswered, and underlines the case for future extension of the network.

### 5.2.1 Second generation network

The immediate priority for the interferometric community is to successfully build, commission and operate the already planned second generation detectors. The further development of ground-based gravitational wave interferometers will be driven by a combination of science opportunities and technology challenges. The first generation of gravitational wave detectors has spurred keen interest in improving the estimates of source strengths and event rates, and while significant uncertainties still remain, a range of target sources stands out, as discussed in the last chapter. The approved second generation detectors (Advanced LIGO and Advanced Virgo) have been designed to allow broad search parameters as the gravitational wave sky is surveyed. In particular, both Advanced LIGO and Advanced Virgo seek higher sensitivity in the 100 Hz regime, and extension to lower frequencies, as the most promising and feasible directions to pursue. A similar strategy with extension to still lower frequency is planned for upgraded second generation and third generation instruments, as they must be able to capitalize on first detections by the advanced (or indeed enhanced) detectors.

Plans and funding for Advanced LIGO, Advanced Virgo and GEO-HF are already approved. Together, these define a major portion of the second generation network. Their development schedules have been coordinated to optimize the scientific output of the network, consistent with their separate obligations to their funding agencies. These include coordinating the schedules for intermediate upgrades to maintain some observational coverage and coordinating shut-downs for major installations to maximize the amount of coincidence operation time. The top lines in figure 5.1 show the planned activities for these detectors.

In the medium term, the high priority for ground-based gravitational wave detector development is expansion of the network to add additional detectors with appropriately chosen intercontinental baselines and orientations to maximize the ability to extract source information. Two important questions arise: what is the required sensitivity for new detectors to make a substantial contribution to the network, and how many more detectors are needed to “complete” the network.

Experience to date has shown that in networks with detectors of different sensitivities, the most substantial gains are made when the detectors sensitivities are within a factor of two. New detectors proposed for the second generation network should therefore match or exceed the sensitivity projected for the other members in the network (at least over a significant frequency band) at the time the new detector becomes operational.

The question of how many detectors a network needs is best framed in the context of the value of additional detectors. As noted, a network consisting only of LIGO-Virgo-GEO has poor angular resolution along directions perpendicular to the line connecting the US and Europe, and correspondingly poor polarization information. A qualitative improvement arises from adding even a single detector at a large distance from this line. Such a network forms a triangle, and will typically have useful angular resolution (~ 1 degree or better in both dimensions) for sources located near the perpendicular to the plane of the triangle. This angular resolution degrades, however, for sources closer to the plane of the triangle. So a triangular network would achieve good resolution for some sources, a qualitative change, but not for all sources.

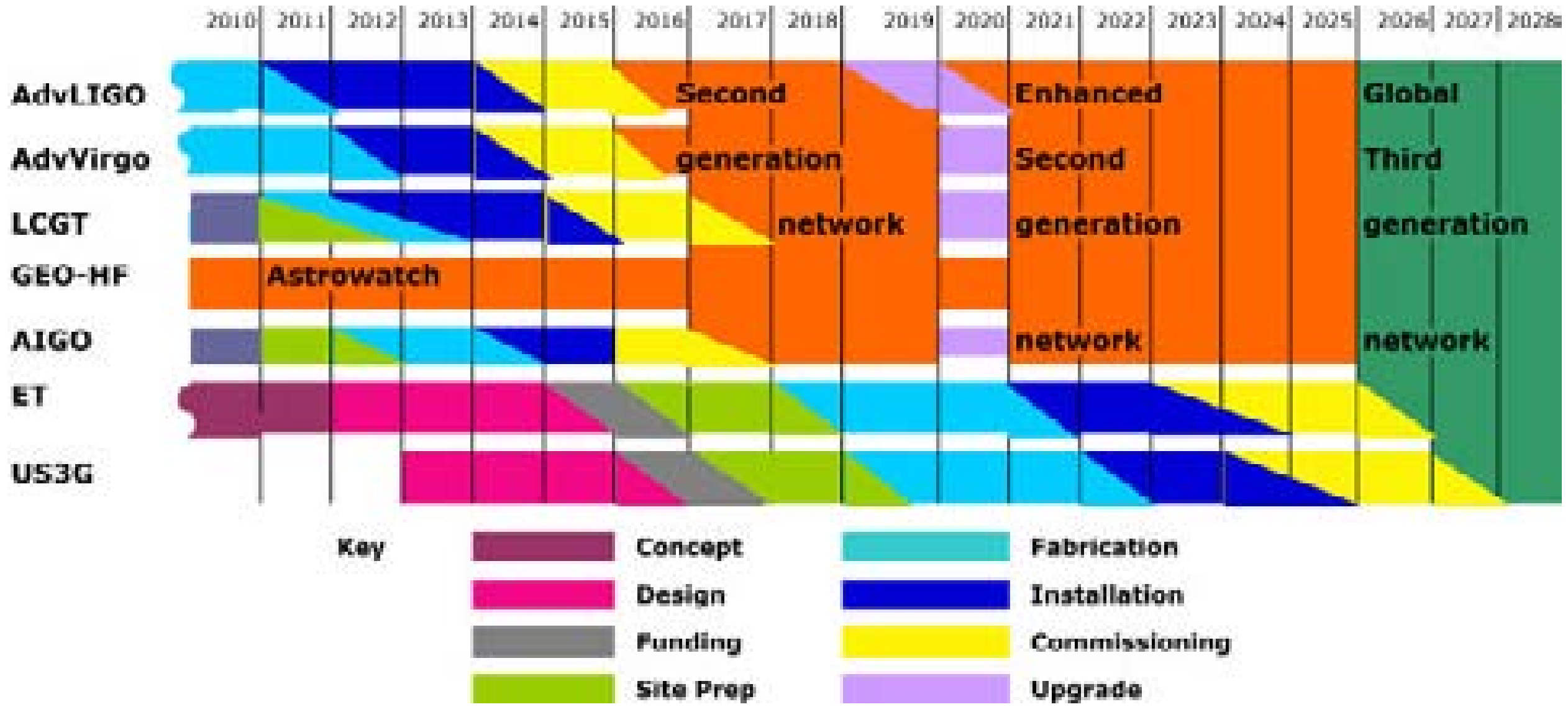

Fig. 5.1 – Time-line for ground based detector developments

The addition of another detector, well out of the plane of the other three, makes a large quantitative change, enabling good angular resolution for sources that would otherwise be poorly resolved. The process is similar to (sparsely populated) coherent aperture synthesis in radio astronomy — the resolution power of the network is determined by the projected area of the network as seen by the source. Preliminary studies indicate that the jump in useful identifications is very significant when going from (effectively) three sites in a plane to four, with one located well out of the plane. As many as three times more sources become locatable to a single galaxy. Further studies are required to quantify this benefit, taking into account all factors, but it is clear that detailed astronomical studies of gravitational wave sources will require an expansion of the network.

To augment the second generation network, the most advanced plans for additional detectors center on the Japanese Large-scale Cryogenic Gravitational-wave Telescope (LCGT), and the Australian International Gravitational Observatory (AIGO). Preliminary discussions have also begun as to the possibility of a detector located in India (INDIGO).

LCGT is a proposed 3-km detector. The interferometer optical configuration is similar to Advanced LIGO and Advanced Virgo, with optical cavities in the arms and power recycling and signal recycling mirrors. However LCGT would consist of two interferometers in one vacuum envelope. The planned laser power is similar to that of Advanced LIGO and Advanced Virgo, giving LCGT comparable sensitivity in the high frequency region.

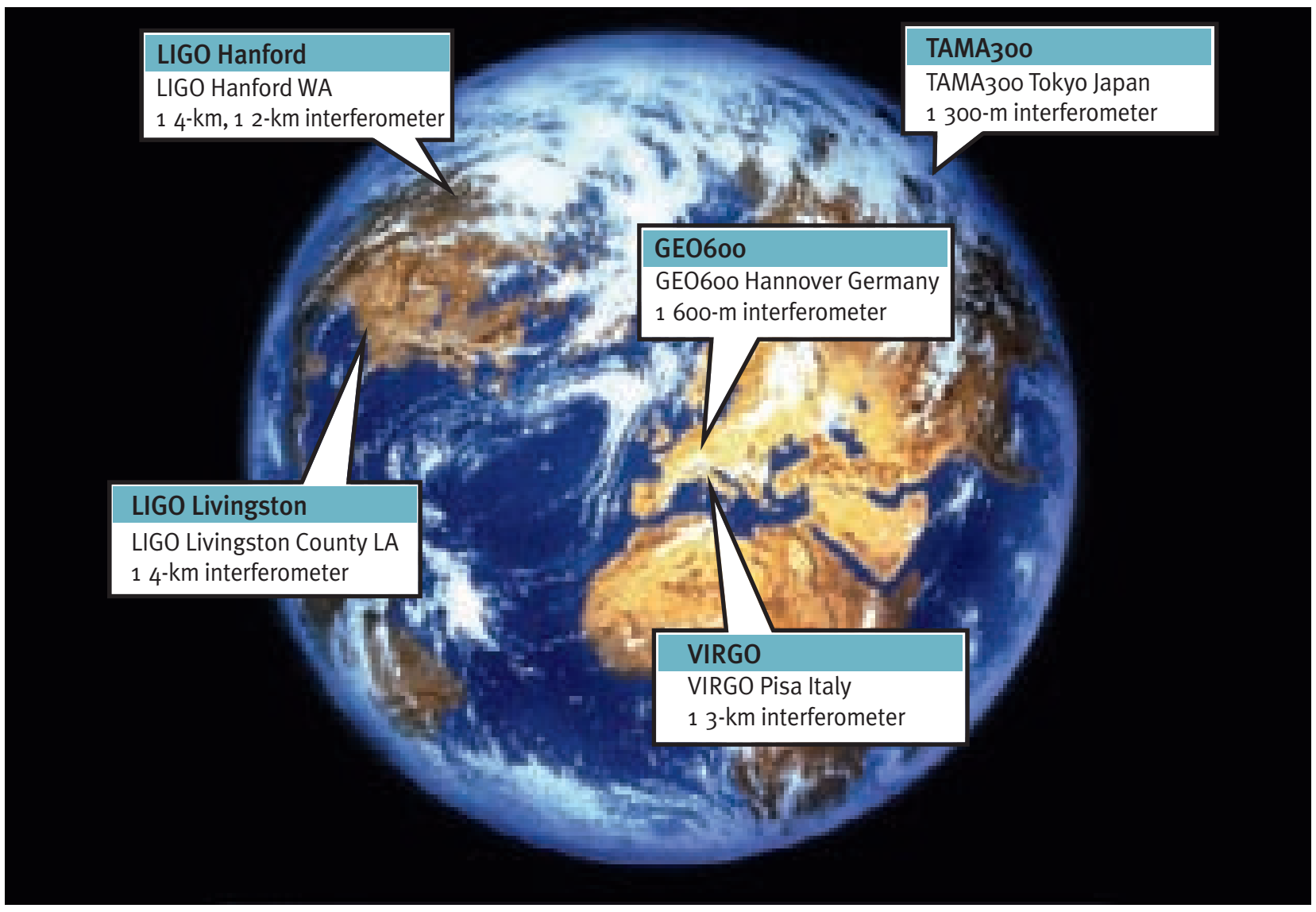

Fig. 5.2 – The current global interferometric network.

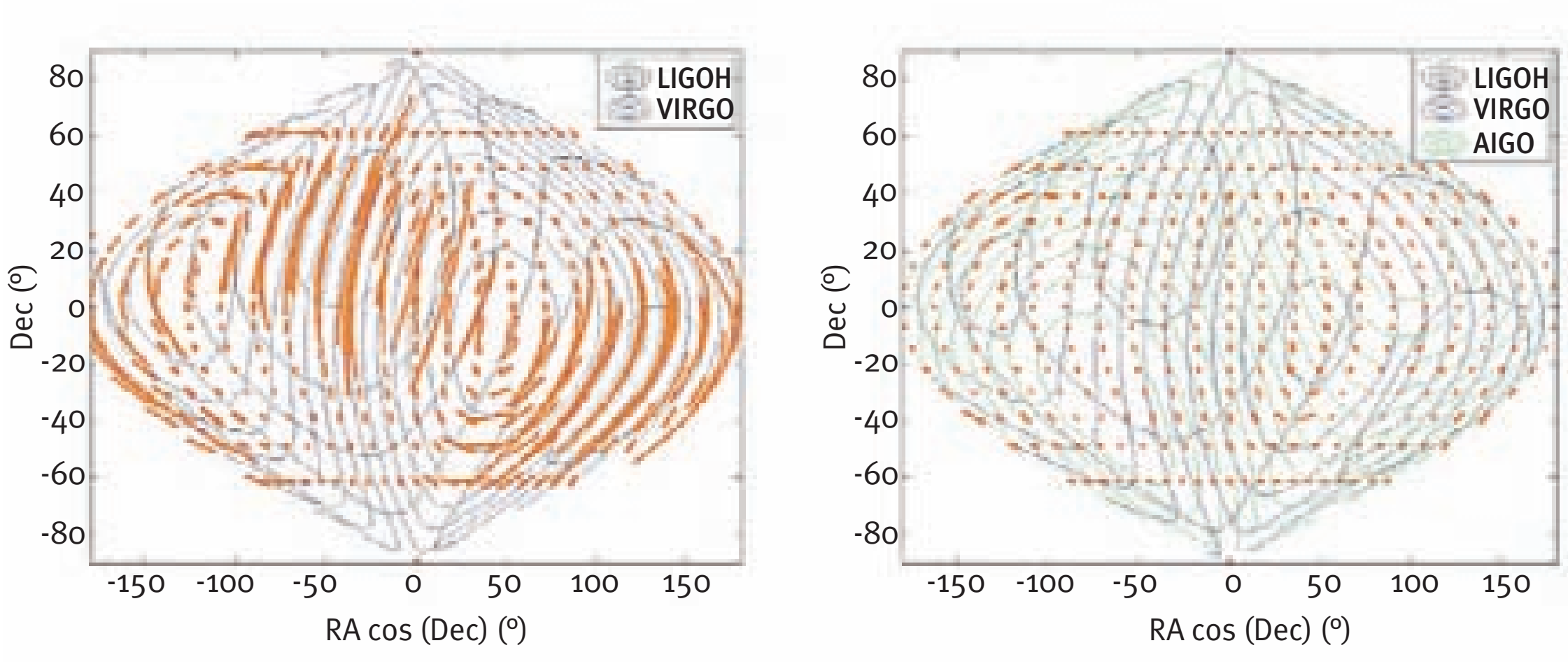

Fig. 5.3 – Angular area maps for world array. The angular uncertainty for each geocentric sky direction is indicated as an ellipse on the sky. These are normally highly elongated. (a) LIGO and Virgo. (b) LIGO, Virgo and AIGO. A further improvement is obtained if LCGT is added to the array.)

However, LCGT is exceptional because it incorporates two unique features: it is to be located underground in the Kamioka mine about 220 km from Tokyo, and it is planned to use cryogenic cooling for the interferometer mirrors (test masses of sapphire). Cryogenic cooling of the interferometer optics offers many benefits (reduced thermal noise, reduced thermal lensing in the substrates), and also presents new challenges. The underground location offers a significantly lower seismic noise level than has been observed at other locations in Japan (notably the TAMA site), and should greatly improve the low frequency performance.

The LCGT design is at an advanced stage. Considerable enabling R&D has been completed and an experienced team of gravitational wave experimenters has developed the concept. The expected sensitivity is comparable to or slightly better than Advanced LIGO and Advanced Virgo, and the quiet site and the early use of cryogenics should make future upgrades easy to perform. A table of the main design parameters is shown opposite.

Funding has not yet been approved[1], but the timeline for LCGT shown in figure 5.1 assumes a favorable decision on funding in the next year or two.

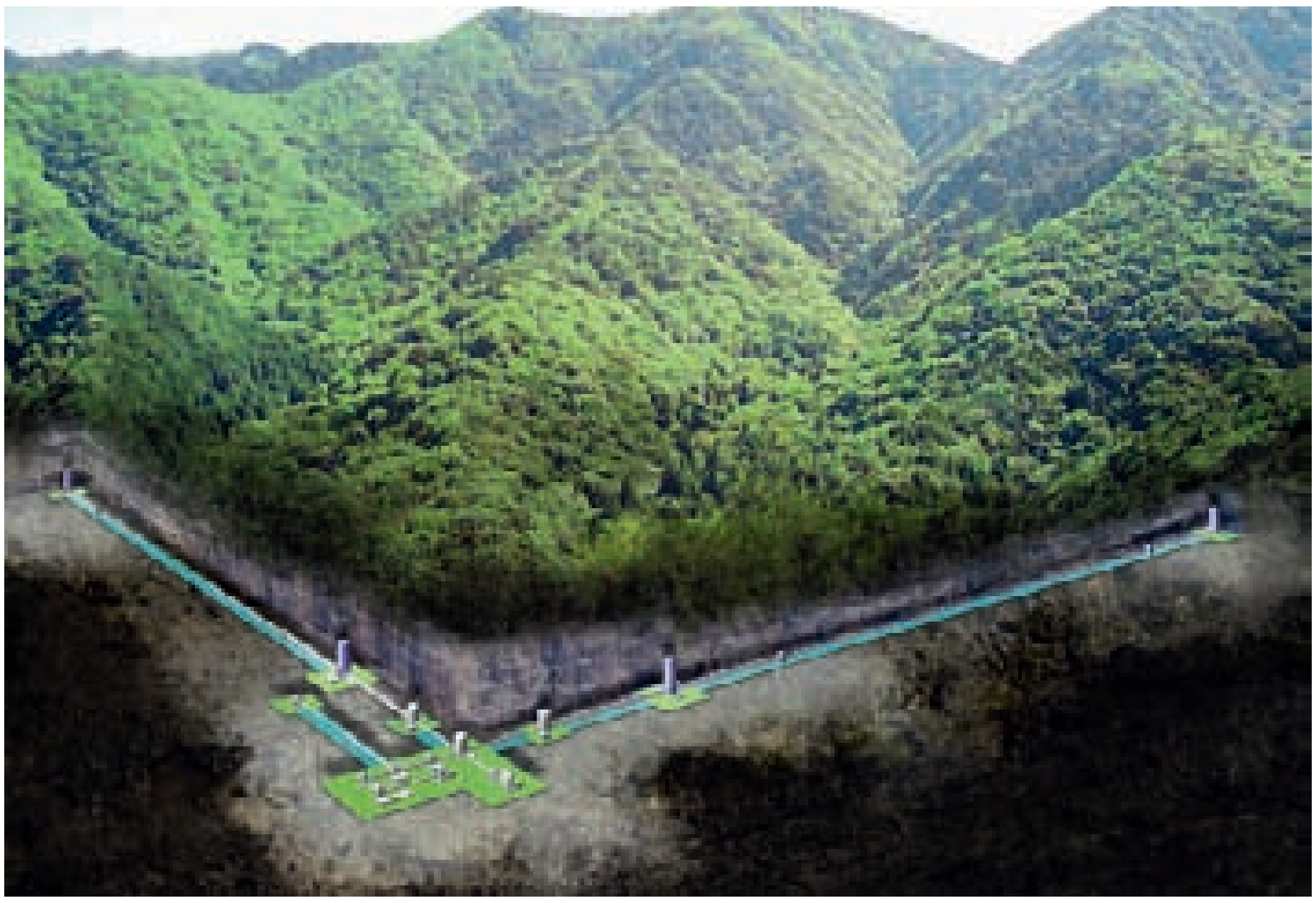

Fig. 5.4 – Artist's impression of the LCGT detector, located underground at the Kamioka mine.

| Item | Parameter |
|---|---|
| Baseline | 3 km |
| Interferometer | Power recycled Fabry-Perot Michelson with RSE |
| Optical Power | Laser: 150 W, Input Power: 75 W, Cavity Finesse: 1550, Power Recycling gain: 11, Effective power: 780 kW |
| Signal bandwidth | 230 Hz, Signal recycling gain: 15 |
| Mirror | Diameter: 25 cm, Thickness: 15 cm Mass: 30 kg, material: Sapphire crystal |
| Pendulum | length: 40 cm, 4 sapphire fibers, 1.8 mm in diameter |
| Acoustic loss angle | Mirror internal: $1 \times 10^{-8}$, Pendulum: $2 \times 10^{-7}$ Optical coating: $4 \times 10^{-4}$ |
| Temperature | Mirror: 20 K, Suspension: 10 K (20 K for noise estimation) |
| Vacuum | $\leq 2\times 10^{-7}$ Pa |

Table 5.1 – Main parameters for the design of LCGT

AIGO is a surface facility. The currently favored location is about 80 km from Perth in Western Australia. The site can accommodate a 5-km detector, although the current design has 4-km arms. The planned configuration is similar to Advanced LIGO and Advanced Virgo and the performance goal is similar.

ACIGA, the lead organization for AIGO, is a member of the LIGO Scientific Collaboration, and thus is familiar with the Advanced LIGO design. Use of Advanced LIGO designs for large portions of AIGO would minimize design cost, and capitalize on lessons learned during Advanced LIGO construction and commissioning. A preliminary cost estimate using this approach has been made, but no funding proposal has been yet submitted. The timeline in figure 5.1 assumes that this approach is adopted and funding becomes available in the next one to two years.

[1] subsequent to the preliminary version of this roadmap being made available internationally, approval of funding for the first phase of construction of LCGT was announced in June 2010.

Based on the arguments outlined above for the desired number and location of additional second generation network detectors, LCGT and AIGO would constitute excellent additions to the global gravitational wave network. Consequently, we make the following recommendations:

**Priority — The construction, commissioning and operation of the second generation global ground-based network comprised of instruments under construction or planned in the US, Europe, Japan and Australia.**

**Recommendation** — We recommend that GWIC provides the forum where international support for efforts to bring about such instruments in Japan, Australia and possibly India, can be coordinated and where the community can work together with the proponents to ensure that the siting, design, orientation, etc. of such instruments is carried out to optimize the scientific capabilities of the global network.

**Recommendation** — We recommend that GWIC organizes a workshop to emphasize the scientific benefits of interferometers in Japan and Australia as a way to encourage further international support and recognition of the potential scientific contributions of these facilities as part of the global network of ground-based gravitational wave detectors.

### 5.2.2 The third generation network — incorporating low frequency detectors

The evolution of the second generation network into the third generation network is envisaged to involve two distinct aspects. Second generation detectors will undergo upgrades in existing facilities, while new detectors aimed at extending the reach of terrestrial detectors at low frequencies will probably require new facilities. This third generation network will operate over a broader spectral band and with greater sensitivity than the second.

#### Low frequency interferometers

The second generation ground-based interferometers combined with the LISA mission provide significant gravitational wave capability at high and low frequencies. Yet the region between 0.1 and 10 Hz is not covered by them. Existing facilities (LIGO and Virgo) can support significant improvements in sensitivity at high frequencies, but still the region below 10 Hz will be difficult if not impossible to probe. Many noise sources increase at low frequencies, but one of the most difficult to overcome is noise due to gravitational forces from a changing mass distribution in the vicinity of the test mass mirrors. For an interferometer located on the earth's surface, the most important sources of changing mass distributions are very low frequency atmospheric waves and surface (Rayleigh) waves. To escape these effects, two options are under consideration and development: performing the measurement in space (such as the DECIGO mission described next chapter), or the construction of one or more underground detectors. Both options entail expense and technological challenges, but also offer target frequency ranges of considerable interest.

A second feature of importance for a low frequency interferometer is longer arm lengths. Most of the significant noise sources at low frequencies – including gravitational noise, thermal noise and radiation pressure noise – are displacement noise sources. Such noise sources cause motions of the test masses which are independent of their separation, and thus when converted to gravitational wave strain, these are reduced in proportion to the arm lengths.

Observations at low frequency may involve different analysis techniques to extract the maximum information from the waves. In particular at low frequencies, determining direction to the source from the phase of the signal at spatially separated detectors becomes less effective (equivalent to a worsening diffraction limit for longer wavelength gravitational waves) and direction information may come instead by analyzing the signals from detectors with different spatial orientation to determine the direction of travel from the transverse polarization of the wave. Further studies are needed to assess the scientific capabilities of configurations, and to settle on the final desired network.

The most advanced concept for a low frequency detector is the Einstein Telescope (ET) project. A conceptual design study for ET was funded by the European Commission beginning in 2008, to assess the feasibility of an underground detector with high sensitivity below 10 Hz. This study will consider a variety of different options, including cryogenics to reduce thermal noise, larger test masses to reduce radiation pressure noise, and squeezing to reduce quantum noise. Initial studies will assume a total length of 30 km for the underground tunnels and will evaluate different configuration including triangle and rotated L's to be sensitive to both polarizations.

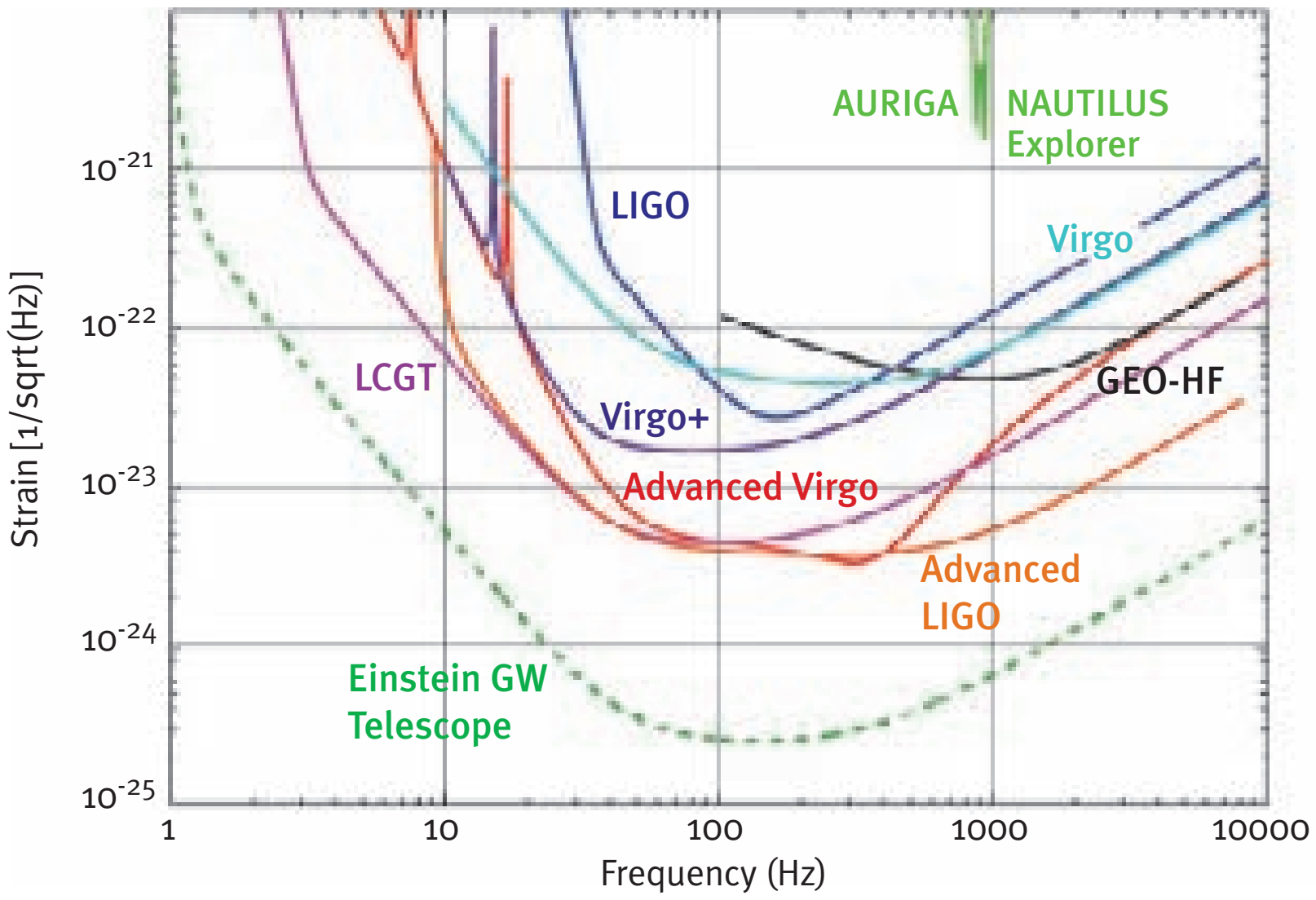

Fig. 5.5 – Schematic sensitivity curves for the evolving gravitational wave detectors.

From a technical point of view, ET could be ready for a start of construction, especially the civil engineering, in 2015. However, funding will require a first detection, which may occur during the eLIGO/Virgo+ era, but is more likely after a year of data collection with Advanced LIGO/Virgo.

This timeline is compatible with the use of the period 2016 - 2018 for preliminary site preparation. Fabrication/construction could commence post 2018. Hence, in this scenario, the start of commissioning of a first version ET detector could be in 2023.

In parallel with this, there is the possibility for a third generation detector to be built in the US, potentially in the Deep Underground Science and Engineering Laboratory (DUSEL).

More generally, other countries such as China and India are seeking to increase their investment in forefront science to enhance their technical capabilities and intellectual capital. Thus the idea of a third generation facility in Asia could materialize within the next decade.

### Surface facilities

The second generation facilities will certainly play a major role in the third generation network. Already, technologies are being investigated which, although not ready for inclusion in the Advanced LIGO and Advanced Virgo detectors, can be used to upgrade these to higher sensitivity. The existing surface facilities (LIGO and Virgo) will not have reached their ultimate sensitivity limits with the Advanced LIGO and Advanced Virgo detectors, except perhaps at the lowest frequencies. Vacuum levels at these facilities will not limit sensitivity at high frequencies until detectors with ~10 times better broadband sensitivity than Advanced LIGO and Advanced Virgo are built.

Improved surface detectors will remain competitive at higher frequency and will provide valuable information for determining the source direction. Thus, they will complement ET and the other potential underground detector(s) which are more sensitive at low frequency. Many of the technologies under development for low frequency detectors (e.g., cryogenics) may also be applicable to improved surface detectors, but other technologies must still be developed. One interesting possibility is the potential development of specialized instruments for particular observations, for example narrow band instruments for LMXB's, or specific pulsars.

The gravitational wave community should also recognize that there may be some evolution in the usage of existing surface facilities to include more than a purely gravitational wave observatory role. The scale and cost of proposed underground interferometers may benefit from some usage of the surface facilities for large scale testing of techniques proposed for underground facilities. In addition, there may be alternative experiments, such as tests of the isotropy of space, which may be proposed to use the unique capabilities of the surface facilities, particularly the long vacuum envelope. The gravitational wave community will need to devise mechanisms to evaluate such usages, and to coordinate them to minimize the disruption to gravitational wave observations.

**Priority — Continued R & D efforts in collaboration with existing design study teams to support the construction, beginning in the post 2018 time frame, of the Einstein Telescope, soon after the expected first gravitational-wave discoveries have been made.**

**Recommendation** — We recommend that GWIC works with the international ground-based gravitational wave community to plan how to optimize the scientific capabilities of a future third-generation network. Specifically, GWIC, in collaboration with any design study groups in the various regions and countries, should organize meetings to assist the community to understand and establish science-driven requirements (e.g. frequency range, sensitivity), possible interferometer designs and configurations, technologies, optical layouts, site configurations and orientations, etc. that would optimize the scientific potential of the network.

**Recommendation** — We recommend that GWIC establishes an international steering body to organize workshops, and promote coordinated R&D efforts in collaboration with existing design study teams to help achieve the goal of an optimized third-generation network.

### 5.2.3 Research and development directions

Technology for achieving the sensitivity required for the second generation detectors is now largely in place. However, technology limitations will continue to play a significant role in the development of third generation instruments. Maintaining a vigorous, forward-looking R&D program to address key technology challenges is crucial to the success of future gravitational wave detectors.

The most important technology limits include
- Gravitational gradients
- Thermal noise
- Quantum noise

The effectiveness of an underground location for suppressing gravitational noise is still not well understood. A significant modeling effort must be combined with a well-considered program of measurements to validate the models and to provide realistic input data to determine whether the required low frequency performance can be achieved. In addition, the community needs to collect seismic and geological data from different locations to aid in planning and siting underground low frequency detectors. Ideally, measurements of such gravitational noise in some frequency range could anchor the models, but would be a challenging goal.

Cryogenics will likely play a key role in third generation detectors. It will be a challenge to develop technologies to minimize mechanical noise that could limit sensitive gravitational wave detection while maintaining adequate cooling power. Much pioneering work in this area has been done by the LCGT Collaboration, but further improvements will be crucial as detectors push towards lower frequencies. In particular, research should be pursued into materials for test masses of significant mass and ultra-low loss at cryogenic temperatures. It will also be important for the broader community to develop expertise in these areas so that the necessary technology is robust and ready to be deployed to all relevant global sites when needed.

Thermal noise due to mechanical losses in coatings is expected to be a main limitation for second generation interferometers. Reducing this will be a critical step toward the third generation network. Improved coatings, either from alternate materials or through improved processing and process control, may help. Interferometer configurations that minimize or even eliminate coatings may also prove effective.

Several paths for quantum noise reduction are under investigation. Quantum noise in the interferometer readout can be reduced by the injection of squeezed vacuum into the interferometer, while radiation pressure noise can be reduced either through squeezing or through increasing the mass of the test mirrors. New interferometer configurations (variable reflectivity recycling mirrors, "optical bars," adaptive controls, etc.) may also be effective in reducing quantum noise.

Already the gravitational wave community cooperates on research in many of these areas, but improved collaboration and communication is still possible. Many areas of technologies apply to both surface and underground interferometers, to space-based detectors, and to other precision measurement experiments. Focused workshops to bring together the researchers from these different communities to share results and promising new ideas will help leverage limited resources.

**Recommendation** — We recommend that GWIC sponsors a series of workshops, each focused on the status and development of a particular critical technology for gravitational wave instruments. Topics in such a series could include cryogenic techniques, coating development for reduced thermal noise, "Newtonian noise," techniques for quantum noise reduction, and overall network configuration. These workshops will help promote exchange of ideas, provide visibility and encouragement to new efforts in critical areas of technology development, and bring to bear the combined resources of the community on important problems.

## 5.3 Acoustic (bar) detectors

Bar detectors were the first gravitational wave detectors in the acoustic band, and have continued for decades to play an important role in high frequency networks. In recent years though, they have been surpassed in sensitivity by laser interferometer based detectors. Already approved second generation interferometric detectors (Advanced LIGO and Advanced Virgo) will make this disparity even more pronounced.

For some time, the excellent duty factor of the bar detectors (the percentage of time they are able to take observational data) has been a major reason for their continued operation, complementing the poorer duty factor of the initial interferometers, particularly in the search for gravitational waves from such rare events as a supernova in our galaxy. However, improvements to the interferometers' duty factor, and growth in the interferometer network, have improved the coverage time. Furthermore, estimates of the gravitational waves from supernovae have become more certain, with many models predicting gravitational waves detectable by the current bar detectors only in a nearby region of our galaxy.

The bar detector groups have explored a number of concepts and techniques for more sensitive detectors: spherical detectors, nested resonators (the DUAL Concept), new materials, ultra cryogenic operation. Unfortunately, this research has not yet identified a clear path toward future sensitivities which can compete with those of the interferometric detectors. Even reasonably optimistic projections for the sensitivity of next generation bar detectors do not approach those of second generation interferometer detectors.

As a result, the supporting agencies have begun to limit or reduce the funding for currently operating bar detectors. However the operation of Nautilus and Auriga will continue until the advanced interferometers are observing, with the operation of Explorer, in CERN, being phased out in the near future.

Many of the technologies and skills developed by the bar detector community may find application to third generation interferometric detectors. Cryogenics, low noise suspensions, thermal noise reduction and data analysis are all topics which need fostering in future projects. Scientists talented in these skills, and who already appreciate their application in the context of gravitational wave detection, can play a vital role in future projects.

**Recommendation** — We recommend that, when the time is appropriate, other projects in the gravitational wave community encourage scientists from the bar detector community to join the ongoing detector projects. The expertise of the bar detector scientists in ultra-sensitive measurement technologies and data analysis is a valuable resource and should be retained within the gravitational wave community.

## 5.4 Data access issues

As the field of gravitational wave astrophysics matures, there is a need to establish a data access policy which will best serve the scientific community.

The current data access model follows the "normal" standard of High Energy Physics. Collaborations have been established to build and operate the instruments, and perform the data analysis. Data are available only to collaboration members, and medium to small groups within each collaboration are responsible for different analysis targets. The collaborations vet the results before publication, and all collaboration members are authors of all collaboration papers. Member groups of the collaboration are expected to perform some degree of service work for the general collaboration (hardware development, operations support, coding, verification of calibration, studies of glitches), in addition to their scientific investigations. New members can join the collaboration by demonstrating a level of commitment to the collaboration's efforts comparable to current members, and normally achieve authorship after a fixed period (typically one year).

The logic behind this choice of model is simple: building, operating and analyzing the data from these instruments is a complex and demanding task, and data analysis rights are a major part of the reward for contributing to this effort. A second aspect is that the data carry numerous instrumental artifacts, and correct data analysis requires knowledge of the instrument to avoid errors in interpretation.

However, achieving the best science output from gravitational wave data will require the engagement of the broader astronomy community, and this data access model runs counter to that of most astronomical facilities, at least in some parts of the world. Many astronomy facilities require data to be released to the broader community, and astronomers interested in using gravitational wave data will expect a similar policy to be in effect.

Questions which need to be addressed in this planning include the following:

- What data products should be produced to adequately serve the user community? What tools and training need to be provided to the user community to enable effective use of gravitational wave data?

- Where does the labor to produce these come from? This is not just a question of funds if the skills needed to produce them (e.g., via interferometer experts) are difficult to attract to this work.
- Is there any proprietary period for the collaborations which operate the gravitational wave detectors to analyze the data? If so, what length of time or what other criteria should trigger the release to the broader community?
- Does the gravitational wave community need to establish any quality control on the use of the data from the gravitational wave network?

The creation of a unified network of detectors complicates the issue of data release, since the data are generated by different collaborations, operating under the rules of different funding agencies.

It will be important that sufficient international dialog takes place before agreements on this topic are finalized between individual projects and their funding agencies.

**Priority — The development of procedures that, beginning in the era of frequent ground-based detection of gravitational waves, will allow the broader scientific community to fully utilize information about detected gravitational waves.**

## 5.5 General background reading relevant to this chapter

*http://gw.icrr.u-tokyo.ac.jp:8888/lcgt/*

*http://www.gravity.uwa.edu.au/*

*http://www.et-gw.eu/*

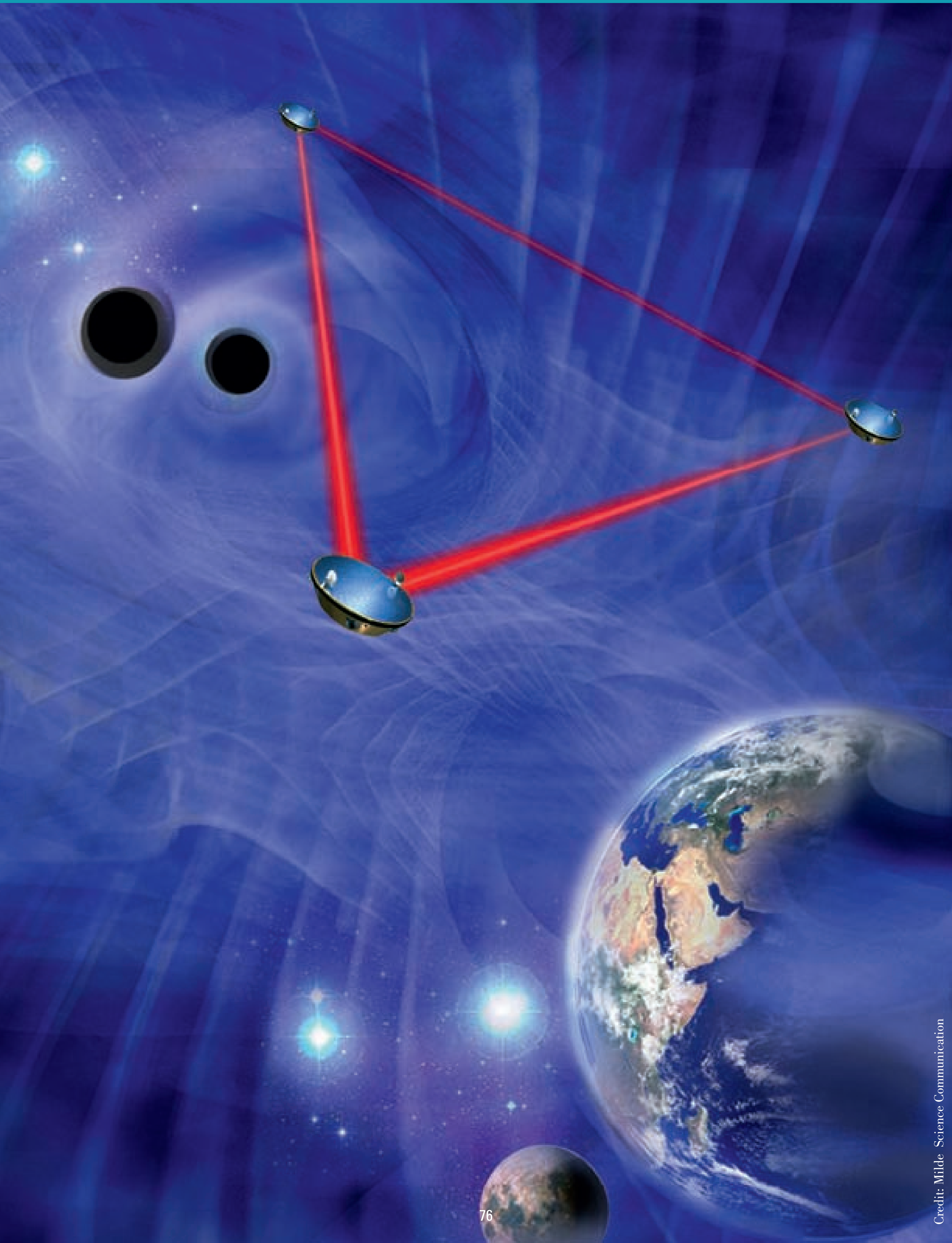

Credit: Milde Science Communication

# 6. The future of the field in response to anticipated scientific opportunities — in space/lower frequencies

## 6.1 Introduction and overview

The range of frequencies from 0.1 Hz down to 1 / (age of the Universe) offers some of the most spectacular phenomena to be studied by gravitational wave science. The scope of gravitational signals comprises mergers of supermassive black hole binaries, extreme mass ratio inspirals into massive black holes, stochastic backgrounds from the early Universe and quantum fluctuations after the big bang. Ground-based detectors have limited sensitivity to gravitational waves with frequencies below about 0.1 Hz because of the unshieldable background of terrestrial gravitational noise. Detection technologies at these frequencies are diverse and range from large baseline laser interferometry through spacecraft tracking and pulsar timing to polarization measurements of the cosmic microwave background.

All of these technologies will eventually be used to observe the complete gravitational wave spectrum covering more than 20 orders of magnitude in frequency. But, for a space mission, the priority of the gravitational wave community is to fly the LISA mission in 2020 to open the low-frequency window from 0.1 mHz to 0.1 Hz. Such a launch is technologically feasible and entirely timely, considering that the technology precursor mission LISA Pathfinder launches in 2012.

Technology development for a LISA follow-on mission is expected to be rather generic across the spectrum of missions now envisaged and should receive focus once the LISA technology is frozen in 2013. Depending on various funding possibilities, an intermediate scale LISA follow-on mission of the type described later in this chapter could yield important scientific payoffs, before a more advanced space gravitational wave observatory is launched. We note that a complementary mission, to study the Cosmic Microwave Background polarization (CMBPOL), is planned, which might be considered a mission of opportunity. Although its prime goal is not the observation of gravitational waves, it will nevertheless furnish interesting gravitational wave observations in the nanohertz region. CMBPOL, however is outside the scope of this document.

No timescale is suggested for the post-LISA developments. Still technology improvements should be planned to allow a smooth transition into the routine multi-wavelength gravitational wave astronomy that is certain to develop after the first detections have been made by the advanced ground-based detectors, and then LISA.

## 6.2 Space Missions

### 6.2.1 LISA – Laser Interferometer Space Antenna

LISA is a joint ESA / NASA space mission planned for launch in 2020. It is the priority space-mission for the gravitational wave community.

The LISA mission uses three identical spacecraft whose positions mark the vertices of an equilateral triangle five million km on a side, in orbit around the Sun. LISA can be thought of as a giant Michelson interferometer in space, with a third arm that provides independent information on the two gravitational wave polarizations, as well as redundancy. The spacecraft separation – the interferometer armlength – sets the range of gravitational wave frequencies LISA can observe (from about 0.1 mHz to above 0.1 Hz).

This range was chosen to reveal some of the most interesting sources: mergers of massive black holes, ultracompact binaries, and the inspirals of stellar-mass black holes into massive black holes.

LISA will be sensitive enough to detect gravitational wave induced strains of amplitude $h = \Delta l / l < 10^{-23}$ in one year of observation, with a signal-to-noise ratio of 5.

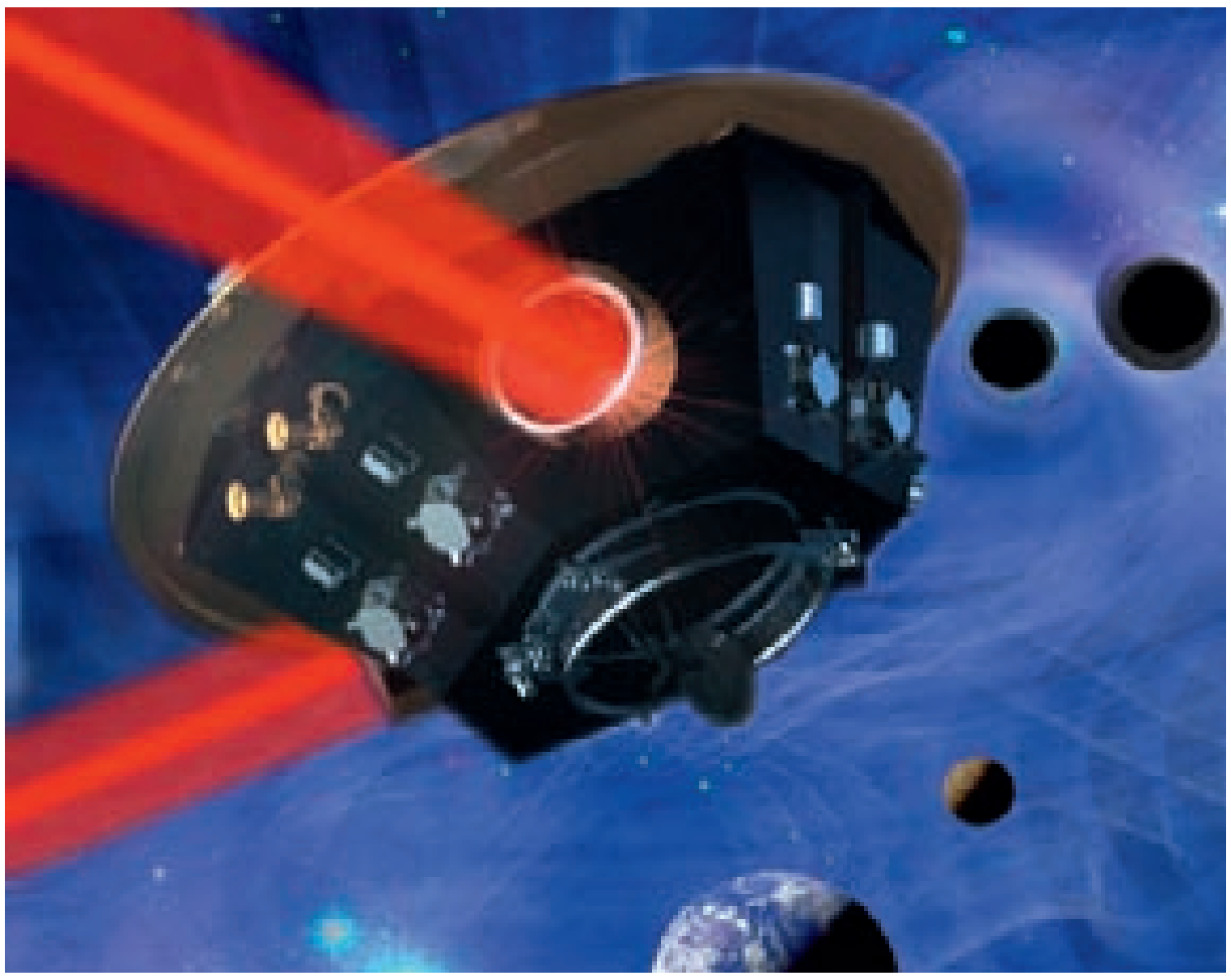

Fig. 6.1 – The LISA triangle will follow 20° behind Earth

The center of the LISA triangle traces an orbit in the ecliptic plane, 1 AU from the Sun and 20° behind Earth, and the plane of the triangle is inclined at 60° to the ecliptic (see Figure 6.2). The natural free-fall orbits of the three spacecraft around the Sun maintain this triangular formation throughout the year, with the triangle appearing to rotate about its center once per year.

The actual implementation used for LISA resembles the technique known as spacecraft Doppler tracking, but realized with infrared laser light instead of radio waves. The laser light going out from one spacecraft to the other corners is not reflected back directly, because diffraction losses over such long distances would be too great. Instead, in analogy with an RF transponder scheme, the laser on the distant spacecraft is phase-locked to the incoming light and transmits a signal back at full intensity.

When the transponded laser light arrives back at the original spacecraft, it is superposed with a portion of the original laser beam, which serves as the local oscillator in a standard heterodyne detection scheme. This relative phase measurement gives information about the length of that interferometer arm, modulo an integer number of wavelengths of the laser light. The difference between the phase measurements for the two arms gives information about the relative changes in the two arms — the gravitational wave signal. A two-arm interferometer can be prone to phase errors due to laser frequency fluctuations. If the arms were exactly equal in length, then laser frequency fluctuations would cancel perfectly in the armlength difference measurement.

Unfortunately, annual variations in the LISA spacecraft orbits prevent perfect cancellation of laser noise. To minimize the measurement error from laser phase noise the lasers are frequency

stabilized — first to an optical cavity, and then to the 5 million km interferometer arm. Any residual laser frequency noise in the LISA measurements will be removed by post-processing on the ground using a technique called Time Delay Interferometry (TDI)

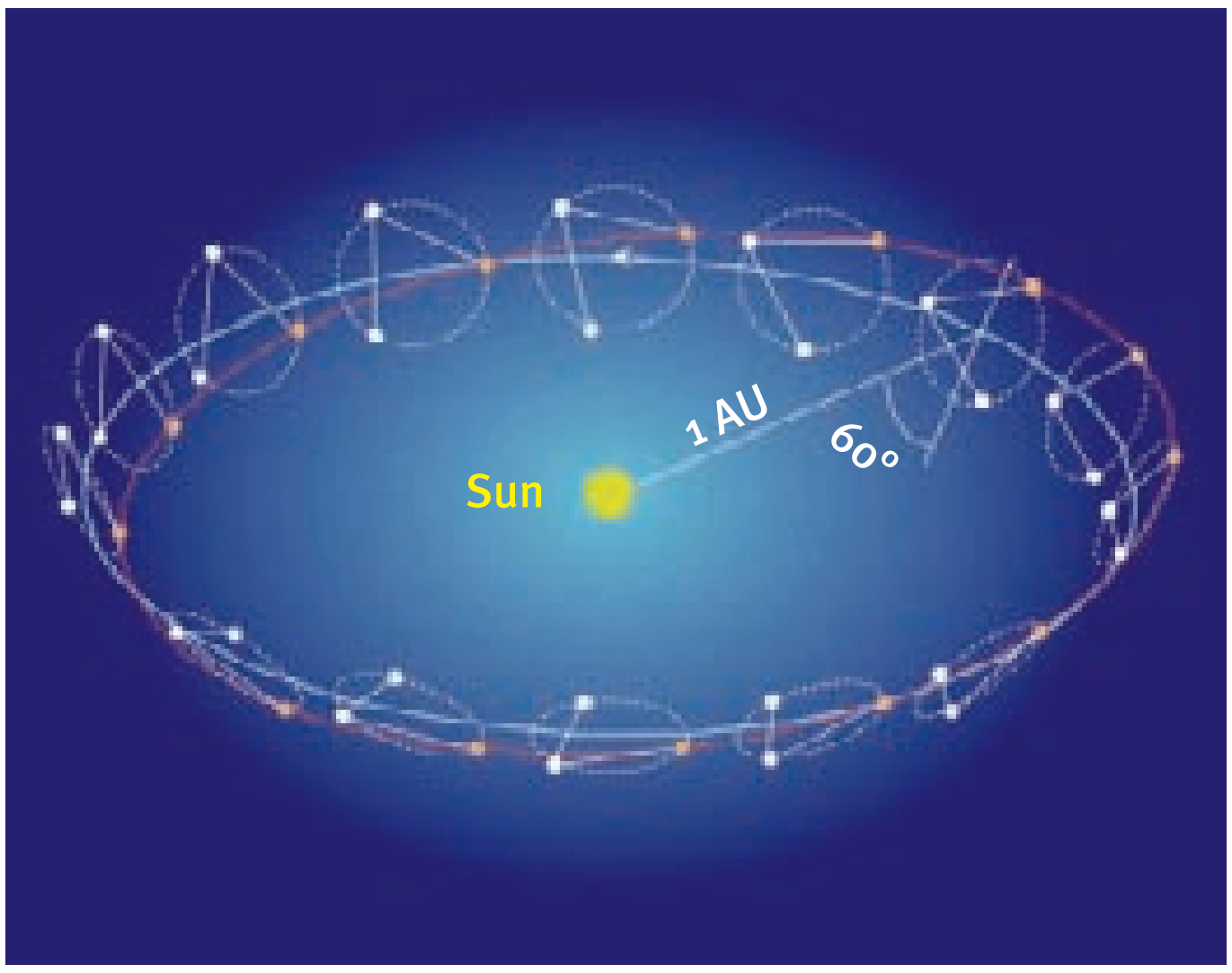

Fig. 6.2 – LISA orbit

Each spacecraft contains a pair of optical assemblies oriented at roughly 60° to each other. Each assembly is pointing toward a similar one on the corresponding distant spacecraft, to form a (non-orthogonal) Michelson interferometer. Through a 40 cm aperture telescope on each assembly, a laser beam from a 1.064 μm Nd:YAG master laser and 2 W Yb-doped fiber amplifier is transmitted to the corresponding remote spacecraft. The same telescope is used to collect the very weak incoming beam (around 100 pW) from the distant spacecraft, and direct it to a sensitive photodetector, where it is combined with a local-oscillator beam derived from the original local laser light. At the heart of each assembly is a vacuum enclosure containing a free flying polished platinum-gold cube, 4 cm in size – the *proof mass* that serves as an inertial reference for the local optical assembly.

The spacecraft surrounding each pair of optical assemblies serves primarily to shield the proof masses from the adverse effects of solar radiation pressure fluctuations; the spacecraft positions do not enter directly into the measurements. Nevertheless, in order to minimize disturbances to the proof masses from fluctuating forces in their vicinities, each spacecraft must be kept moderately centered around the proof masses (to about 10 nm/√ Hz in the measurement band). This is achieved by a *drag-free* control system based on small electric thrusters and displacement sensors.

All three spacecraft can be launched together by a single Atlas V. Each spacecraft carries a small, steerable antenna used for transmitting the science and engineering data in the Ka-band to the NASA Deep Space Network. The nominal mission lifetime is five years.

There is already a robust community effort in the development, simulation and demonstration of LISA data analysis (see the bibliographic database at www.srl.caltech.edu/lisa). LISA has an active program of data simulation and “mock data challenges” to refine the understanding of its science capabilities. Techniques have been developed for detection and parameter estimation for all classes of LISA sources: massive binary black hole coalescences, galactic white-dwarf binaries, extreme mass-ratio inspirals, and stochastic backgrounds. These techniques have a rich heritage, not only from ground-based gravitational wave detectors, but also from related analysis fields such as sonar, radar, seismology, radio astronomy, and voice recognition.

LISA directly addresses many of the research priorities and big questions raised by recent astronomy and physics decadal and community reports such as *Astronomy and Astrophysics in the New Millennium*, and *Connecting Quarks with the Cosmos*.

In Europe, LISA and LISA Pathfinder are integral components of ESA’s Cosmic Vision Scientific Programme announced in 2005 (ESA 2005). In the US, LISA has been endorsed as a high priority mission in several influential reports. In particular, most recently in 2007, the *Beyond Einstein Program Assessment Committee, commissioned by the National Research Council of the National Academies*, gave a glowing recommendation for LISA.

**Priority — The completion of LISA Pathfinder and a timely selection of LISA in 2012/2013 for a launch in 2020 to open the low frequency gravitational wave window from 0.1 mHz to 0.1 Hz.**

### 6.2.2 Other space missions

There are several possible intermediate scale follow-on missions to LISA discussed below, missions which have a variety of fruitful scientific outcomes including characterization of dark energy, elucidation of the formation mechanism of supermassive black holes in the center of galaxies, potential detection of the cosmological background and verification and characterization of inflation.

These would bridge the frequency gap between LISA and terrestrial detectors such as Advanced LIGO, Advanced Virgo and LCGT. They can be valuable as a follow-up to LISA by observing inspiral sources that have moved above the LISA band, and can also play a role as a predictor for terrestrial detectors by observing inspiral sources that have not yet moved into the terrestrial detector band.

Further, an important advantage of specializing in the frequency band between ~0.01 Hz and 10 Hz is that the confusion limiting noise caused by irresolvable gravitational wave signals from many compact binaries is expected to be very low above 0.1 Hz, and thus improving potential searches for a stochastic gravitational wave background.

### 6.2.2.1 Intermediate Frequency Missions (Example Mission Concepts: ALIA, DECIGO, BBO)

ALIA is a proposed intermediate LISA follow-on mission aimed at observing the frequent merger events of 10 solar mass black holes with Intermediate Mass Black Holes (IMBHs) of 50 to 50,000 solar masses at redshifts up to z = 10. This will shed light on the important question of how massive black holes formed initially and then grew in mass. ALIA may achieve roughly 30 times more sensitivity than LISA and a sweet spot of the sensitivity curve for signal frequencies between 30 mHz and 100 mHz via use of 500,000 km arms and by employing 30 W lasers and 1.0 m mirrors. This is completely viable technology and a launch could be scheduled within a few years after LISA.

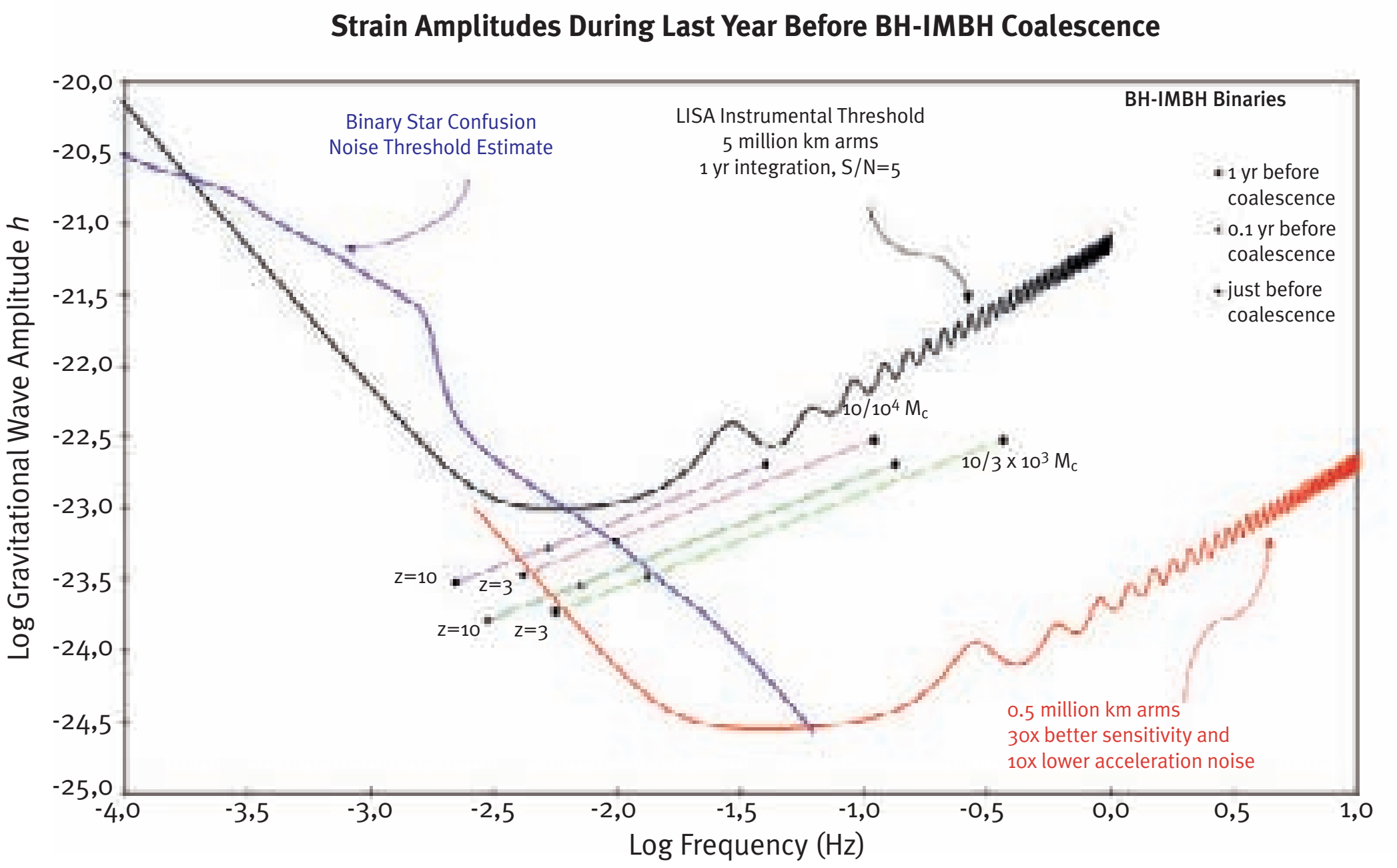

Fig. 6.3 – Sensitivity for LISA and a possible intermediate LISA follow-on mission.

The DECi-hertz Interferometer Gravitational wave Observatory (DECIGO) is a higher sensitivity mission, with a maximum sensitivity in the 1 – 10 Hz range.

The DECIGO pre-conceptual design consists of three drag-free spacecraft, whose relative displacements are measured by a differential Fabry–Perot (FP) Michelson interferometer. The arm length of 1,000 km was determined to realize a finesse of 10 with a 1 m diameter mirror and 0.5 µm laser light. The mass of the mirror is 100 kg and the laser power is 10 W. Three sets of such interferometers sharing the mirrors as arm cavities comprise one cluster of DECIGO. The constellation of DECIGO is composed of four clusters of DECIGO located separately in a heliocentric orbit.

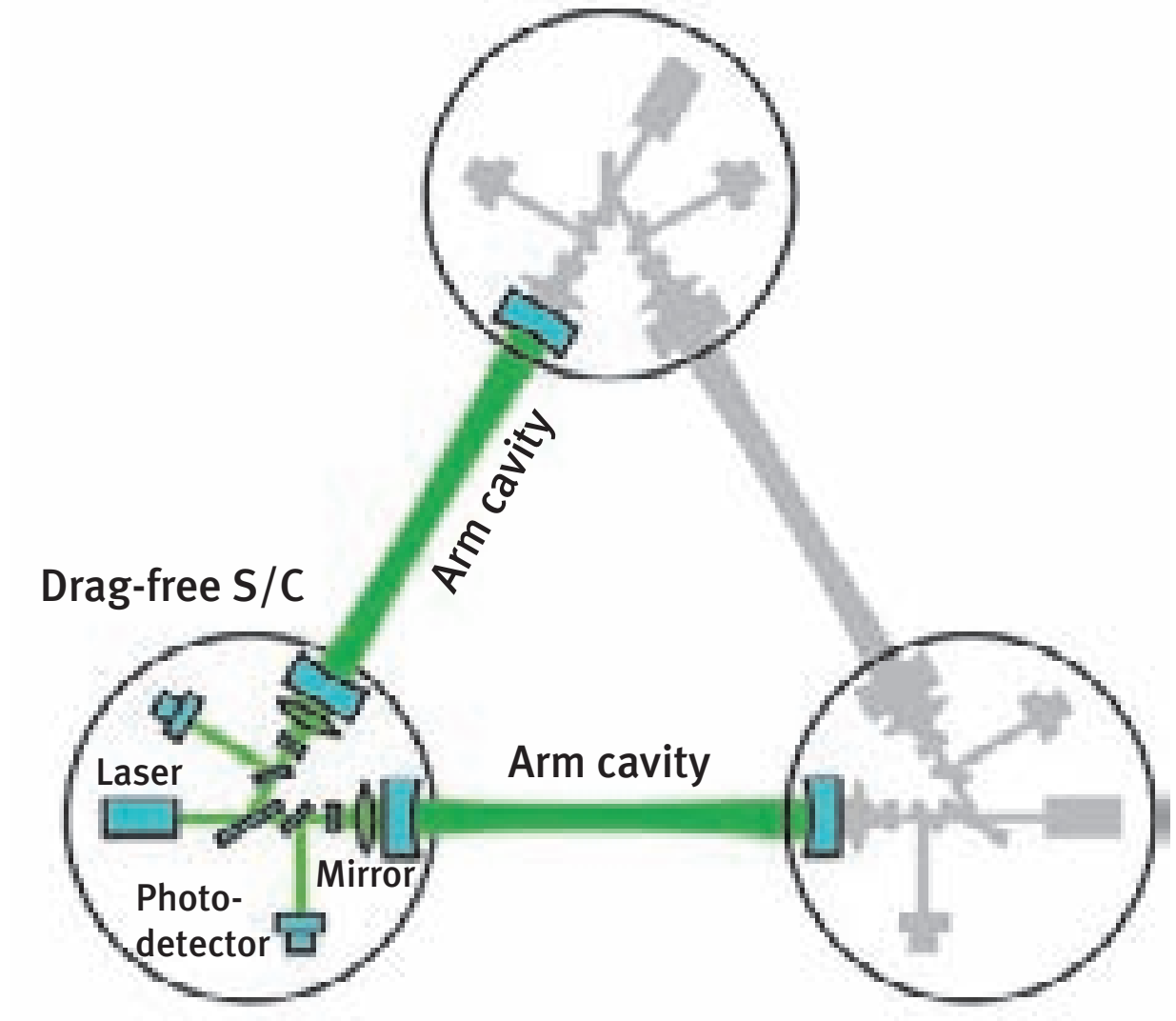

Fig. 6.4 – Pre-conceptual design of DECIGO.

Ideally its sensitivity goal will be limited by radiation pressure noise below 0.15 Hz, and by shot noise above 0.15 Hz. To attain this, all the technical noise should be suppressed well below this level, which imposes stringent requirements for the DECIGO subsystems.

The plan is to launch two missions before DECIGO: DECIGO pathfinder (DPF) and pre-DECIGO. DPF will test the key technologies with one spacecraft, and has recently been selected as one of the five important mission candidates for the small satellite series run by JAXA/ISAS with a potential launch date of 2015. Pre-DECIGO, a mission with three spacecraft forming a single two-arm interferometer, is targeted at detection of gravitational waves with minimum specifications. Ideally pre-DECIGO would be launched six years after DPF. Finally, full DECIGO could be launched in another six years after Pre-DECIGO, helping to open a new intermediate window of observation for gravitational wave astronomy.

The Big Bang Observer (BBO) is an additional proposed LISA follow-on mission, targeted at detecting stochastic gravitational waves from the very early Universe. It will also be sensitive to the final year of binary compact body (neutron stars and stellar mass black holes) inspirals out to $z < 8$, mergers of intermediate mass black holes at any $z$, rapidly rotating white dwarf explosions from Type 1a supernovae at distances less than 1 Mpc and ~1 Hz pulsars with non-axisymmetric magnetic fields of $B > 3 \times 10^{14}$ G.

The mission is proposed in a number of stages. The first will consist of three spacecraft in solar orbit separated from each other by 50,000 km. Ultimately, in the final stage, there would be three such constellations of spacecraft, separated by 120° from each other in solar orbit.

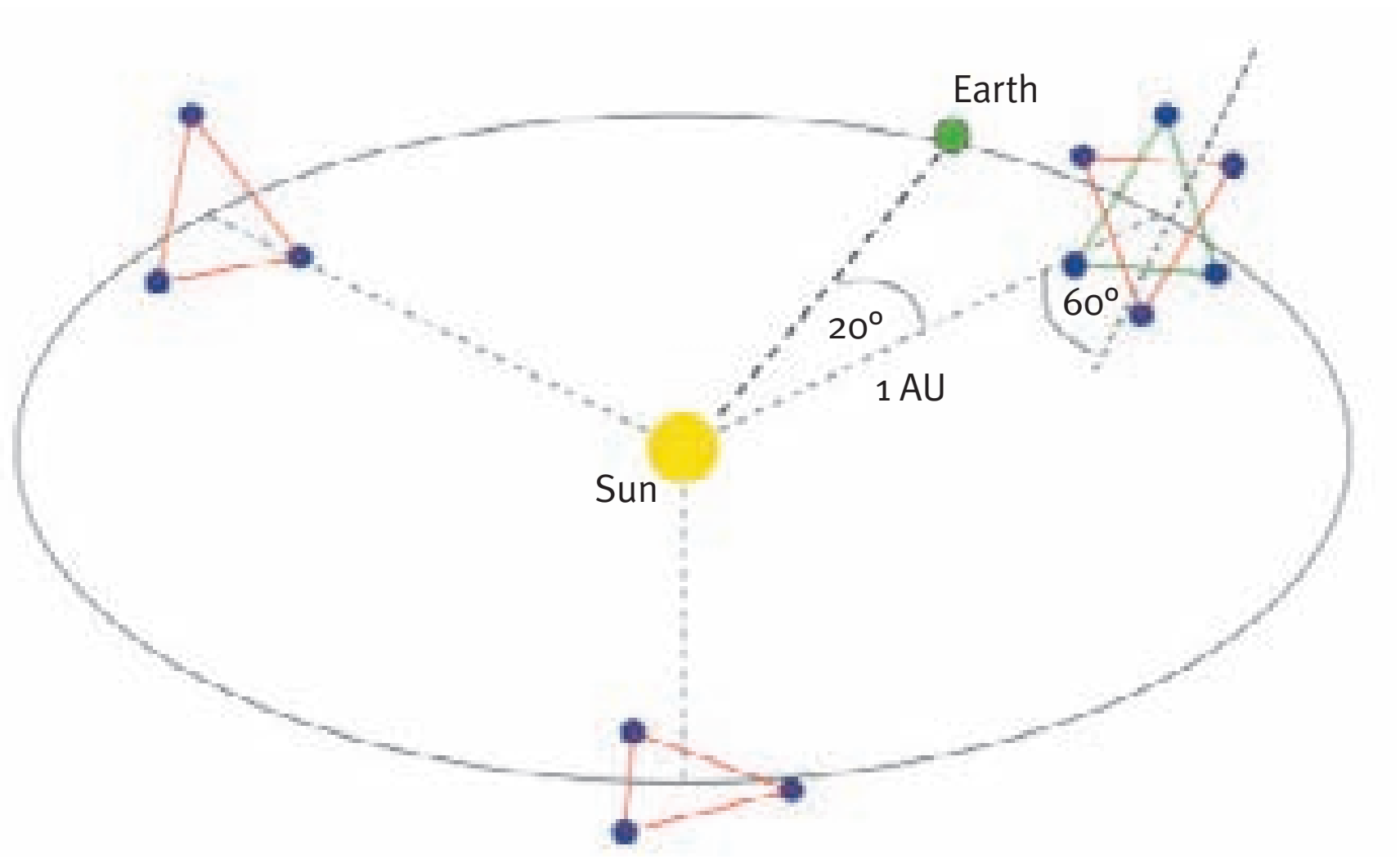

Fig. 6.5 – BBO layout (from J. Crowder, N. J. Cornish: Beyond LISA: Exploring Future Gravitational Wave Missions. arXiv:gr-qc/0506015 v3 17 Oct 2005)

The proposed sensitivity near 0.5 Hz will be about $1 \times 10^{-24}$ $Hz^{-1/2}$ in strain or $5 \times 10^{-17}$ m $Hz^{-1/2}$ in displacement. This is to be compared to $1 \times 10^{-20}$ $Hz^{-1/2}$ and $5 \times 10^{-11}$ m $Hz^{-1/2}$ for LISA. The noise spectra of BBO, LISA and Advanced LIGO are compared in figure 6.6.

To reach the required level of sensitivity without using resonant cavities in the arms, the laser power must be high and/or the wavelength must be kept small. Each spacecraft will have two 300 W lasers with a wavelength of 355 nm, obtained by frequency tripling Nd:YAG lasers. About 8 W will be detected at the far spacecraft after diffraction losses from a 2.5 m diameter collecting mirror on the far spacecraft, $5 \times 10^7$ m away.

The need for substantial technology development is key for any and all of the possible post LISA intermediate frequency missions. Successful development of suitable photodiodes, high power and reduced noise lasers, appropriate mirror shapes, high power transmission optics and low thermal noise materials is crucial and needs to be pursued in a timely way.

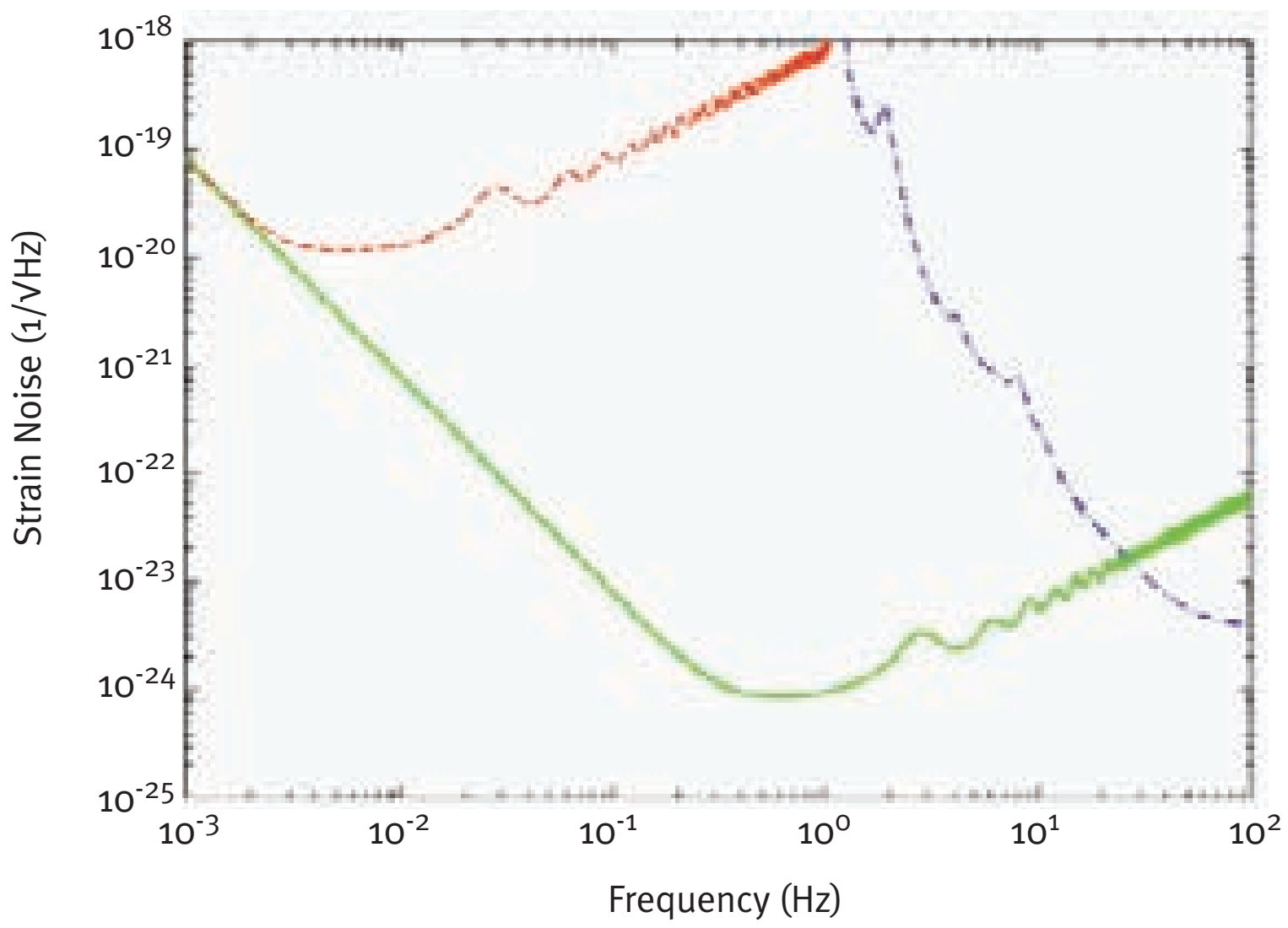

Fig. 6.6 – Strain noise, as the square root of strain power spectral density, versus frequency for three advanced interferometric gravitational-wave detectors: LISA, Advanced LIGO and BBO. LISA is the dashed (red) curve, Advanced LIGO is the dash-dotted (blue) curve and BBO is the solid (green) curve. In all three cases, the noise has been averaged over all possible source locations on the sky and all source polarizations. (From E. S. Phinney)

**Priority — The timely start for technology development for LISA follow-on missions.**

## 6.3 Pulsar timing

**The International Pulsar Timing Array Collaboration is making excellent progress towards detecting signals in the nano-hertz range.**

Looking ahead, the European Very Large Array (EVLA) will provide another highly sensitive telescope in the Northern Hemisphere on a timescale of two years, and the proposed Allen Telescope Array 350-dish build out (USA) would add a similarly sensitive telescope within five years. The proposed Square Kilometer Array (SKA) will have very high sensitivity and is expected to observe 100 millisecond pulsars with rms residuals of the order of 50 ns for observing periods of 10 years, pushing the detection limit for the stochastic background at 3 nHz to $\Omega_{GW} \sim 10^{-13}$. The site of the SKA is not yet chosen but it is planned that it will be operational by the year 2020. Given the importance of the IPTA's science, the community may seriously consider building a dedicated pulsar timing facility.

**Priority — The continued development of an international pulsar timing array for the study of gravitational waves in the nano-Hertz band. This effort requires continued development of algorithms and data acquisition systems, and access to substantial amounts of time on the world's largest radio-telescopes.**

## 6.4 General background reading relevant to this chapter

For literature on LISA and related science see:

*www.srl.caltech.edu/lisa*

*http://universe.nasa.gov/new/program/bbo.html*

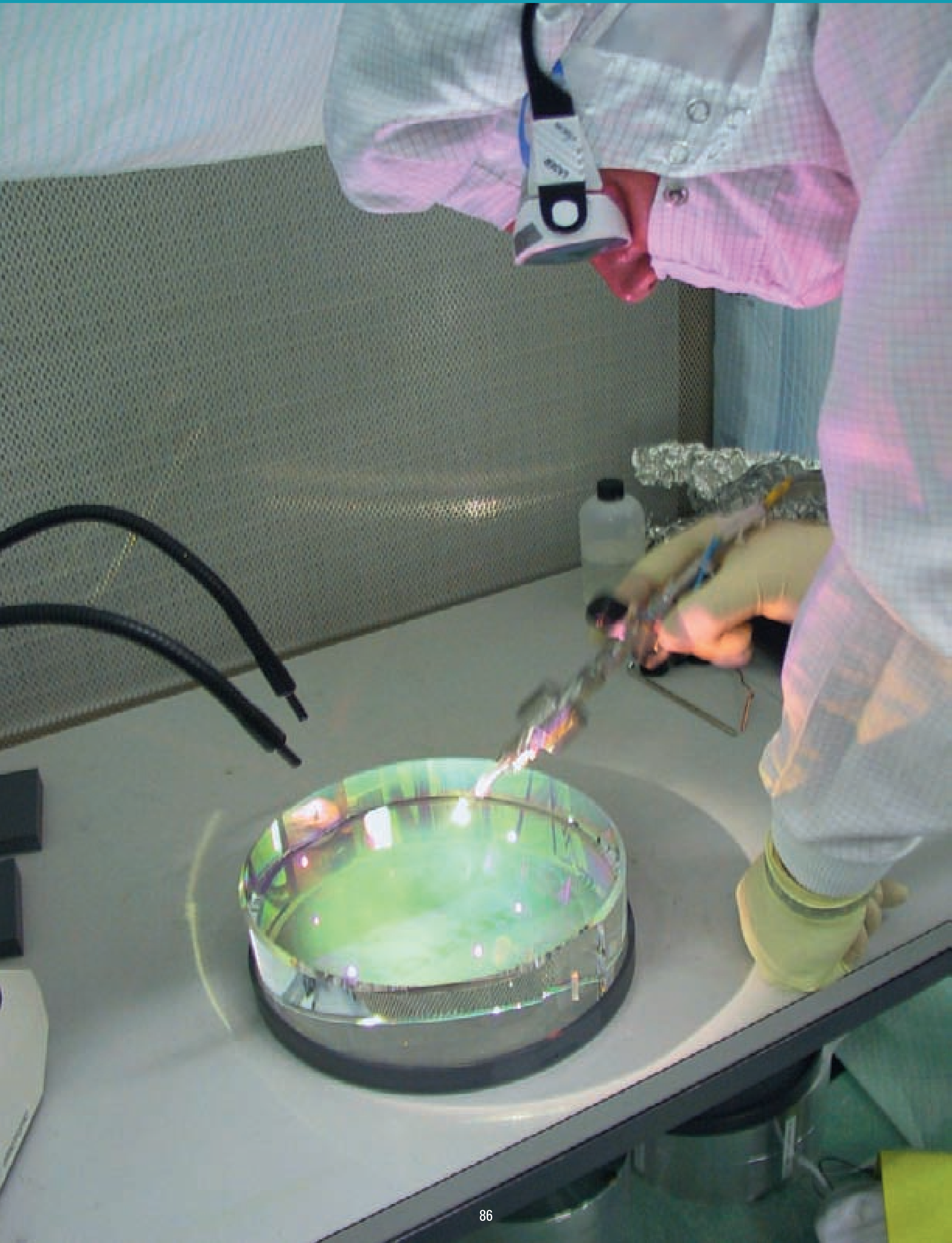

# 7. Impact of gravitational wave science on other fields

## 7.1 Introduction

As it progresses, gravitational wave research will evolve into a vibrant new sector of astrophysical observation of the Universe. As such it is expected to provide a powerful investigative tool in astrophysics, cosmology and fundamental physics.

These aspects of gravitational wave research have been treated in Chapter 4. The focus of this section is on the impact that the development of gravitational wave observatories has had, is having and promises to have, on other fields of scientific research, including:

- The science of classical and quantum measurements and high-precision spacetime metrology at large
- Optics, quantum optics and laser systems
- Space science and technology
- Geology and geodesy
- Material science and technology
- Cryogenics and cryogenic electronics
- Computing
- Methods in theoretical physics.

## 7.2 The science of measurement

Since its beginning, gravitational wave research has required an understanding of the meaning of basic spacetime measurements, for example the relative positions of macroscopic bodies when measured on a sub-femtometer scale. This need led quickly to the investigation and understanding of the quantum limit in the measurements of "macroscopic" degrees of freedom. Gravitational wave research has produced a wealth of relevant results like the fundamental concepts of Quantum Non-Demolition, the Standard Quantum Limit and Back-Action Evasion, methods for analyzing general opto-mechanical systems (like the two photon method), and formal theories of quantum measurement.

The general strategy of Back-Action Evasion has since been adopted in various quantum optics experiments. More recently, optical rigidity and damping/anti-damping have been used extensively in experiments involving mechanical oscillators, with the aim of reaching a Heisenberg-Limited mechanical quantum state.

The original strong push from the gravitational wave field and a corresponding push from quantum optics, has now led to various quantum measurement experiments involving photons, ensembles of atomic or nuclear spins, nano-mechanical-oscillators, and others. Future gravitational wave detectors, on their path for greater sensitivity, will also need quantum measurement schemes and will continue as leading players in this evolving field.

In these applications, large interferometric gravitational wave detectors, and perhaps the most advanced resonant detectors still in operation, will serve as ideal test benches to observe how quantum mechanical laws apply to macroscopic objects.

When the classical noise budget of gravitational wave detectors drops below the standard quantum limit, test masses have the potential to be put into states that are nearly Heisenberg limited, where they can demonstrate macroscopic entanglement and test such theories as teleportation and non-conventional de-coherence, among others.

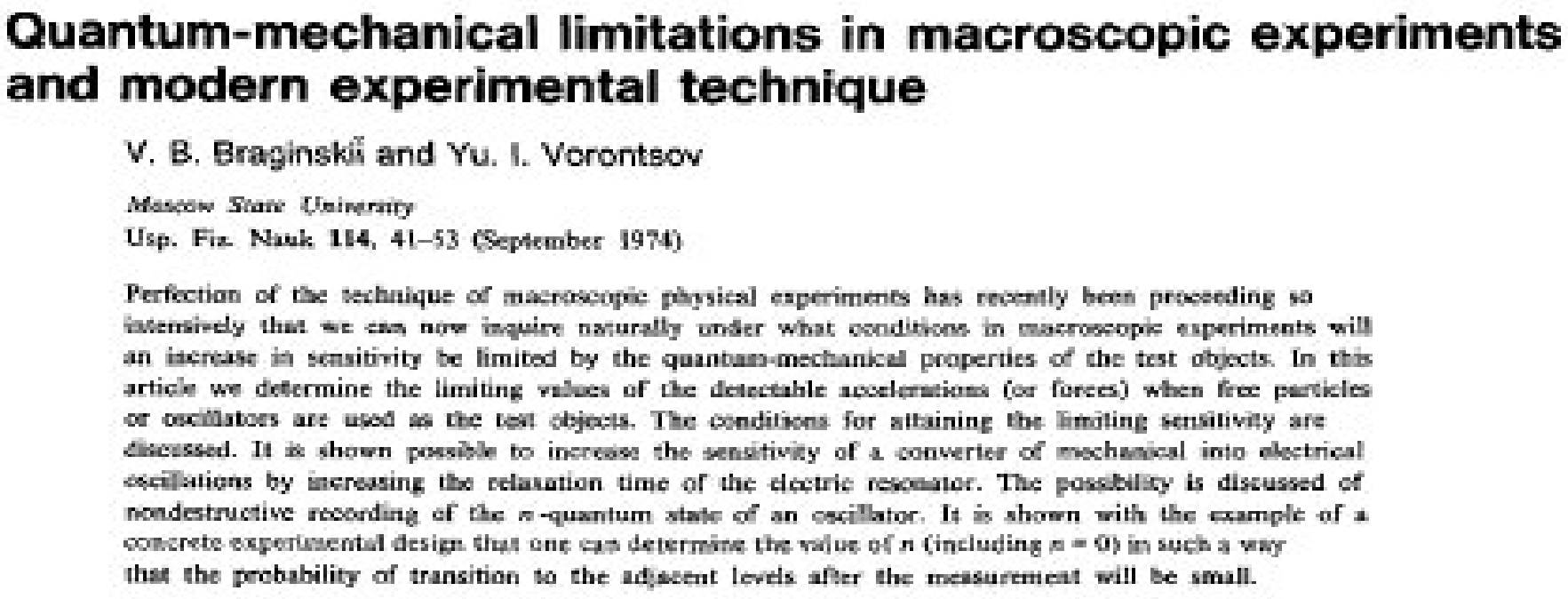

**Quantum-mechanical limitations in macroscopic experiments and modern experimental technique**

V. B. Braginskiĭ and Yu. I. Vorontsov

*Moscow State University*

Usp. Fiz. Nauk 114, 41–53 (September 1974)

Perfection of the technique of macroscopic physical experiments has recently been proceeding so intensively that we can now inquire naturally under what conditions in macroscopic experiments will an increase in sensitivity be limited by the quantum-mechanical properties of the test objects. In this article we determine the limiting values of the detectable accelerations (or forces) when free particles or oscillators are used as the test objects. The conditions for attaining the limiting sensitivity are discussed. It is shown possible to increase the sensitivity of a converter of mechanical into electrical oscillations by increasing the relaxation time of the electric resonator. The possibility is discussed of nondestructive recording of the $n$-quantum state of an oscillator. It is shown with the example of a concrete experimental design that one can determine the value of $n$ (including $n = 0$) in such a way that the probability of transition to the adjacent levels after the measurement will be small.

Fig. 7.1 – The original paper from Braginsky and Vorontsov on QND measurements

Finally, the expertise of gravitational wave scientists and their technologies can easily be used to build small-scale optomechanical systems, which might enter the quantum domain more easily, and could be justified as prototypes for future gravitational wave detectors as well as providing tests of quantum mechanics.

Besides the study of sophisticated measurement schemes, gravitational wave research has produced a series of devices and instruments of direct application to metrology. Some of these are described in the following sections. Here it is worth mentioning a few.

Gravitational wave detection has required the development of ultra-stable oscillators since there is a fundamental relationship between length measurement and frequency measurement (df/f = dL/L). Thus sapphire microwave oscillators were developed as frequency stabilized sources for the motion sensing of a particular resonant mass detector (namely NIOBE at the University of Western Australia), and these were subsequently utilized as flywheel oscillators for atomic clocks, and for measurement of the time dependence of fundamental constants through comparison of different types of clocks. In all cases the direct application of gravitational wave clock technology led to state of the art measurements.

Laser frequency stabilization has been important for the same reason. RF reflection locking (Pound-Drever-Hall) for lasers developed for gravitational

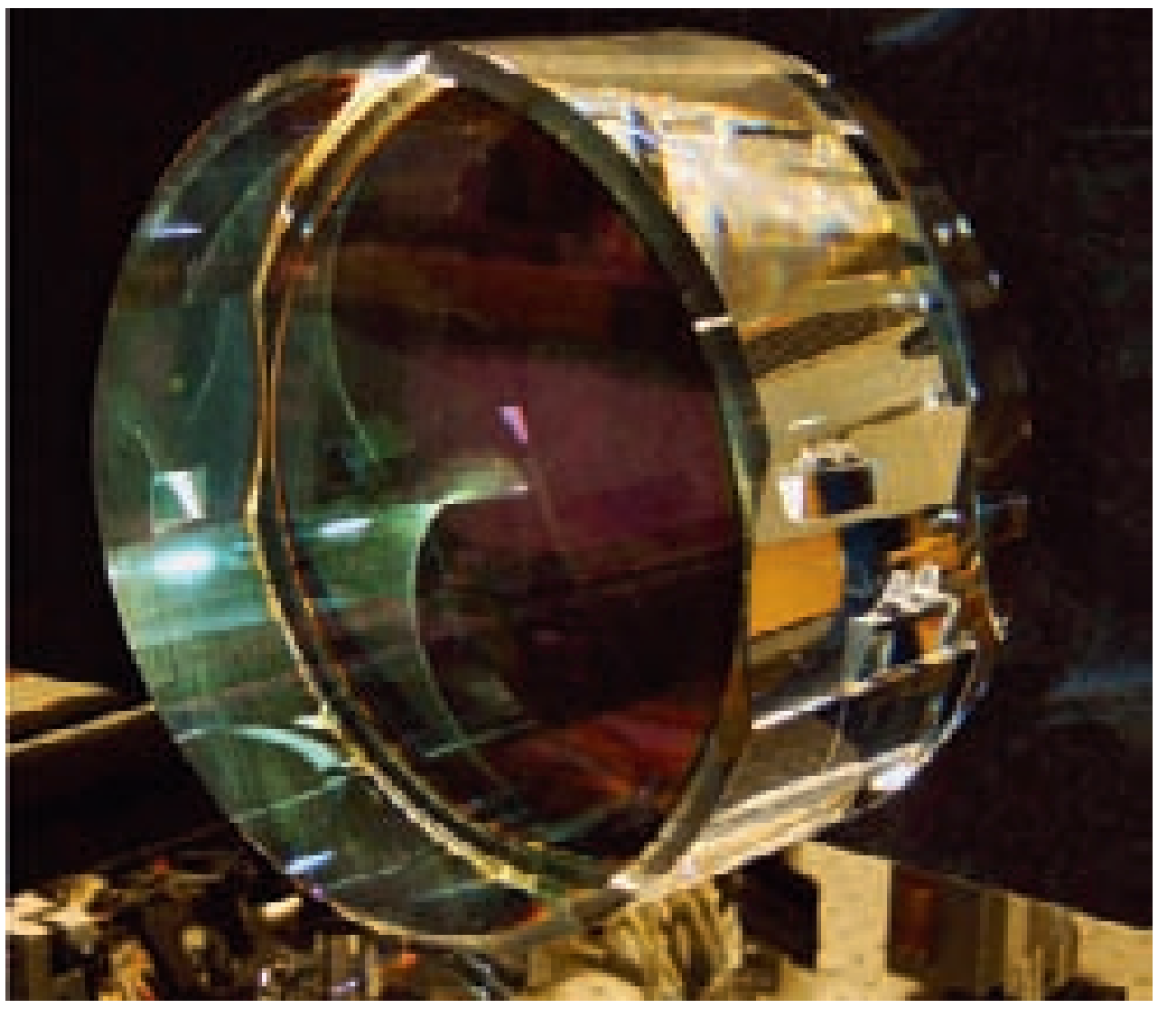

Fig. 7.2 – Beamsplitter for the GEO600 gravitational wave detector with low-loss ion-beam-sputtered dielectric coating and attachments jointed to the mirror using the technique of hydroxide-catalysis bonding, discussed in section 7.6

wave detectors with Fabry-Perot cavities in the arms has become standard stabilization method for obtaining low line widths for spectroscopic and frequency standards applications. Cryogenic sapphire optical cavities and room temperature stabilization cavities based on ULE have found important roles in metrology and in the development of optical clocks.

The study of parametric instabilities in laser interferometers has led to the definition of new methods of cooling mechanical resonators to the quantum ground state through 3-mode parametric interactions. Other cooling techniques based on 2-mode interactions are a direct application of gravitational wave parametric transducers developed in the 1980's and 1990's.

Research on mirror coatings by the gravitational wave community has led to the conclusion that thermal noise in coatings results in a major limitation to the performance of state of the art frequency standards.

Finally resonant detectors with electro-mechanical transducers have pushed the sensitivity limit of audio-band SQUID devices coupled to actual loads very near to the quantum limit.

## 7.3 Optics

Separating optics from measurement science is a bit arbitrary. However optics is one of the sectors where gravitational wave research has consistently contributed a series of devices and methods that need specific mention.

First and foremost the development of high power single frequency 1.06 micron lasers was and continues to be driven by the gravitational wave field, and is now preparing the way for new lasers for free space communication. A laser of this type is, for instance, in operation on the TerraSAR spacecraft testing a long range laser link for optical communication.

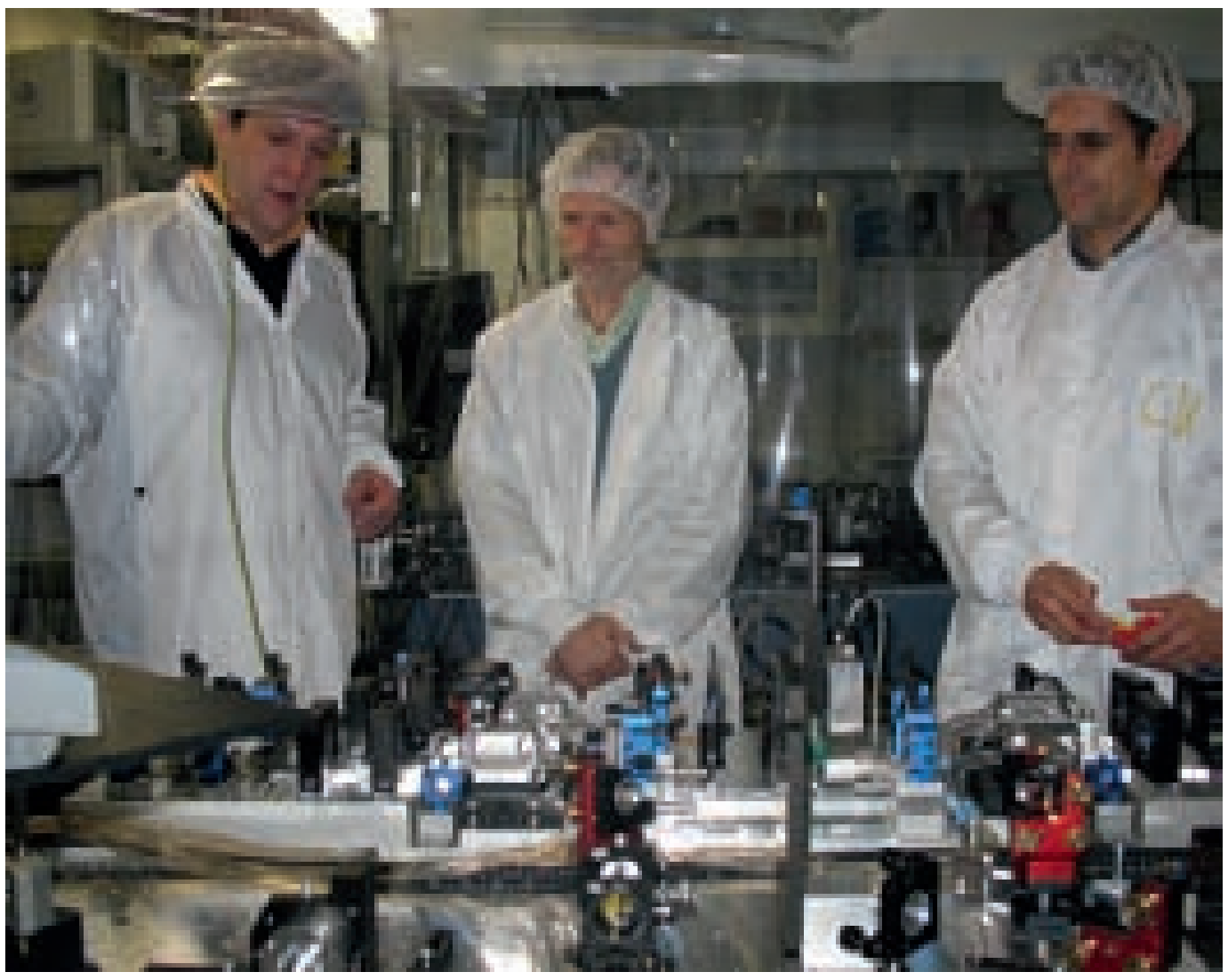

Fig. 7.3 – The Advanced LIGO laser shaping up.

Both space-based and ground-based projects have stimulated the development of fiber based laser systems that eventually will benefit fiber laser technology for remote sensing and coherent laser radar applications.

Low optical absorption at visible and IR wavelengths has resulted in the technique of hydroxy-catalysis bonding (see below) being transferred from the space gravitation measuring community through the gravitational wave research to industry in the US, for use in the fabrication of high-power, single-mode, fiber-amplifier systems. Obviously the already mentioned laser frequency stabilization is a key contribution to optics proper.

The simultaneous requirements for power and amplitude and frequency stability have never been met before, and while there are no other fields that require the same level of overall performance, spin-offs in various subsets have been very useful in remote sensing, laser radar and Na guide star development. In each of these applications spectral control by injection locking or injection seeding was used, and lessons learned from gravitational wave laser research have been useful, especially at higher powers.

Specific devices connected with high quality wave front testing may prove relevant for other fields of optics. For instance high performance Hartmann sensors ($<\lambda/10{,}000$) may be used in optometry and optical diagnostics for both technology and science. Other devices that offer valuable applications are moderate scale (35 cm diameter) optics with $\lambda/1000$ polishing, coating and metrology.

Space-borne interferometry for LISA and LISA Pathfinder is inaugurating an array of new technologies. Fully monolithic, high stability optical benches working in the mHz range may find applications in a series of other space missions and thereby be extended to incorporate steerable mirrors and telescopes with picometer stability, as discussed in the next section.

## 7.4 Space science and technology

LISA is an entirely new concept of space mission. It is probably the first system that consists of a single instrument comprising three spacecraft in which the spacecraft are part of the instrument itself. The extent of the new technologies required by LISA is exceptional and has convinced space agencies to test many of them on the dedicated LISA Pathfinder mission.

In summary these new technologies are:

- Local and large baseline interferometric tracking of free-falling bodies
- Drag-free navigation with inertial sensors and micro-thrusters
- Spacecraft mechanical stability and the control of self-gravity to very high accuracy

The first of these has no space-heritage at all. However in the development phase of LISA Pathfinder – to test the local part of the test-mass to test-mass tracking –monolithic optical assemblies tracking test-mass motion at 10 pm/$\sqrt{\text{Hz}}$ in the mHz range have been demonstrated.

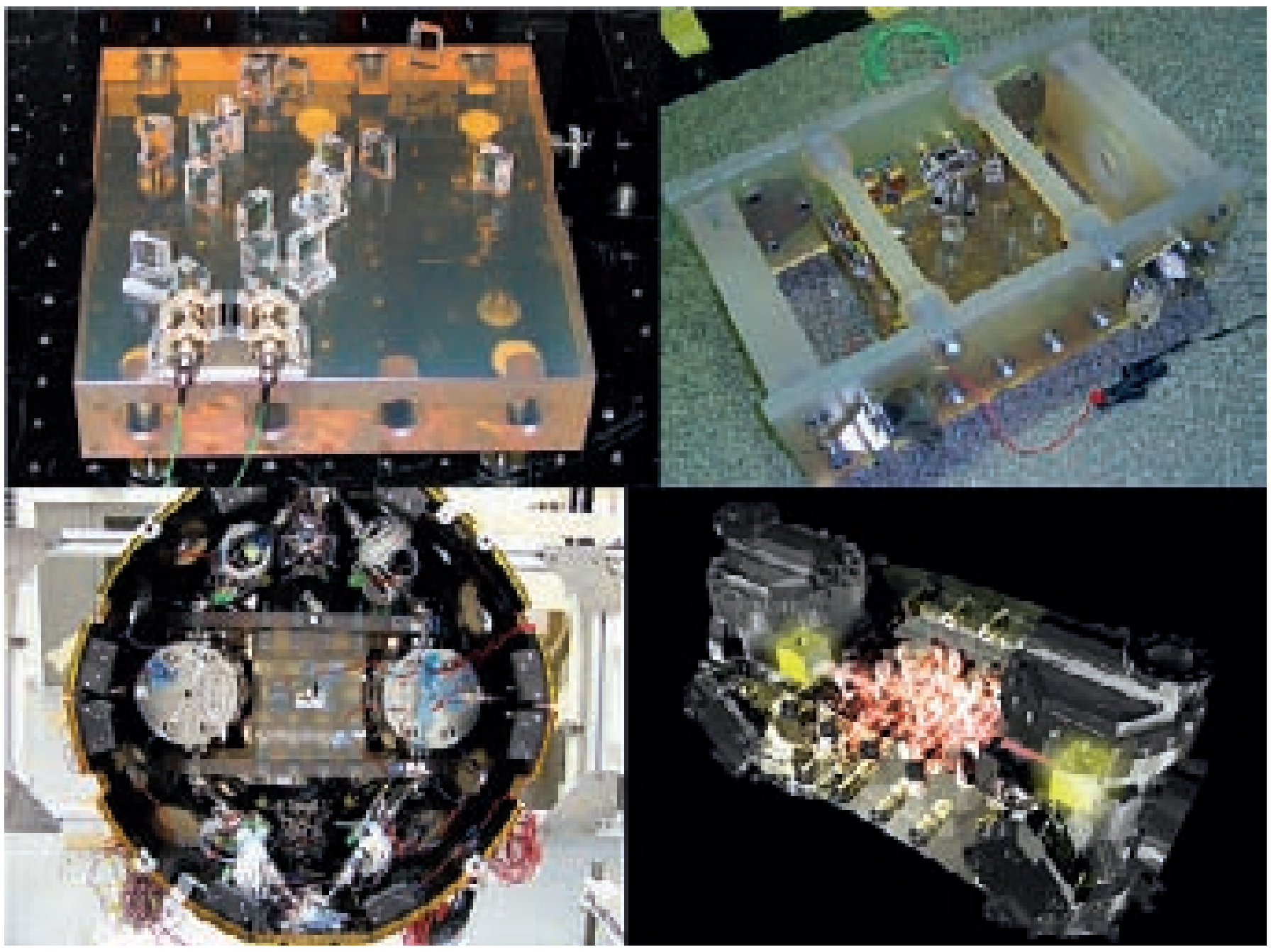

Fig. 7.4 – LISA Pathfinder and the high-accuracy interferometer tracking of free-falling bodies. From top left clockwise: the monolithic optical bench obtained by hydroxy-catalysis bonding. The same assembled in the Zerodur supporting structure. 3D view of the LISA Technology Package showing the two free-floating test-masses and the interferometer displacement readout. A structural model of the LTP assembled into the spacecraft supporting structure for qualification tests

The long baseline tracking has seen a wealth of new concepts and techniques to cope with limited laser stability and limited baseline accuracy. Synthetic interferometers with time delay interferometry, and the use of a large two-spacecraft arm as an optical reference are prime examples.

One obvious prerequisite of the above is that test-masses are in free-fall. This is achieved by a non-contacting spacecraft that uses micro-thrusters to follow the test-masses, a scheme known as drag-free navigation. First LISA Pathfinder, and later LISA, is expected to push the purity of free-fall of small (kg size) test-masses by two and then three orders of magnitude relative to other missions expected to fly in the same time frame, the improvement relative to already flown missions being much larger. All this has required the development of new inertial sensors and micro-Newton thrusters.

Achieving the performance in drag-free means controlling spacecraft features never considered before. Time resolved thermo-mechanical analyses were developed to keep the local gravitational noise under control. Large local gravity is incompatible with the required performance so that gravitational balance has been developed with tens of pico-g accuracy.

All the above technologies have uses in other fields. As previously mentioned high-power, long-distance laser links have an obvious application to laser telecommunication, indeed LISA will likely implement the first interplanetary spacecraft-to-spacecraft telecommunication system. The LISA laser system is being developed with other industrial applications in mind.

Large size optics with pico-meter mechanical stability are shared with other space missions (eg Gaia) and cross-fertilization is expected in that area as well.

The Micro-Newton thruster is a generic technology for high precision formation flying and drag-free

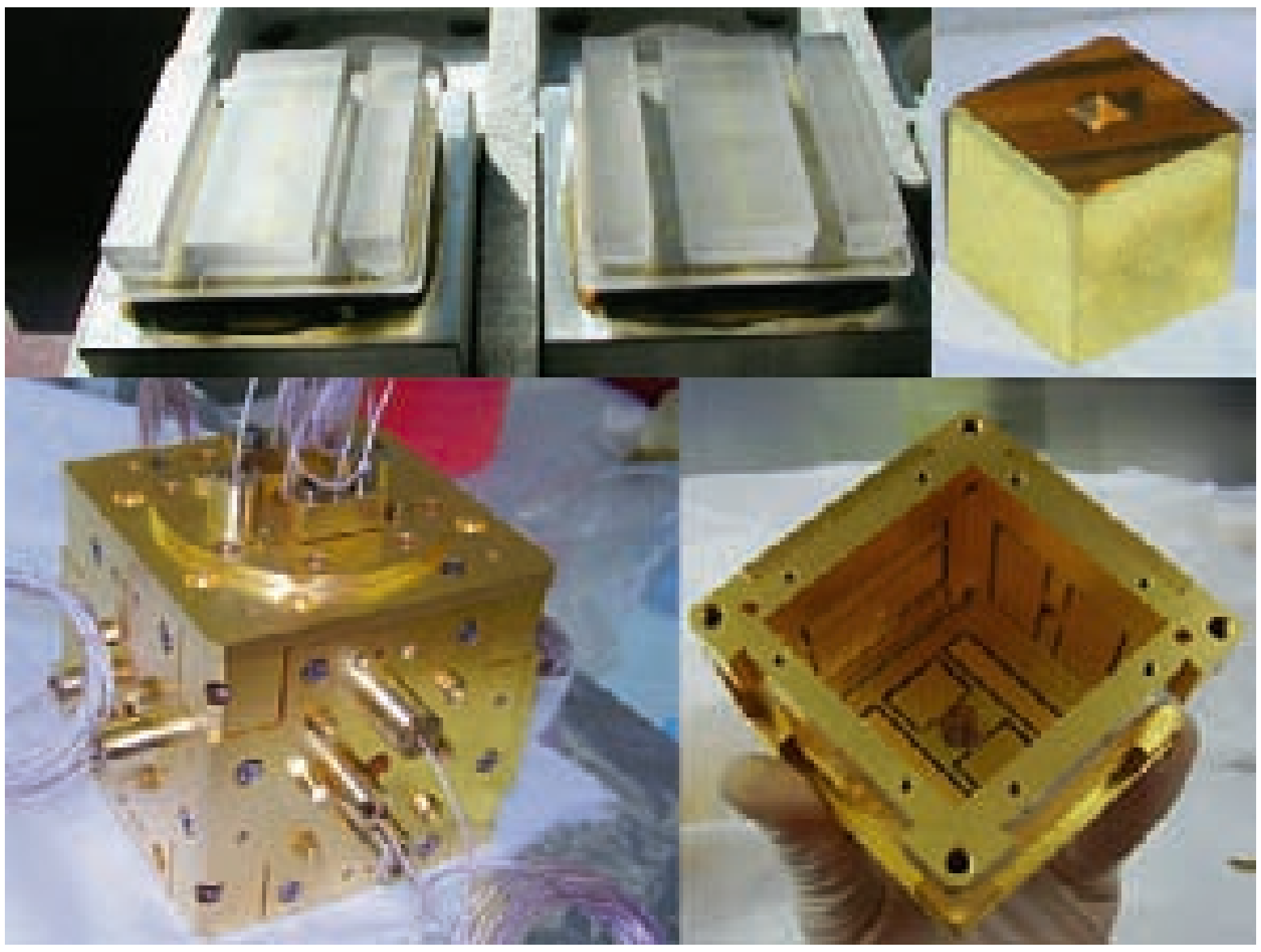

Fig. 7.5 – LISA Pathfinder Inertial sensor. From top left clockwise: Sapphire electrodes for capacitive sensing; the gold-platinum test-mass; the partially assembled electrode-housing; an integrated prototype

operation, and a large utilization base for it is foreseeable including missions already in implementation such as Microscope.

The LISA Pathfinder two-test-mass payload is the most sensitive local gravity gradiometer concept ever planned. A two spacecraft test-mass to test-mass laser tracking system, like a small size LISA single arm, is another high sensitivity gradiometer. The obvious application for these is to Space Geodesy, in particular to the subfield of measurement of changes in the Earth's gravity field due to changes in water storage on land, deep ocean currents, etc. As is widely known, the GRACE mission has made major advances in this field, and a drag-free version of GRACE which also has laser interferometric measurements between the two spacecraft is a possibility for a future follow-on to GRACE, if the extra cost can be kept low. The LISA Pathfinder Mission is likely to demonstrate drag-free satellites that more than meet the geodesy requirements, and reduce the cost considerably.

The entire field of so called "Fundamental Physics in Space" is waiting for LISA Pathfinder results as a significant proof of concept groundbreaking event. For instance tests of relativity, and specifically much improved measurement of the gravitational time delay for electromagnetic waves passing close to the Sun, may be performed by making use of further improved versions of the Gravitational Reference Sensors being developed for LISA and tested on LISA Pathfinder. The improvements needed involve extension of the operational frequency range down to roughly 0.4 microhertz. Improving these technologies beyond the needs of LISA, for missions like BBO, will certainly require major advances, including two-frequency laser links, much more powerful lasers and much better drag-free systems.

## 7.5 Geodesy and geophysics

This is a neighboring field dealing with gravity, space measurements and mass distribution that has much in common with the gravitational wave developments. It is not surprising then that many technologies can be, and have been, shared.

We have already mentioned space-borne gravitational geodesy, but other gravitational wave techniques have also found significant application in the development of geophysical exploration instruments. Several gravity gradiometer projects around the world are using many aspects of gravitational wave technology, particularly in the area of cryogenic motion sensing. Several non-gravitational airborne instruments have benefitted from gravitational wave vibration isolation technology.

Development of underground experimental sites, in high stability hard rock environments, presents a good opportunity for geophysical measurements. The temperature stability and low seismic noise are commonly beneficial to both geophysics and gravitational wave detection. Currently there is no strong motivation to construct km-scale tunnels for the geophysics community, and their present 100 m baseline may be sufficient for their needs. However once km-scale tunnels are available, geophysicists are likely to request space inside for their interferometers also. Their interferometers are simple and easily maintained, and will not interfere with the gravitational wave experiments.

## 7.6 Material science and technology

More and more gravitational wave development hinges on the development of special materials and material technology.

Hydroxy-catalysis bonding, already mentioned, though originally developed in Stanford for the Gravity Probe B experiment, was further developed by the Glasgow/Stanford collaboration for utilization in GEO and will be implemented in Advanced LIGO and possibly in Advanced Virgo. Its application to fabrication of high-power, single-mode, fiber-amplifier systems has already been mentioned. The technology has also been transferred to CSIRO (Australia) under a consulting arrangement, for use in the construction of precision optical systems. Further, the bonding technology is being applied to the development of glass integral field units (IFU) for astronomical spectroscopy and for use in prototype mirror actuators under development for the European Extremely Large Telescope (ELT).

Extensions of the bonding technique to joint silicon carbide have resulted in a patent application, with expressions of interest from Astrium in Germany to adapt this technology for application in space qualified stable optical systems.  Also, collaboration was initiated with TNO-TPD in the Netherlands for use of the technology for other space missions.

On the coatings side, we already mentioned the transfer of knowledge to the field of frequency standards. Given the critical role that coatings play within the field it is highly likely that further transfer of knowledge to other optical applications will occur.

## 7.7 Cryogenics and cryogenic electronics

Cryogenics has been at the basis of operation of resonant detectors since the 1980s. Detectors of increasing sophistication have been built with the last generation (AURIGA and NAUTILUS) operating with dilution refrigerators below 1 K. The Nautilus bar, having a mass of 2.3 tons, reached a temperature as low as 95 mK, cooled by special soft copper link. The novelty for these cryogenic applications has been the need for vibration isolation within a cryogenic environment including low mechanical coupling to the potentially noisy refrigerator. This work is of direct relevance in the field of optical frequency standards where lasers are stabilized to cryogenic cavities which have to be undisturbed by mechanical noise.

Indeed a particularly challenging area is the necessity for quiet refrigeration with flexible heat links. Several patents have been submitted in this area and quiet refrigerators are in manufacture by companies for general research in material science. It is interesting to note that one of these companies has joint personnel with a cryogenic gravitational wave detector group.

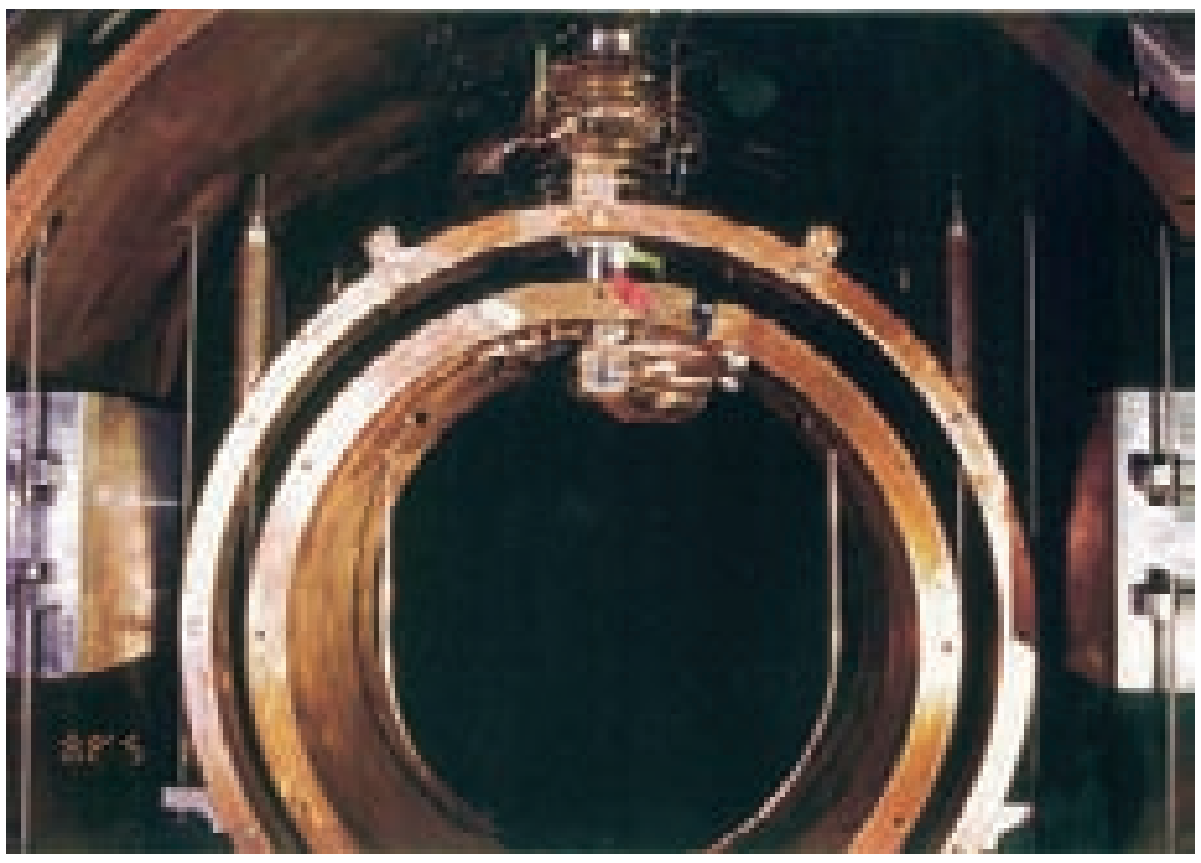

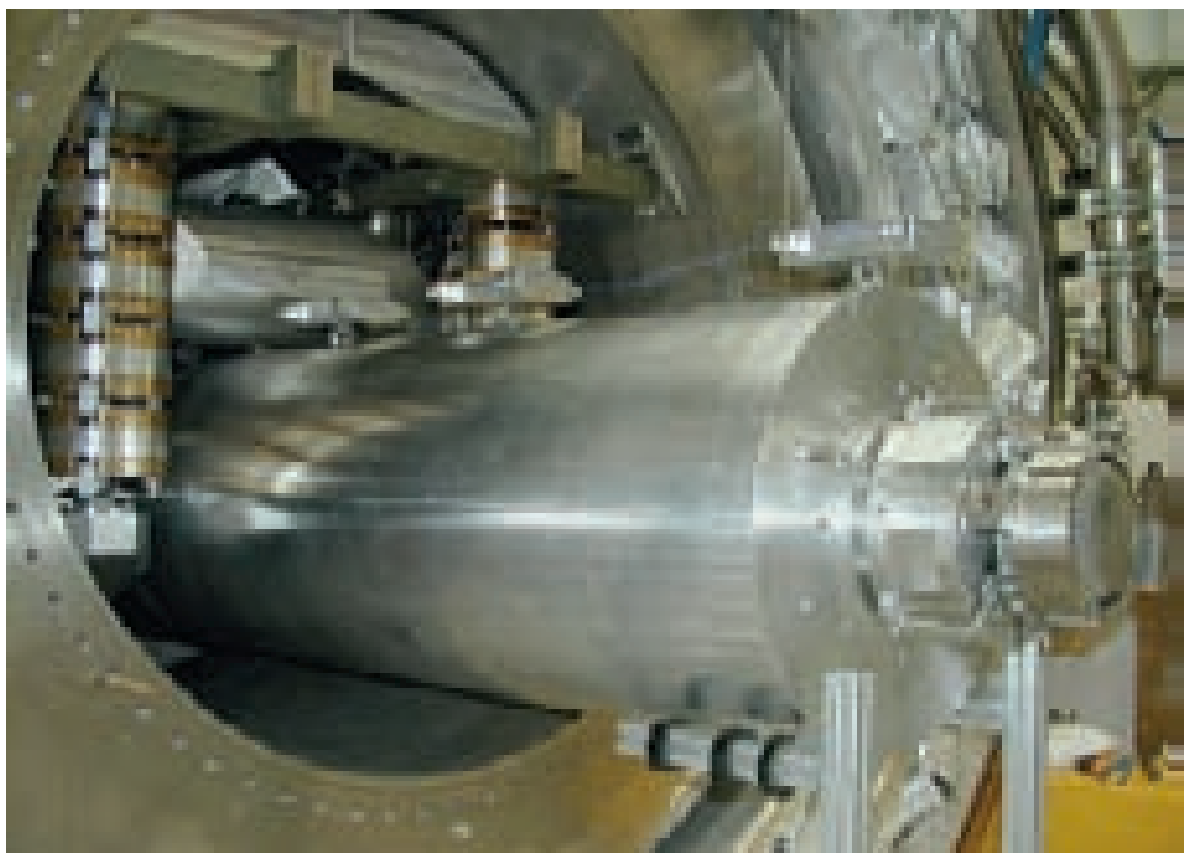

Fig. 7.6 – NAUTILUS present, showing the thermal links between the dilution refrigerator and the suspension system (left); AURIGA present, showing the transducer assembly at one end of the Al rod (right).

The activity on resonant detectors has also produced an entire series of low noise electromechanical transducer readout systems based on specially developed SQUID devices achieving, as already mentioned, the lowest energy resolution to date.

The two-stage SQUID, at 16 $h$ true "noise" energy, developed for the AURIGA detector is still the leading device in SQUID electronics. Such electronics have a variety of applications including NMR spectrometry.

## 7.8 Computing

Because gravitational wave research requires computer-intensive data analysis it has driven grid computing research in a number of directions such as workflow planning, workflow management and execution, volunteer computing, data replication and data placement for computation.

More specifically gravitational wave research has driven new functionality and development with the following specific grid computing tools:

**BOINC**
*(http://boinc.berkeley.edu/)* The Einstein@Home effort, aimed at analyzing gravitational wave interferometer data to detect signals from continuous wave sources such as pulsars, builds on BOINC and made significant contributions to it.

**Condor and Condor DAGman**
*(http://www.cs.wisc.edu/condor/)* The LIGO Data Grid relies heavily on Condor for managing complex workflows. This has driven significant new functionality now available to other communities in official releases of Condor.

**Globus GridFTP and Replication Location Service**
*(http://www.globus.org/)* The LIGO Data Replicator (LDR) is used to replicate in bulk, interferometer data to analysis sites. The newer Globus Data Replication Service is based on LDR and is available to other communities.

**Pegasus**
(http://pegasus.isi.edu/) The Compact Binary Coalescence (CBC) search group, of LIGO and Virgo, is using Pegasus for large, complex, and data intensive workflow planning. This is driving new functionality, scalability, and robustness in the Pegasus suite of tools.

**TclGlobus**
*(http://tclglobus.ligo.caltech.edu/)* The TclGlobus project developed functionality to allow the Tcl/Tk scripting language to use the Globus Toolkit, and enabled grid interfaces in Tcl software simply by including the TclGlobus package. TclGlobus was developed entirely by the LIGO Scientific Collaboration but is available to other communities.

As gravitational wave data analysis matures, scientists are analyzing more data, terabytes and soon petabytes, using more complex and sophisticated workflows. Their demands will continue to drive developments of grid and distributed computing in the future.

Thus computer science colleagues involved with the projects described above eagerly await the deployment and commissioning of advanced terrestrial detectors.

Within both the U.S. and in Europe a number of funded research projects have clearly demonstrated the benefits of teaming computer scientists working on issues of grid and distributed computing with domain specific scientists and in particular gravitational wave researchers. Projects within the U.S. include GriPhyN *(http://www.griphyn.org/)*, iVDGL *(http://www.ivdgl.org/)* and Open Science Grid (http://www.opensciencegrid.org/). In Europe, EGEE – Enable Grid for E-Science – is an EU funded project to develop a large scale eScience infrastructure for Europe. One of the project's goals is to develop a generic middleware stack, known as gLite, to provide a high level of infrastructure for Virtual Organizations using the grid to tackle their problems.

## 7.9 Methods in theoretical physics

The perspective of inaugurating the gravitational sector in the observation of the Universe has triggered a large quantity of theoretical studies that have an impact on physics at large. A few examples follow.

The main impact so far has been the development of highly accurate post-Newtonian solutions of Einstein's equations for binary inspirals (accompanied by a deeper understanding of the theory itself) motivated by the requirement for accurate templates. Another impact has been the development of theoretical techniques (still incomplete) to calculate the self-force on small objects orbiting massive black holes (EMRIs).

A further example is in stellar dynamics in the presence of black holes, especially central black holes; new analysis methods are being developed to understand the behavior of stellar clusters in this situation.

Collective effects such as phase transitions in fundamental theory are difficult to study directly at the Terascale and beyond, even with colliders such as the Large Hadron Collider, because only a small number of particles interact at a time. Gravitational wave backgrounds from early Universe phase transitions are an active area of study in many theories including signatures of symmetry breaking, extra dimensions and some forms of inflationary reheating. Like superstrings, detecting such effects would have a profound influence on our understanding of physics beyond the Standard Model and its unification with gravity.

For the last few years there has been nuclear and materials science research spurred by links to possible neutron star observations. Also over the last decade there has been work on qcd color superconductivity relating to r-modes of neutron stars.

Finally it has been noted recently that interferometer technology is capable of measuring positions with such precision that they reach halfway, geometrically, to the Planck scale: the ratio of size to displacement is larger than the ratio of displacement to Planck length. Passing this threshold means that certain classes of holographic theories of quantum gravity can in principle be studied by direct experiment. Development of this technology may open a field of direct quantum gravity experiments in addition to the study of classical gravitational waves.

## 7.10 Conclusions and impact on the roadmap

Gravitational wave research has always been an "instrument technology intensive" field, requiring the frontiers to be pushed forward simultaneously in many fields. This strong technological drive has certainly helped to consolidate the case for the field over the years of low first-detection probability.

Pushing instrumentation limits will remain a necessity for ground-based observatories heading towards the third generation era, and for LISA and its successors. It seems certain that this development will continue to cross-fertilize nearby fields at a good rate, mostly in the field of optical devices at large, of sophisticated optical measurement schemes, and of laser sources and material investigations. At the same time, as has always been the case, the field will be alert to technology developments in other fields that could have a productive impact on instrumentation and other tools needed to maintain the scientific progress of the gravitational wave field.

# 8. Recommendations to GWIC to guide the development of the field

## 8.1 Introduction

The GWIC Roadmap Committee has identified a set of priorities and recommendations that address specific short-term activities that GWIC and the gravitational wave community should undertake to enhance progress towards the most important goals of this roadmap or to improve the focus of the field on key scientific and technical issues. This chapter provides the context or background for each of these priorities and recommendations and lists the priorities and recommendations themselves.

## 8.2 Completion of the second-generation global network

**Background**—robust ground-based gravitational wave astronomy based on the projected capabilities of second generation interferometers (e.g. Advanced LIGO and Advanced Virgo) requires a fully global array of instruments spaced at continental distances to provide good pointing accuracy over the whole sky. Therefore, instrumentation of comparable sensitivity to Advanced LIGO and Advanced Virgo is highly desirable in the Southern Hemisphere and in Asia.

We therefore emphasize the importance of implementing interferometers with sensitivity comparable to Advanced LIGO and Advanced Virgo, in both Asia and in the Southern Hemisphere. It is essential for the field that these instruments become operational relatively early in the observational lifetime of Advanced LIGO and Advanced Virgo.

**Priority — The construction, commissioning and operation of the second generation global ground-based network comprised of instruments under construction or planned in the US, Europe, Japan and Australia.**

**Recommendation** — We recommend that GWIC provides the forum where international support for efforts to bring about such instruments in Japan, Australia and possibly India, can be coordinated and where the community can work together with the proponents to ensure that the siting, design, orientation, etc. of such instruments is carried out to optimize the scientific capabilities of the global network.

**Recommendation** — We recommend that GWIC organizes a workshop to emphasize the scientific benefits of interferometers in Japan and Australia as a way to encourage further international support and recognition of the potential scientific contributions of these facilities as part of the global network of ground-based gravitational wave detectors.

## 8.3 Gravitational wave detectors in space

**Background** — The range of frequencies from 0.1 Hz down to 1 / (age of the Universe) offers some of the most spectacular gravitational wave science. The scope of gravitational signals comprises mergers of supermassive black hole binaries, extreme mass ratio inspirals into massive black holes, galactic binaries, stochastic backgrounds from the early Universe and quantum fluctuations after the big bang. Ground-based detectors will never be sensitive to gravitational waves with frequencies below about ~ 0.1 Hz because of the unshieldable background of terrestrial gravitational noise.

Detection technologies at these frequencies are diverse and range from pulsar timing and large baseline laser interferometry through spacecraft tracking to polarization measurements of the cosmic microwave background. All of these technologies will eventually be used to observe the complete gravitational wave spectrum covering more than 20 orders of magnitude in frequency.

LISA, a space-based interferometer funded by NASA and ESA will open the low-frequency gravitational wave window from 0.1 mHz to 0.1 Hz. The goal of a launch of LISA in 2020 is technologically feasible and entirely timely, considering that the technology precursor mission LISA Pathfinder will launch in 2012.

DECIGO is a space-based mission under consideration in Japan to explore the gravitational wave window from 0.1 to 10 Hz. The DECIGO Pathfinder mission is expected to be launched in the middle part of the next decade. The DECIGO Pathfinder was selected as one of the five important mission candidates for the small-science satellite series run by JAXA/ISAS and is a potential candidate for the third launch shot.

Technology development for post-LISA missions should be planned to make a smooth transition into the routine multi-wavelength gravitational wave astronomy that is certain to develop after the first detections have been made by LISA and advanced ground-based detectors.

Technology development for such missions should be pursued in a timely manner. Any undue delays would push back the time-scale for post-LISA missions.

Pulsar Timing is making major progress with the formation of the International Pulsar Timing Array collaboration.

**Priority — The completion of LISA Pathfinder and a timely selection of LISA in 2012/2013 for a launch in 2020 to open the low frequency gravitational wave window from 0.1 mHz to 0.1 Hz.**

**Priority — The timely start for technology development for LISA follow-on missions.**

**Priority — The continued development of an international pulsar timing array for the study of gravitational waves in the nano-Hertz band. This effort requires continued development of algorithms and data acquisition systems, and access to substantial amounts of time on the world's largest radio-telescopes.**

## 8.4 Theory, data analysis and astrophysical model building

**Background** – To fully exploit the scientific potential of gravitational-wave observatories will require the ability to analyze and interpret the data obtained. To accomplish this, a continuing collaboration among data analysts, astrophysical theorists, and analytical and numerical general relativists will be essential. Continuing interactions between these theoreticians and the instrument builders will also be important. Because many of the most important sources of gravitational radiation involve strong-field dynamical general relativity, large scale, high-performance numerical computations will continue to play a critical role.

**Priority — A strong and ongoing international program of theoretical research in General Relativity and astrophysics directed towards deepening our understanding of gravitational wave observations.**

## 8.5 Toward a third-generation global network

**Background**— The scientific focus of a third-generation global network will be gravitational wave astronomy and astrophysics as well as cutting edge aspects of basic physics. Third-generation underground facilities are aimed at having excellent sensitivity from ~1 Hz to ~$10^4$ Hz. As such, they will greatly expand the new frontier of gravitational wave astronomy and astrophysics. In Europe, a three year-long design study for a third-generation gravitational wave facility, the Einstein Telescope (ET), has recently begun with funding from the European Union. In order to optimize the scientific contributions of the third-generation network it will be important that the network as a whole (including the second-generation instruments) constitute a well-conceived global array that is optimized to provide the best science. This will require coordination and collaboration by the designers of third-generation facilities across the globe taking into account the capabilities of the second-generation facilities.

**Priority — Continued R & D efforts in collaboration with existing design study teams to support the construction, beginning in the post 2018 time frame, of the Einstein Telescope, soon after the expected first gravitational-wave discoveries have been made.**

**Recommendation** — We recommend that GWIC works with the international ground-based gravitational wave community to plan how to optimize the scientific capabilities of a future third-generation network. Specifically, GWIC, in collaboration with any design study groups in the various regions and countries, should organize meetings to assist the community to understand and establish science-driven requirements (e.g. frequency range, sensitivity), possible interferometer designs and configurations, technologies, optical layouts, site configurations and orientations, etc. that would optimize the scientific potential of the network.

**Recommendation** — We recommend that GWIC establishes an international steering body to organize workshops, and promote coordinated R&D efforts in collaboration with existing design study teams to help achieve the goal of an optimized third-generation network.

## 8.6 Development of key technologies for third generation ground-based instruments

**Background** — Successful deployment of third generation, underground gravitational wave instruments will require development of a number of new technologies by the gravitational wave community. The use of cold test masses and associated cryogenic technology with its pumps, moving cryo-fluids and other possible sources of mechanical noise is extremely challenging. Development of low loss coatings will require a major systematic program of sensitive measurements. "Squeezing" and other techniques for manipulating the quantum properties of light are still primarily pursued in small scale laboratory environments. Many of these (and other) development programs take place in a small number of places and with limited coordination and communication. It is important that these developments are well understood by the rest of the community and that additional efforts take place in other regions of the world so that robust technologies are ready when required for the third-generation facilities.

**Recommendation** — We recommend that GWIC sponsors a series of workshops, each focused on the status and development of a particular critical technology for gravitational wave instruments. Topics in such a series could include cryogenic techniques, coating development for reduced thermal noise, "Newtonian noise," techniques for quantum noise reduction, and overall network configuration. These workshops will help promote exchange of ideas, provide visibility and encouragement to new efforts in critical areas of technology development, and help bring to bear the combined resources of the community on these problems.

## 8.7 Outreach and interaction with other fields

**Background**— A critical goal for the field in the coming decade is to increase the interactions with other fields of astronomy so that the scientific use of gravitational wave signals, both by themselves and in concert with other signals (electromagnetic, neutrino) yields the highest scientific return. Given the timescale in which gravitational wave astronomy will produce results, we now have an excellent opportunity for our field to foster the interest of the current and coming generations of astronomers and astrophysicists in the scientific potential of our field, as well as communicating the excitement of gravitational wave astronomy to the public.

**Recommendation** — We recommend that GWIC plans an outreach campaign focused on engaging public, school and political audiences with the excitement, promise and gains to society of the science and technology of gravitational wave astronomy.

**Recommendation** — We recommend that GWIC plans increasing interactions with other areas of astronomy and, astrophysics, as well as other relevant scientific communities. The primary goal of this campaign would be to engage the interest of such scientists in the work underway in gravitational wave astronomy and astrophysics, and to foster collaboration and scientific exchanges. We recommend, for example, that GWIC encourages and facilitates multi-messenger collaborations between the gravitational wave, electromagnetic and neutrino astronomy communities. GWIC sponsored workshops with attendees from these communities would be an effective step in this direction. The outreach campaign should be developed and organized by a standing GWIC subcommittee.

## 8.8 Public release of data

**Background** — In the next decade the field of gravitational wave astronomy will become an important component of astronomy worldwide. As such, it will be important for data from gravitational wave instruments to be made available to the scientific community and the public as is now the case for publicly funded astronomical observatories in many countries. This will require a carefully implemented study on the part of the international gravitational wave community to determine the steps required to meet this goal, the appropriate deliverables, and the effort and cost involved, taking into account the requirements of funding agencies in different countries.

**Priority — The development of procedures that, beginning in the era of frequent ground-based detection of gravitational waves, will allow the broader scientific community to fully utilize information about detected gravitational waves.**

## 8.9 Continued involvement of the bar detector community in the field

**Background** — Recent years have seen a shift in the leadership in gravitational wave observations, as the first generation of large gravitational wave interferometers has begun operation near their design sensitivities, taking up the baton from the bar detectors that pioneered the search for the first direct detection of gravitational waves. The current generation of these bar detectors, until relatively recently the main observational instruments, has been surpassed in sensitivity by laser interferometer based detectors capable of extended observations. Research by the bar detector groups has not yet identified a clear path toward future sensitivities which can compete with the interferometric detectors. As a result, the supporting agencies have begun to limit or reduce the funding for bar detectors. However the operation of Nautilus and Auriga will continue until the advanced interferometers are observing, with the operation of Explorer, in CERN, being phased out in the near future.

**Recommendation** — We recommend that, when the time is appropriate, other projects in the gravitational wave community encourage scientists from the bar detector community to join the ongoing detector projects. The expertise of the bar detector scientists in ultra-sensitive measurement technologies and data analysis is a valuable resource and should be retained within the gravitational wave community.

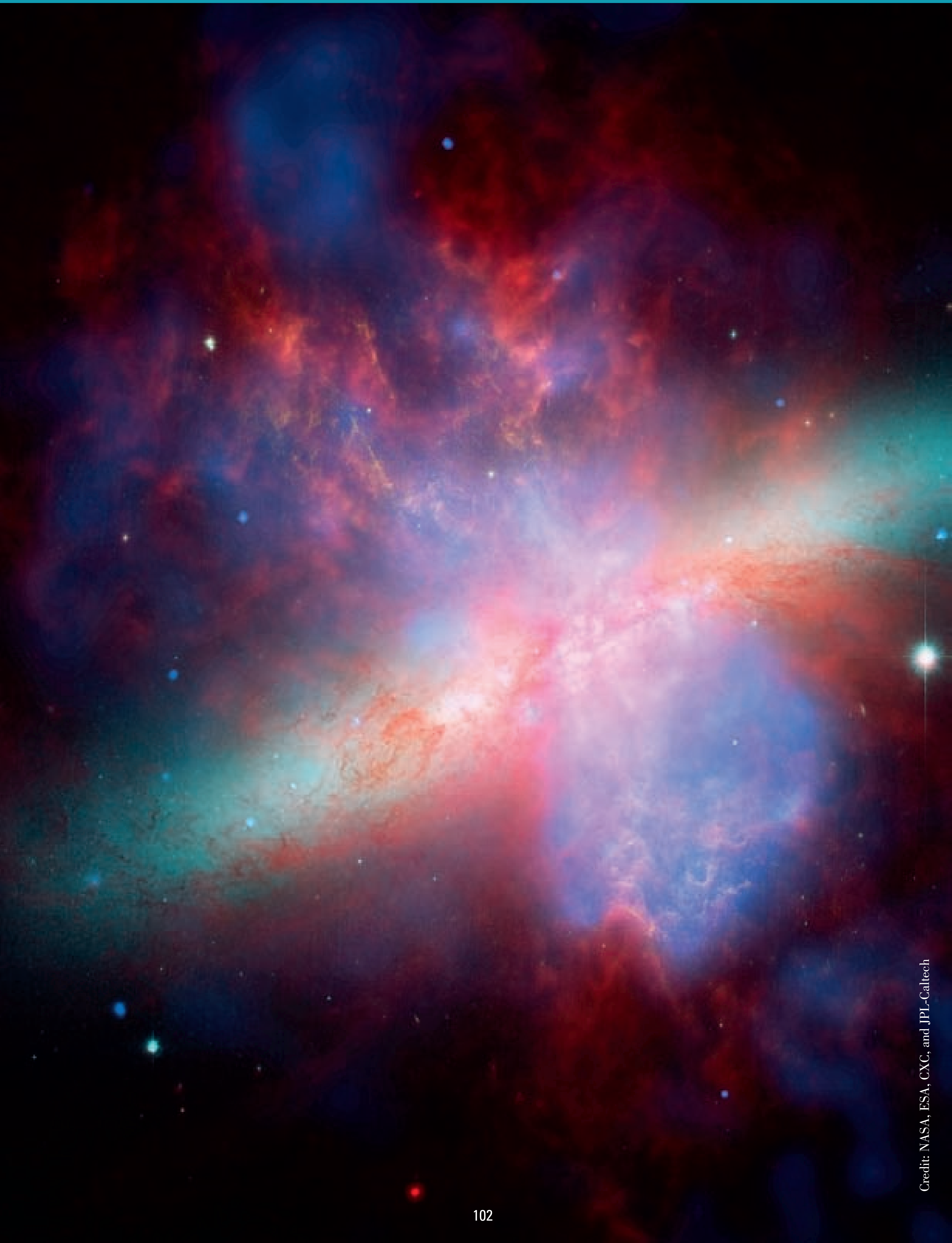

Credit: NASA, ESA, CXC, and JPL-Caltech

# 9. Conclusions

The field of gravitational wave science is on the verge of direct observation of the waves predicted by Einstein's General Theory of Relativity and opening the exciting new field of gravitational wave astronomy. In the coming decades, ultra-sensitive arrays of ground-based instruments and complementary spaced-based instruments observing the gravitational wave sky will discover entirely unexpected phenomena while providing new insight into many of the most profound astrophysical phenomena known. This new window into the cosmos could radically change humanity's understanding of the Universe.

Recognizing that the field has reached this historic brink, the Gravitational Wave International Committee initiated the development of a strategic roadmap for the field with a 30-year horizon. The goal of this roadmap is to serve the international gravitational wave community and its stakeholders as a strategic tool for the development of capabilities and facilities needed to address the exciting scientific opportunities on the intermediate and long-term horizons. It represents the synthesis of best judgment of many of the leaders in the field and has the support of the gravitational wave community.

The ground-based gravitational wave detectors currently taking data and the advanced detectors that will be brought on-line over the next five to ten years will be critical in establishing the field of gravitational astronomy through the detection of high luminosity gravitational wave sources such as the merger of binary neutron stars and black holes. These detectors are akin to early telescopes, and future, more sensitive detectors will follow to make it possible to observe a greater variety of phenomena, and expand our knowledge of fundamental physics, cosmology and relativistic astrophysics. Gravitational wave observations will afford us a unique tool to gaze into the heart of some of the most violent events in the Universe and will help us unravel current mysteries, particularly when combined with other astronomical observations.

To fully capitalize on this important scientific opportunity, a true global array of gravitational wave antennae separated by continental distances is needed to pinpoint the sources on the sky, and to extract all the information about the source's behavior that is encoded in the gravitational wave signal. In the medium term, this means the first priority for ground-based gravitational wave detector development is expansion of the network begun by LIGO, Virgo and GEO and their planned upgrades. Additional detectors with appropriately chosen intercontinental baselines and orientations are needed to produce a high sensitivity, all-sky network that will maximize the ability to extract source information from observed gravitational waves. In the longer term, there is an important opportunity to further expand the scientific reach of this ground-based network by developing underground interferometers with even better sensitivity and frequency coverage.

Many gravitational wave sources only radiate at much lower frequencies. The most promising route to access these is to fly a laser interferometer in space. The scientific reach of such a space-based instrument would be complementary to the ground-based array in the same way that optical and x-ray astronomy are complementary and have provided information about different types of astrophysical objects and phenomena. In the intermediate term, the highest priority space gravitational wave project is LISA. This joint NASA/ESA mission promises to detect sources to the edge of the Universe and to shed new light on the evolution of galaxies. Mission design and technology development for LISA are proceeding well.

A critical goal for the field in the coming decade is to increase the interaction with other fields of astronomy so that the scientific use of gravitational wave signals, both by themselves and in concert with other signals (electromagnetic, neutrino) realizes its full potential. Given the timescale in which gravitational wave astronomy will produce results, we now have an excellent opportunity for our field to foster the interest of the current and coming generations of astronomers and astrophysicists in the scientific potential of our field, as well as communicating the excitement of gravitational wave astronomy to the public.

Finally, we note that gravitational wave research has always been a technology intensive field, driving technology in some areas while benefiting from developments in others. The need to stretch the limits of sensitivity has resulted in developments that have profited other fields of science and technology, notably precision measurement, metrology, optics, lasers, stable optical fabrication techniques, space technology, and computational techniques. At the same time, the field has been alert to technology developments in other realms that could enable new scientific directions in gravitational wave science. There is no doubt that this highly productive interchange between gravitational wave science and other spheres of science and technology will continue, providing tangible benefits to society, in addition to the intellectual satisfaction of an improved knowledge of our Universe.

# Appendices

## A. 1 Matrix of scientific opportunities in gravitational wave science

| **Time Frame** | **Decade 2010 - 2020** | | **Decade 2020 - 2030** | | | **Decade 2030 - 2040** | |
|---|---|---|---|---|---|---|---|
| **Frequency Band** | 10 Hz - 1 kHz | 3 - 30 nHz | 1 Hz - 1kHz | 0.1 - 100 mHz | 1 - 30 nHz | 1 Hz-1kHz | 0.1-10Hz |
| **Instruments Techniques** | Enhanced and Adv LIGO-VIRGO GEO-HF | Pulsar timing (PPTA, EPTA, NanoGrav) | ET and other third generation | LISA | Pulsar timing (IPTA, SKA) | ET & other third generation | BBO |
| **Fundamental Physics, General Relativity** | test speed of GW relative to light, test polarizations of GW | | test NS equation of state with pulsar GW | test polarizations of GW | test polarizations of GW | constrain the mass of the graviton from GW phasing of inspirals | cosmic backgrounds – tests of quantum era in universe |
| | effects of frame dragging, precessing BH mergers, black hole spectroscopy | | Stochastic GW back-ground at $\Omega_{GW}$ $10^{-12}$ | bound graviton mass | bound graviton mass | high precision tests of GR through waveform phasing | high precision tests of GR through waveform phasing |
| | test merger waveforms-strong field GR | | strong field tests of gravity; models of black hole merger | black hole spectroscopy – no- hair tests | properties of cosmic strings | Stochastic GW background at $\Omega_{GW}$ of $10^{-13}$ | |
| | test NS equation of state with NS-BH inspirals | | test NS equation of state with NS-BH inspirals | EMRIs – trace BH spacetime | | | |
| | complete dynamics of BH mergers | | black hole spectros-copy, no-hair theorem, frame dragging | bound post inflation phase transitions | | | self-consistency test of slow-roll inflation |
| **Cosmology** | precision Hubble Const (NS-NS or NS-BH mergers with EM counterparts) | | high precision cosmology (Hubble constant, dark matter, etc.) | precision Hubble Const (NS-NS or NS-BH mergers with EM counterparts) | Limit/detect GW from inflation era | precision Hubble Const (NS-NS or NS-BH mergers with EM counterparts) | precision Hubble Const (NS-NS or NS-BH mergers with EM counterparts) |
| | | | dark energy param. to 5-10% from 1000's of γ-bursts | dark energy parameters to 1% from 2-10 sources | | Intermediate black holes and if they were galactic seeds | dark energy parameters to <1% from $10^5$ sources |
| **Astronomy** | z=0 Merger rate of NS-NS systems (not found via EM) | merger rate of $10^8$-$10^9$ $M_S$ BH at z<2 | Merger rate of BNS, BBH and BH-NS systems and test of population synthesis | merger rate of $3.10^4$-$10^7$ $M_S$ BH up to z=25 | stochastic GW background, galactic merger rates, BH mass functions | merger rate of NS-NS, NS-BH and BH-BH systems at very high redshifts | z=0-5 evolution of merger rates of NS-NS and NS-BH |
| | merger rate of NS-BH systems (not yet discovered via EM); test binary stellar evolution theory (BSE) | | correlations of BNS and NS-BH mergers with GRBs; GRB progenitors understanding short and long bursts | Spins and masses of SMBH in galactic nuclei growth mechanisms | GW from individual binary SMBH | GW from supernovae; coincident observation with neutrinos and cosmic rays | z=0-10 evolution of merger rates of BH-BH |
| | merger rate of BH-BH systems; test BSE | | possible detection of unexpected sources; a new window on the Universe | EMRI rate for $3.10^4$-$10^7$ $M_S$ BH at z~0 – stellar remnants in galactic nuclei | | Systematic study of glitching pulsars, flaring magnetars & em other transients | z=0-30 merger rate history of IMBH |
| | bound non-axisymmetries in msec pulsars | | detect non-axisymmetries in msec pulsars | EMRI rate for $3.10^5$-$10^7$ $M_S$ BH at z<2 – Population III stellar remnants? | | | detect non-axisymmetries in 1-10s magnetars |
| | do IMBH exist? | | do IMBH exist? Could they be seeds of galaxy formation | Masses, positions of all short-period accreting and non-accreting white dwarfs in Milky Way; test BSE | | Star formation history as a function of redshift, stellar mass function | Direct detection of expanding quadrupole in SNIa. |
| | detect non-axisymmetries in msec pulsars | | do bars form in core collapse supernova events? | Effects of tides in white dwarfs | | | Non-axisymmetries in accreting WDs |
| | correlations of NS-NS mergers with GRBs | | glitching pulsars, flaring magnetars | | | | prediction of GRBs with 1yr notice. |

## A.2 Membership of the GWIC Roadmap Subcommittee

**Chair:** Jay Marx, Caltech

Karsten Danzmann, AEI Hannover
James Hough, University of Glasgow (ex-officio)
Kazuaki Kuroda, Institute for Cosmic Ray Research, Tokyo
David McClelland, Australian National University
Benoit Mours, LAPP Annecy
Sterl Phinney, Caltech
Sheila Rowan, University of Glasgow
Flavio Vetrano, University of Urbino
Stefano Vitale, University of Trento
Stan Whitcomb, Caltech
Clifford Will, Washington University, St Louis

**Advisory member:**
B. Sathyaprakash, Cardiff University

## A.3 List of scientists and funding agency representatives consulted in developing the roadmap

**Barry C. Barish**
Linde Professor of Physics, Emeritus at the California Institute of Technology

**Eugenio Coccia**
Director of the INFN Gran Sasso Laboratory and Professor of Astronomy & Astrophysics at the University of Rome "Tor Vergata"

**Adalberto Giazotto**
INFN Director of Research, INFN Pisa

**Craig Hogan**
Director of the Center for Particle Astrophysics at the Department of Energy's Fermi National Accelerator Laboratory and Professor of Astronomy & Astrophysics at the University of Chicago

**Anneila Sargent**
Benjamin M Rosen Professor of Astronomy at the California Institute of Technology

**Bernard Schutz**
Director, Max Planck Institute for Gravitational Physics, Potsdam

**Rainer Weiss**
Professor of Physics, Emeritus, Massachusetts Institute of Technology

**Beverly Berger**
Gravitational Physics Program Director, National Science Foundation

**Benedetto D'Ettorre Piazzoli**
Vice-President, INFN

**Stavros Katsanevas**
Deputy Director, In2p3/CNRS

**John Womersley**
Director, Science Programmes, Science and Technology Facilities Council

The GWIC roadmap committee would like to thank the above and is also grateful to Jim Faller (JILA), Eric Gustafson (California Institute of Technology), Martin Hendry, (University of Glasgow), Tom Prince (CalTech), Norna Robertson (California Institute of Technology), Ken Strain (University of Glasgow) and Benno Willke (AEI Hannover), for their comments and input to this document.

## A.4 Status of scientific results from ground-based interferometers

The most sensitive band of current ground-based gravitational wave interferometers extends between 50 Hz and 1500 Hz. In this band, the strongest sources of gravitational wave signals are expected to be from compact binary systems during their inspiral, coalescence and merger phases and from the oscillations of the object that forms after the merger. We expect gravitational waves to be emitted during supernova collapse events; we also expect emission of continuous gravitational waves from rapidly rotating neutron stars and possibly a stochastic gravitational wave background. Significant upper limits have been obtained for each of these sources.

The first interferometer to make extended observations with a sensitivity for broadband bursts comparable to the bar detectors was the TAMA detector. This detector collected 1000 hours of data in 2001, and the data were used to set limits on bursts of gravitational waves incident on the earth in a 500 Hz wide band centered on 1 kHz. Although generic searches such as this are difficult to interpret in terms of sources, the sensitivity level was typical of what might be emitted by a binary inspiral in our galaxy or a very nearby supernova. No sources were found, setting a rate limit for such strong sources.

Recent searches with LIGO, Virgo, and GEO600 have greatly improved on these early results. LIGO's two-year long S5 science run, the longest run so far, included the participation of GEO600. The latter part of this run was joined by Virgo in its first Science Run (VSR1). Results from S5/VSR1 are summarized below. However, progress is rapid. In July 2009 LIGO and Virgo started their next run (S6/VSR2) with "enhanced" LIGO and Virgo detectors, which have achieved somewhat better sensitivity and add to the chance of detecting the first signal in the near term.

Binary systems of compact objects evolve in orbits that gradually shrink in time due to the emission of gravitational radiation. The exact time-frequency evolution of the signal depends on a number of parameters, but systems with masses up to 200 solar masses are expected to emit signals with significant energy content in the frequency band of current ground-based detectors. The sensitivity of a search for binary inspiral signals may be characterized by its horizon distance. This is the distance at which an optimally located and oriented equal mass binary system is expected to produce a signal with matched-filter SNR = 8[2]. Figure A.4.1 shows the horizon distance as a function of mass of the binary system during the S5/VSR1 run, which comprise hundreds of galaxies for relatively low mass systems and many thousands of galaxies for higher-mass systems. Still, the detection of a gravitational wave signal from a binary inspiral signal is not at all ensured in S5/VSR1 data: the expected detection rates are 1 event per 5000 to 5 years for 1.4-1.4 solar mass systems; 1 event every 10000 to 5 years for 1.4-10 solar mass systems and 1 event every 5000 to 1 years for 10-10 solar mass systems. Enhancements to the current detectors are expected to achieve an improvement in strain sensitivity of a factor of ~2. With a horizon distance of 60 Mpc to neutron star systems the expected rates grow to 1 event every 60 to 4 years of actual observing time. Advanced detectors operating at a horizon distance of 450 Mpc to neutron star systems bring the expected detection rates between several to order hundred events per year of observing time.

[2] An average over source position on the sky and orientation of the binary system will reduce the average detected signal by a factor of approximately 2.26.

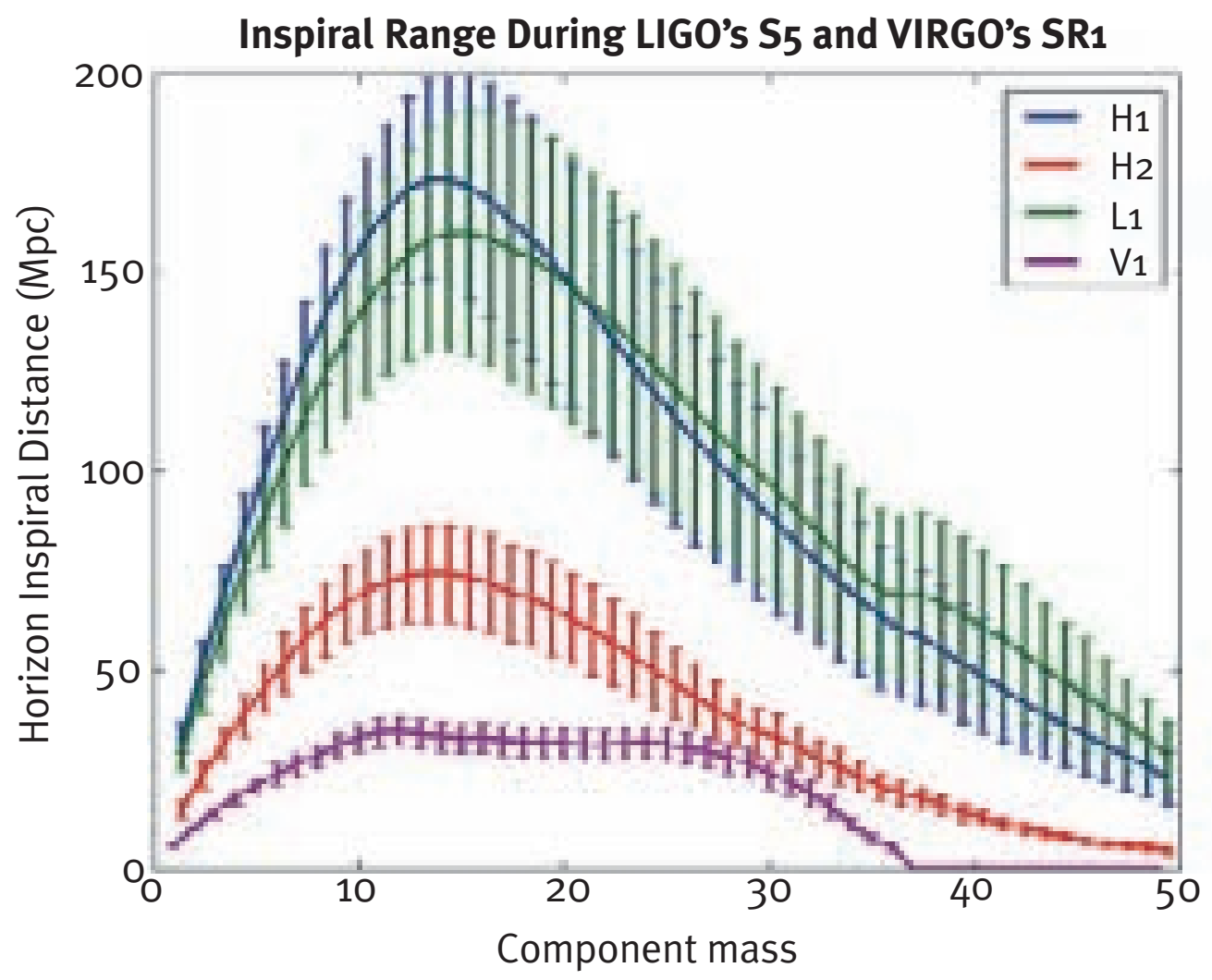

Fig. A.4.1 – Average horizon distance during the S5 run as a function of the total mass of the binary system, for equal mass systems. H1 and H2 are the Hanford 4- and 2-km baseline detectors. L1 is the 4-km detector in Louisiana.

Triggered searches that take place when an independent electromagnetic observation is available can be more sensitive than blind searches. In February 2007 a short hard GRB, GRB 070201, was detected and localized within an area which includes one of the spiral arms of the M31 Galaxy (only ~770 kpc distant). Since some short hard GRBs are thought to be produced in the merger phase of binary neutron star systems (BNS) or neutron star-black hole binaries (NSBH) this particular GRB was of great interest. An inspiral search was carried out on the available gravitational wave data for systems with component masses in the range 1-3 and 1-40 solar masses respectively but no signal was found. This null result excluded the possibility that this GRB could be due to a binary neutron star or NSBH inspiral signal in M31 with very high confidence (greater than 99%). This finding, along with other observational evidence, suggests that a soft gamma ray repeater (SGR) flare event in M31 is the most likely origin of this GRB. Over 170 other GRBs have also been examined, and upper limits placed on the possible gravitational-wave energy emitted by each.

A systematic search for gravitational wave signals associated with 191 SGR bursts was carried out using S5 data and prior data in coincidence with the 27 Dec. 2004 giant flare from SGR 1806-20. No signals were found and upper limits on the isotropic gravitational wave energy were placed. At a nominal distance of 10 kpc these upper limits overlap with the range of electromagnetic isotropic energy emission in SGR giant flares ($10^{44}$ - $10^{46}$ erg) and some of the upper limits on the ratio of the gravitational wave and electromagnetic energies are within the range of theoretically possible values. SGR flares and GRBs continue to be prime targets for gravitational wave searches as more are detected.

In general there are many circumstances in which short bursts of gravitational waves are expected, lasting from a few milliseconds to a few seconds such as the merger phase of a binary system or the collapse of a stellar core. Blind searches for these types of events are routinely carried out using robust algorithms which can detect arbitrary burst signals in the data. No such signals have been detected yet despite improvements in sensitivity. In the S5/VSR1 run, signals generated by converting of order 5% of a solar mass at the distance of the Virgo cluster, or $\sim 2\times10^{-8}$ of a solar mass at the Galactic center, would have been detected with 50% efficiency. Estimates of the expected amplitude of burst signals vary quite widely and scenarios exist which predict emission that is detectable at greater distances. For example, numerical simulations predict the emission of up to 3.5% of total mass in gravitational waves for black hole mergers. A system of this type formed by two 50 solar mass black holes at 180 Mpc would produce gravitational waves which could have been detected with 50% efficiency in S5.

An isotropic stochastic background of gravitational radiation is expected due to the superposition of many unresolved signals, both of cosmological and astrophysical origin. The background is described by a function $\Omega_{GW}(f)$, which is proportional to the energy density in gravitational waves per logarithmic frequency interval. The most recent results from a search for isotropic backgrounds come from the analysis of the S5/VSR1 data from LIGO and Virgo. For a flat gravitational wave spectrum this search puts a 95% Bayesian upper limit at

$$\Omega_{GW}\left[\frac{H_0}{72 kms^{-1}\ Mpc^{-1}}\right]^2 \le 6.5\times10^{[illegible]}$$

in the frequency range 49-169 Hz. This limit has surpassed the bounds that may be inferred from measurements of light-element abundances, WMAP data and the big bang nucleosynthesis model, and is starting to rule out some of the theoretical models of the stochastic gravitational-wave background (such as some cosmic (super)string models).

Fast rotating neutron stars are expected to emit a continuous gravitational wave signal if they present a deviation from a perfectly axisymmetric shape, if their r-modes are excited, or if their rotation axis is not aligned with their symmetry axis. In all cases the expected signal at any given time is orders of magnitude smaller than any of the short-lived signals that have been described above. However, since the signal is present for a very long time (to all practical purposes, in most cases, one may consider it there all the time), one can increase the SNR by integrating for a suitably long time.

LIGO and Virgo have performed searches for continuous gravitational waves from known radio pulsars, but no gravitational waves have been detected. This is not unexpected because for most systems the indirect upper limit on the amplitude of gravitational waves that one may infer from the measured spin-down rate of the systems is more constraining than the limit determined by the gravitational wave observations. In one case, however, gravitational wave observations are actually beating the electromagnetic spin-down limit and probing new ground. This is the case of the Crab pulsar. LIGO observations from the full S5 data beat the spin-down upper limit (in amplitude) by a factor of about 7, assuming phase coherence between the gravitational wave and radio signals. Hence, the gravitational wave luminosity is constrained to be less than about 2% of the observed spin-down luminosity. Limits on other known pulsars, albeit not beating the spin-down upper limits, reach values as low as a few $10^{-26}$ in the intrinsic gravitational wave amplitude $h_0$ and several $10^{-8}$ in ellipticity $\varepsilon$. These results show that at the current sensitivity, it is possible that LIGO and Virgo could, in principle, detect a continuous gravitational wave signal, coming from an unusually nearby object, unknown electromagnetically and rotating close to 75 Hz. The extended sensitivity of Virgo to lower frequencies makes it of great value in searching for gravitational waves from pulsars.

The most promising searches look for previously unknown objects, and are often referred to as blind searches. To search every direction and frequency with a 2-year integration time is computationally intractable because of the need to make extremely fine, direction-dependent Doppler modulations due to the Earth's motion. So sensitivity must be sacrificed via incoherent methods, such as summing many strain powers measured over short integration times. All-sky upper limits on strain for circularly polarized continuous waves (optimal orientation) are shown in Fig. A.4.2 for the 50-1100 Hz band, based on the first eight months of S5 data using thousands of summed strain powers with coherence times of 30 minutes (PowerFlux method). The deepest blind searches require an enormous amount of computational power and in fact are carried out by Einstein@Home, a distributed computing project that uses compute cycles donated by the general public. Einstein@Home is the second largest public compute project in the world and delivers an average 100Tflops of compute power continuously. The increased computational power of Einstein@Home permits the use of a hierarchical algorithm that starts with coherence times of 25 hours. Fig. A.4.2 show the resulting expected full-S5 sensitivity from Einstein@Home.

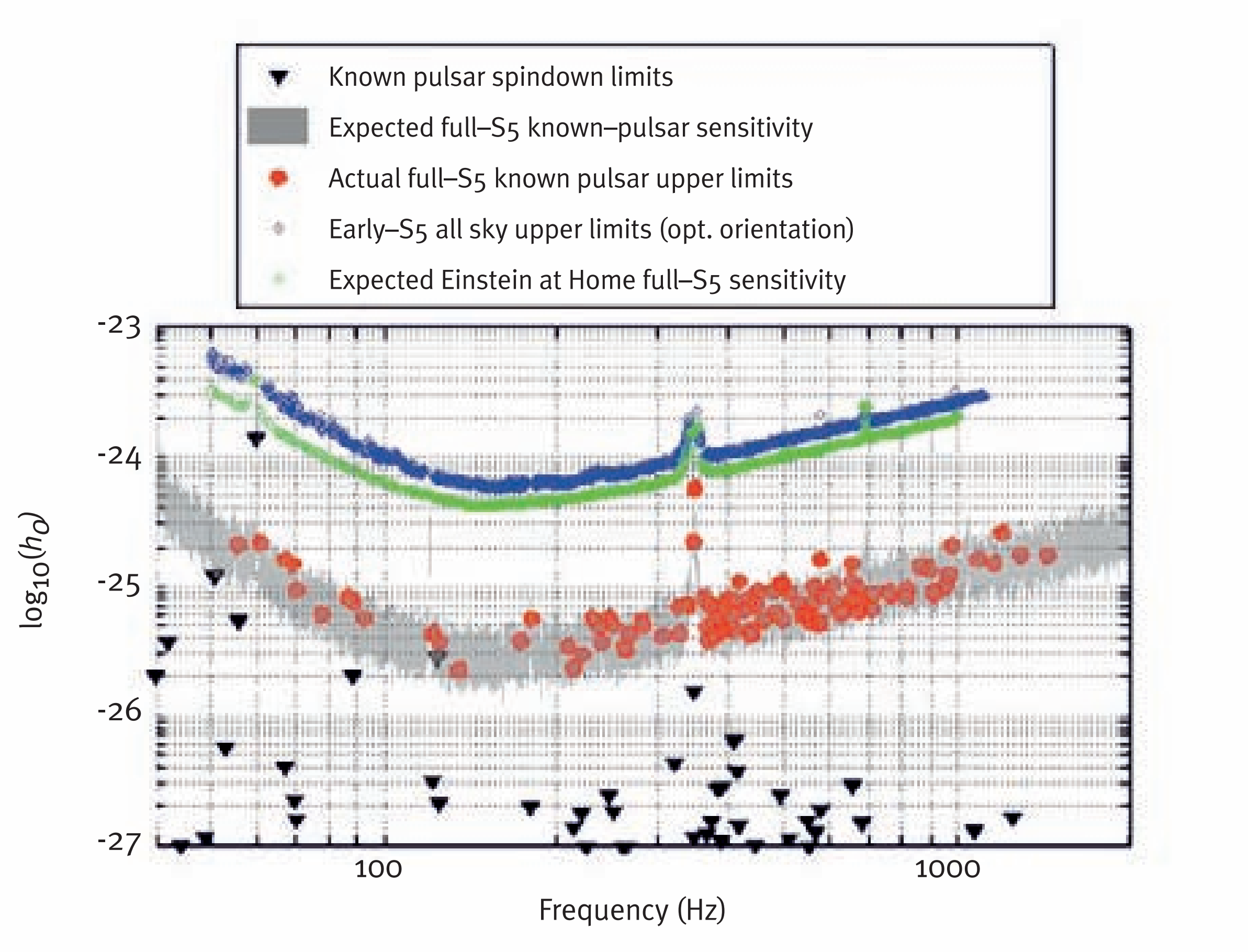

Fig. A.4.2 – Spindown limits for known pulars and LIGO S5 strain search sensitivities for known and unknown pulsars

## A.5 List of Acronyms

**AEI** – Albert Einstein Institute

**AIGO** – Australian International Gravitational Observatory

**ALIA** – Advanced Laser Interferometer Antenna – one proposal for a space-based detector for the frequency range 30mHz to 1 Hz

**ALLEGRO** – resonant bar detector based in Louisiana, USA and operated by Louisiana State University

**ATNF** – Australian Telescope National Facility

**AURIGA** – resonant bar detector based in Legnaro, Italy operated by INFN

**BBH** – Binary Black Hole

**BBO** – Big Bang Observer

**BEPAC** – Beyond Einstein Programme Advisory Committee

**BH** – Black Hole

**BNS** – Binary Neutron Star

**BOINC** – Berkeley Open Infrastructure for Network Computing

**BSE** – Binary Star Evolution

**CASSINI** – joint NASA/ESA/ASI space mission to explore Saturn and its moons

**CBC** – Compact Binary Coalescence

**CLIO** – Cryogenic Laser Interferometer Observatory

**CMB** – Cosmic Microwave Background

**CMBPol** – Cosmic Microwave Background Polarimeter

**CNRS** – Centre National de la Recherché Scientifique

**CSIRO** – Commonwealth Scientific and Industrial Research Organisation

**CXC** – Chandra X-ray Centre

**D** – Germany

**DAGMan** – Directed Acyclic Graph MANager

**DC** – Direct Current

**DECIGO** – DECi-hertz Interferometer Gravitational-wave Observatory for space

**DPF** – DECIGO Path Finder

**DUAL** – two nested resonant bar detectors

**DUAL Mo** – two nested resonant bar detectors, the interior made from Molybdenum

**DUSEL** – Deep Underground Science and Engineering Laboratory

**EBEX** – E and B EXperiment

**EGEE** – Enable Grid for E-sciencE

**ELT** – Extremely Large Telescope

**EM** – Electromagnetic

**EMRI** – Extreme Mass Ratio Inspiral

**EVLA** – European Very Large Array

**EPTA** – European Pulsar Timing Array

**ESA** – European Space Agency

**ET** – Einstein Telescope

**EXPLORER** – resonant bar detector based at CERN and operated by INFN

**F** – France

**FP** – Fabry-Perot

**FTP** – File Transfer Protocol

**GB** – Great Britain

**GEO600** – German-British gravitational-wave detector

**gLite** – middleware

**GR** – General Relativity

**GRB** – Gamma Ray Burst

**GriPhyN** – Grid Physics Network

**GW** – Gravitational Wave

**GWIC** – Gravitational Wave International Committee

**HF** – High Frequency

**I** – Italy

**IFU** – Integral Field Units

**IGEC** – International Gravitational Event Collaboration

**IMBH** – Intermediate Mass Black Hole

**INDIGO** – concept for a gravitational-wave detector in India

**INFN** – Instituto Nazionale di Fisica Nucleare

**IPTA** – International Pulsar Timing Array

**IR** – Infrared

**ISAS** – Institute of Space and Aeronautical Science

**IUPAP** – International Union of Pure and Applied Physics

**iVDGL** – international Virtual Data Grid Laboratory

**JAXA** – Japanese Aerospace eXploration Agency

**JDEM** – Joint Dark Energy Mission

**JILA** – An interdisciplinary institute for research and graduate education in the physical sciences, located on the campus of the University of Colorado, Boulder

**LAGEOS** – LAser GEOdetic Satellite

**LCGT** – Large-scale Cryogenic Gravitational-wave Telescope

**LDR** – LIGO Data Replicator

**LIGO** – Laser Interferometric Gravitational-wave Observatory

**LISA** – Laser Interferometer Space Antenna

**LMXB** – Low-Mass X-ray Binary

**LOFAR** – LOw Freqeuncy ARray

**LSC** – LIGO Scientific Collaboration

**M31** – Messier 31 Andromeda Galaxy

**miniGRAIL** – mini GRavitational Antenna In Leiden

**MIT** – Massachusetts Institute of Technology

**MSPs** – MilliSecond Pulsars

**NANOGrav** – North American Nanohertz Observatory for Gravitational-waves

**NAOJ** – National Astronomical Observatory of Japan

**NASA** – National Aeronautics and Space Administration

**NAUTILUS** – resonant bar detector at Laboratori Nazionali di Frascati Italy and operated by INFN

**Nd:YAG** – Neodymium-doped Yttrium Aluminium Garnet

**NIOBE** – resonant bar detector in Perth, Australia and operated by the University of Western Australia

**NL** – Netherlands

**NMR** – Nuclear Magnetic Resonance

**NS** – Neutron Star

**POLARBEAR** – project designed to measure the polarisation of the cosmic microwave background

**PPTA –** Parkes Pulsar Timing Array

**QND** – Quantum Non-Demolition

**QUIET** – the Q/U Imaging ExperimenT

**R&D** – Research & Development

**RF** – Radio Frequency

**RSE** – Resonant Sideband Extraction

**S5** – 5th LIGO Science Run

**S6** – 6th LIGO Science Run

**Sco-X1** – Scorpius X-1

**SGR** – Soft Gamma Repeater

**SGR A** – Sagittarius A

**SKA** – Square Kilometre Array

**SMBH** – Super Massive Black Hole

**SNR** – Signal to Noise Ratio

**SPIDER** – balloon-borne experiment to search for the inflationary cosmic gamma-ray background

**SFR** – stellar formation rate

**SQL** – Standard Quantum Limit

**SQUID** – Superconducting QUantum Interference Device

**TAMA** – Japanese gravitational-wave detector

**TDI** – Time Delay Interferometry

**TNO** – Netherlands Organisation for Applied Scientific Research

**TPD – TNO** – Institute for applied physics, part of TNO

**ULYSSES** – joint NASA/ESA space mission that will make measurements of the space above the Sun's poles

**US** – United States

**Virgo** – Italian-French - Dutch gravitational-wave detector

**VSR1** – Virgo Science Run 1

**WDBs** – White Dwarf Binaries

**WMAP** – Wilkinson Microwave Anisotropy Probe

**Yb** – Ytterbium

# Masthead

The GWIC Roadmap is printed on 100% recycled paper certified by the Forest Stewardship Council and awarded the EU-Ecolabel.

GWIC

GWIC

GRAVITATIONAL WAVE INTERNATIONAL COMMITTEE

www.gwic.ligo.org